\documentclass[twocolumn,times,tighten]{aastex701}

\hypersetup{linkcolor=blue,citecolor=blue,filecolor=cyan,urlcolor=blue}

\def\smallplottwo#1#2{\centering \leavevmode
\includegraphics[width=89.5mm]{#1}
 \hfil \includegraphics[width=89.5mm]{#2} }

\def\smallplotthree#1#2#3{\centering \leavevmode
\includegraphics[width=88.8mm]{#1} \includegraphics[width=88.8mm]{#2}
\includegraphics[width=88.8mm]{#3}}

\def\smallplotfour#1#2#3#4{\centering \leavevmode
\includegraphics[width=89.5mm]{#1} \includegraphics[width=89.5mm]{#2}
\hfil \includegraphics[width=89.5mm]{#3} \includegraphics[width=89.5mm]{#4}}

\def\smallplotsix#1#2#3#4#5#6{\centering \leavevmode
\includegraphics[width=89.5mm]{#1} \includegraphics[width=89.5mm]{#2}
\hfil \includegraphics[width=89.5mm]{#3} \includegraphics[width=89.5mm]{#4}
\hfil \includegraphics[width=89.5mm]{#5} \includegraphics[width=89.5mm]{#6}}

\def\smallploteight#1#2#3#4#5#6#7#8{\centering \leavevmode
\includegraphics[width=89.5mm]{#1} \includegraphics[width=89.5mm]{#2}
\hfil \includegraphics[width=89.5mm]{#3} \includegraphics[width=89.5mm]{#4}
\hfil \includegraphics[width=89.5mm]{#5} \includegraphics[width=89.5mm]{#6}
\hfil \includegraphics[width=89.5mm]{#7} \includegraphics[width=89.5mm]{#8}}

\def\largeplotone#1{\centering \leavevmode
\includegraphics[width=150mm]{#1} }

\graphicspath{{./}{figures/}}

\shorttitle{Properties of O-rich AGB stars in the LMC and the Milky Way}
\shortauthors{Suh} 

\begin{document}

\title{Properties of Oxygen-rich Asymptotic Giant Branch stars in the LMC and the Milky Way}

\author[orcid=0000-0001-9104-9763,sname='Suh']{Kyung-Won Suh}
\affiliation{Department of Astronomy and Space Science, Chungbuk National
University, Cheongju, 28644, Republic of Korea}
\email[show]{kwsuh@chungbuk.ac.kr}

\begin{abstract}
We investigate the properties of oxygen-rich asymptotic giant branch (OAGB)
stars in the LMC and the Milky Way using infrared (IR) color–magnitude
diagrams, two-color diagrams, spectral energy distributions (SEDs), and
variability data. The observed SEDs are analyzed with radiative transfer models
to constrain the properties of the central stars and their circumstellar dust
shells and to estimate distances. For OAGB stars in the LMC, whose distance is
well established, a set of representative OAGB models is derived by fitting
theoretical SEDs to the observations. These models are subsequently applied to
Galactic OAGB stars, for which Gaia DR3 parallaxes are often uncertain, to
estimate distances from SED fitting. For Galactic Mira variables, distances are
also determined using a period–magnitude relation calibrated with Mira stars in
the LMC. We find that OAGB stars in the two galaxies exhibit broadly similar IR
properties, and that the theoretical models successfully reproduce the observed
IR characteristics, despite the extreme rarity of heavily obscured objects with
optically thick silicate dust envelopes in the LMC. We further examine Galactic
OAGB stars associated with OH, H$_2$O, and SiO masers, including the
relationship between absolute OH maser fluxes and IR colors. In addition, we
investigate the IR properties of Mira-type OAGB stars and characterize their
long-term variability using archival observations together with 16 years of
WISE photometric data. Finally, we present a new catalog of 70,877 Galactic
OAGB stars, providing a more reliable and comprehensive dataset for future
studies.
\end{abstract}



\section{Introduction} \label{sec:intro}

The asymptotic giant branch (AGB) stage consists of two distinct phases: the
early AGB (E-AGB) and the thermally pulsing AGB (TP-AGB) (e.g.,
\citealt{iben1983}). All E-AGB stars are generally believed to be oxygen-rich AGB
(OAGB) stars, while carbon-rich AGB (CAGB) stars are thought to evolve as the
successors of intermediate-mass OAGB stars ($\sim$2$-$4.5 $M_{\odot}$ for solar
metallicity) when thermal pulses occur (e.g., \citealt{karakas2014}). More
massive OAGB stars enter the TP-AGB phase without changing the chemistry, where
they experience higher mass-loss rates and develop thicker dust envelopes. Almost
all AGB stars are long-period variables (LPVs; e.g., \citealt{hofner2018}). In
this paper, the term LPV is used collectively for OGLE small-amplitude red giants
(OSARGs), semiregular variables (SRVs), and Mira variables (e.g.,
\citealt{swu13}), whose light variations are not necessarily strictly periodic.

IR observational data from several missions, including the Wide-field Infrared
Survey Explorer (WISE), Two-Micron All-Sky Survey (2MASS), Akari, Infrared Space
Observatory (ISO), Midcourse Space Experiment (MSX), and Infrared Astronomical
Satellite (IRAS) have been instrumental in identifying OAGB stars and studying
their properties. Furthermore, the Near-Earth Object WISE Reactivation
(NEOWISE-R) mission (\citealt{mainzer2014}) conducted photometric observations at
3.4 and 4.6 $\mu$m from 2013 to 2024.

The Gaia Data Release 3 (DR3) provides valuable optical data for more than a
billion stars (\citealt{rimoldini2023}). The updated parallax distances
(\citealt{bailer-jones2021}) and extinction estimates (e.g.,
\citealt{lallement2022}) derived from Gaia DR3 enable more accurate
determinations of the absolute luminosities of many Galactic OAGB stars.

Although Gaia DR3 has substantially improved distance estimates for many Galactic
OAGB stars, reliable parallaxes remain unavailable or highly uncertain for a
large fraction of the sample, mainly because their extended circumstellar dust
envelopes leave them faint at optical wavelengths and because their optical
photocenters may have been displaced by large convective cells (e.g.,
\citealt{suh2024a}; see Section~\ref{sec:dist-ext}). In this study, we use the
well-established distance to the LMC to calibrate the luminosities of our OAGB
models and to quantify the uncertainty of the SED fitting. The calibrated models
are then applied to Galactic OAGB stars, whose distances are derived by fitting
their observed SEDs. For those identified as Mira variables, we also derive
alternative distance estimates from the period–magnitude relation (PMR)
established for LMC Miras.

We investigate the properties of OAGB stars in the LMC and the Milky Way.
Section~\ref{sec:sample} introduces catalogs of known OAGB stars in both
galaxies. Section~\ref{sec:agbmodels} describes the radiative transfer models of
circumstellar dust shells and the calibration of the model parameters using LMC
OAGB stars with well-determined distances. In Section~\ref{sec:seds}, the
observed spectral energy distributions (SEDs) are compared with theoretical
models to constrain the properties of the central stars and their circumstellar
dust envelopes and to estimate distances.

Section~\ref{sec:irpro} compares the observed IR two-color diagrams (2CDs) and
color-magnitude diagrams (CMDs) with theoretical model predictions.
Section~\ref{sec:comoc} discusses the differences between OAGB and CAGB
populations in the LMC and the Milky Way. Section~\ref{sec:maser} examines the
properties of OH, H$_2$O, and SiO maser sources associated with Galactic OAGB
stars. Section~\ref{sec:spatial} examines the spatial distribution of Galactic
OAGB stars in the Milky Way. Section~\ref{sec:pul} presents the PMR of Mira-type
OAGB stars. Section~\ref{sec:catalog-oagb} presents a new catalog of Galactic
OAGB stars. Finally, Section~\ref{sec:sum} summarizes the main results.

\section{Sample stars}\label{sec:sample}

We compiled samples of OAGB stars in the LMC and the Milky Way from published
catalogs (see Table~\ref{tab:tab1}). For each object, counterparts were
identified in Gaia DR3, 2MASS, IRAS, MSX, ISO, AKARI, WISE, Spitzer, SIMBAD, and
the American Association of Variable Star Observers (AAVSO; International
Variable Star Index, version 2026 March 22; \citealt{watson2006}). All sample
objects are referenced by their unique SIMBAD identifiers, and their Gaia DR3
counterparts were verified through SIMBAD.

To better understand the nature and evolutionary pathways of OAGB stars in both
galaxies, it is essential to compare their properties with those of other closely
related classes of evolved stars. Therefore, this study also includes CAGB stars
in both galaxies, as well as post-AGB (PAGB) stars and planetary nebulae (PNe)
within the Milky Way.

\begin{table*}
\footnotesize
\centering
\caption{Sample of AGB stars in the LMC and AGB, PAGB, and PNe objects in the Milky Way\label{tab:tab1}}
\begin{tabular}{llllllllll}
\hline \hline
Group & Reference & Number & Selected & IRAS  & Gaia DR3   & AllWISE & OH$^f$ & H$_2$O & SiO \\
\hline
LMC-OAGB   &\citet{riebel2012}  & 26,232  & 13,070$^a$  & 7  & 12996  & 12769 & & & \\
LMC-OAGB-S$^b$ &\citet{jones2017}  & 77   & 41  & 5  & 38  & 40  &   & & \\
LMC-OAGB-OH/IR$^b$  &\citet{suh2020}  & 10  & 6 & 5  & 4   & 6   & 6 & & \\
\hline
LMC-CAGB  &\citet{suh2026}  & 7347  & 7347  & 86  & 7322  & 7306  &  & & \\
\hline
MW-OAGB\_IC  & \citet{suh2024b}  & 18,684 & 19,784$^c$ & 19,784 & 18,103  & 18,792  & 1938  & 531  & 2488 \\
MW-OAGB\_IC-S$^{d}$ &  &   & 228   &  228   & 172   & 206  & 136 & 67  & 119 \\
MW-OAGB\_IC-EOH$^{d}$    &  &   & 34   &  34  & 1  & 29 & 34  & 12 & 24 \\
MW-OAGB\_NI   &\citet{suh2024b} & 45,413 & 51,093$^c$ & -  &49,976  &50,205 &  462 & 1 & 487\\
\hline
MW-CAGB\_IC & \citet{suh2024a}  & 4909 & 4909  & 4909 & 4862  & 4752  & & & \\
MW-CAGB\_NI & \citet{suh2024a}  & 2254 & 2254  & -    & 2254  & 2254  & & & \\
\hline
MW-PAGB  & SIMBAD & 311  & 310$^{e}$      &201  &242  &309  & 51 & 7 & 2 \\
MW-PNe   & SIMBAD & 12,936  & 2607$^{e}$  &1092 &1618 &2607 & 17 & 1 & 1 \\
\hline
\end{tabular}
\begin{flushleft}
$^a$Only bona fide LMC-OAGB stars were selected (see Section~\ref{sec:lmc-agb} for details).
$^b$Subgroups of LMC-OAGB objects (see Section~\ref{sec:lmc-agb}).
$^c$Completely updated and revised with data from SIMBAD, literature, and lists of maser sources (see Section~\ref{sec:gagb}).
$^d$Subgroups of OAGB\_IC objects (see Section~\ref{sec:gagb}).
$^e$See Section~\ref{sec:pagb}.
$^f$For the Galactic objects, sources detected in the OH 1612 MHz maser line in the database of \citet{engels2015}
(see Section~\ref{sec:omaser} and Table~\ref{tab:tab2}).
\end{flushleft}
\end{table*}

\begin{table*}
\footnotesize
\setlength{\tabcolsep}{4pt}
\centering
\caption{OH, H$_2$O, and SiO Maser sources in Galactic OAGB stars\label{tab:tab2}}
\begin{tabular}{lllllllllll}
\hline  \hline
Maser  &Reference  &   Original    & Observed$^a$  & Detected   &   Duplicate   &   Remaining  &  MW-OAGB  &MW-PAGB & MW-PNe & Other$^b$\\
\hline
OH  & \citet{engels2015}$^c$ & 16,879$^d$ & 6598 & 2730 & - & 2730 & 1938+462 & 51 & 17 & 262 \\
\hline
H$_2$O & \citet{ks2012} & 902  &  902 &  452 &  0 &  452  &  &&  &\\
 & KVN$^e$              & 953  & 953 &   297  & 197  &  100  &  &&  & \\
 & \citet{yung2014}     & 106  & 106  &  24   &  5  &  19  &  && &\\
 & \citet{wu2022}       & 175  & 175  &  20   &   1  &   19  &  &&  &\\
total & -  &  - &  -   &  - & - &  590  & 531+1  & 7 & 1 & 50\\
\hline
SiO & \citet{ks2012}   & 1774 &  1627  & 1627  &   0 &  1627  &  &&& \\
 & \citet{deguchi2012}$^f$  & 510 &  510  &  313   &  309  &  4  &  &&&\\
 & KVN$^e$                & 953 & 953  & 522   &  333  &  189  &   &&& \\
 & \citet{messineo2018}   & 571  & 570  &  361   &  16  &  345  &  &&&  \\
 & \citet{wu2018}    & 221 & 221 &   43   &   14  &   29  &  &&&  \\
 & \citet{stroh2019}   &   1427  & 1420 &  1025   &  23  &  1002  &  &&&   \\
 & \citet{rizzo2021}   &   67 & 67  &  51   &  45  &  6  &   &&&  \\
 total & -  &  - &  -   &  - &  - & 3202  & 2488+487 & 2  &  1 & 224\\
\hline
\end{tabular}
\begin{flushleft}
$^a$The number of observed sources after removing duplicates based on SIMBAD
identifiers. $^b$Detected sources that belong to none of the three groups (e.g.,
red supergiants, young stellar objects, and unclassified objects). $^c$Version
2.5 (updated on March 19, 2023). The database lists individual
observations rather than sources: its 16,879 entries refer to 6900 distinct
objects, or 6874 sources after the names were resolved to SIMBAD identifiers and
the observations of one object listed under different identifiers were merged;
6598 of them were observed at 1612 MHz. A source is counted as detected when it
was detected at 1612 MHz in at least one observation according to the detection
flags of the database; sources observed only in the 1665 and 1667 MHz main lines
are not counted (see Section~\ref{sec:omaser}). $^d$The number of
observations, not of sources. $^e$Korean VLBI Network (\citealt{kim2010};
\citealt{cho2012}; \citealt{kim2013}; \citealt{yoon2014}; \citealt{cho2017}).
$^f$Includes the data from \citet{deguchi2010}.
\end{flushleft}
\end{table*}

\subsection{AGB stars in the LMC}\label{sec:lmc-agb}

For the LMC, \citet{riebel2012} and \citet{jones2017} used data from the Spitzer
Space Telescope legacy program Surveying the Agents of Galaxy Evolution (SAGE) to
compile a catalog of 26,232 OAGB stars (LMC-OAGB; Table~\ref{tab:tab1}). Among
these, 77 sources (LMC-OAGB-S) were confirmed through Spitzer IRS spectroscopy to
exhibit prominent silicate dust features, forming a particularly reliable subset
of TP-AGB OAGB stars.  For each object, we retrieved the corresponding SIMBAD
entry. The LMC-OAGB-S sample is a subset of the LMC-OAGB catalog (see
Table~\ref{tab:tab1}). The LMC-OAGB-OH/IR objects, which exhibit OH maser
emission (see \citealt{suh2020}), constitute a further subset of the LMC-OAGB-S
sample. OH/IR stars are typically regarded as more massive OAGB stars,
characterized by denser dust envelopes and elevated mass-loss rates (e.g.,
\citealt{engels1983}; \citealt{justtanont2015}; \citealt{suh2021}).

\citet{Jiménez-Arranz2023} performed a comprehensive kinematic study of the LMC
using Gaia DR3 proper motions and developed an efficient method for
distinguishing LMC stars from Milky Way foreground contaminants. To improve the
purity of our LMC-OAGB sample, we applied their membership probabilities to the
Gaia DR3 counterparts and retained only sources with $PM > 0.9$, where $PM$
denotes the LMC membership probability. This conservative threshold effectively
removes foreground interlopers while preserving the majority of genuine LMC
members.

After excluding stars classified as non-members and removing objects with SIMBAD
types inconsistent with OAGB classifications, we obtained a final sample of
13,070 bona fide LMC-OAGB stars from the initial set of 26,232 candidates (see
Table~\ref{tab:tab1}). The Optical Gravitational Lensing Experiment III
(OGLE-III) presented 91,995 LPVs in the LMC \citep{sus09}. There are 175 OGLE-III
Mira variables in the LMC-OAGB sample.

For LMC-CAGB stars, we use the catalog of CAGB stars in the LMC compiled by
\citet{suh2026}.

To derive the absolute magnitudes of AGB stars in the LMC, we adopt a distance of
49.6 kpc, consistent with the precise measurement of 49.59 $\pm$ 0.54 kpc
reported by \citet{pietrzynski2019}.

For LMC-OAGB stars, we did not make corrections for interstellar extinction. The
foreground and internal extinction is typically $A_V \lesssim 0.5$~mag (e.g.,
\citealt{skowron2021}), corresponding to $A_K \lesssim 0.04$~mag. Since the
scaling distances are determined from photometry at $\lambda \ge 2~\mu$m
(Section~\ref{sec:seds}), this extinction is negligible compared with the IR
color range of the sample. The Gaia-band extinction ($A_G \lesssim 0.4$~mag) is
comparable to the model--observation discrepancies at visual wavelengths
(Section~\ref{sec:limit}).

\subsection{AGB stars in the Milky Way}\label{sec:gagb}

\citet{suh2021}, \citet{suh2024a}, and \citet{suh2024b} compiled catalogs of AGB
stars in our Galaxy, classifying them into two groups: those with IRAS
counterparts (AGB\_IC objects), identified using the IRAS source catalog for
brighter or more isolated stars, and those without IRAS counterparts (AGB\_NI
objects), derived from the AllWISE or Gaia DR3 catalogs, which focus on fainter
stars or those in densely populated regions. The majority of AGB\_IC objects also
have entries in the AllWISE or Gaia DR3 catalogs (see Table~\ref{tab:tab1}).

For Galactic OAGB stars, we build upon the OAGB catalog of \citet{suh2024b},
incorporating substantial updates based on data from SIMBAD, recent literature,
and catalogs of OH, H$_2$O, and SiO maser sources (see Section~\ref{sec:omaser}).
In this work, we adopt a sample of 70,877 MW-OAGB stars, consisting of 19,784
MW-OAGB\_IC and 51,093 MW-OAGB\_NI objects (see Table~\ref{tab:tab1}).

The classification of OAGB stars is based on the following criteria: (1) they
display spectroscopic features characteristic of O-rich stars (see
Section~\ref{sec:specdata}) or are identified as OH, H$_2$O, and SiO maser
sources (see Section~\ref{sec:omaser}); (2) their IR properties, including 2CDs,
CMDs, and SEDs, are consistent with those of established OAGB stars; (3) their
absolute magnitudes in W3[12] are brighter than -5.2 mag (refer to
Section~\ref{sec:lum-loagb}); and (4) Mira variables qualify as AGB stars (see
Section~\ref{sec:pul}). The classification criteria were applied using either AND
or OR operations, depending on the availability of the required data. In most
cases, the AND operation was used. In certain cases, the OR operation was applied
with great caution, guided by information from SIMBAD and relevant literature.

A subset of 228 MW-OAGB\_IC stars that have been well studied and possess
detailed observed SEDs, including IR spectral data (see Section~\ref{sec:seds}),
is designated as MW-OAGB\_IC-S. Among these, 34 objects that show 1612 MHz OH
maser emission (Section~\ref{sec:omaser}) are classified as MW-OAGB\_IC-EOH,
representing extreme OH/IR stars with thick dust envelopes ($\tau_{10} \geq 10$;
see Table~\ref{tab:tab1}).

For the Galactic CAGB sample, we adopt 7163 MW-CAGB stars (4909 MW-CAGB\_IC and
2254 MW-CAGB\_NI; see Table~\ref{tab:tab1}) from the catalog of \citet{suh2024a},
using the revised distance estimates derived from CAGB model SEDs by
\citet{suh2026}.

Many previous catalogs of AGB stars contain red supergiants (RSGs). We have
carefully checked all sample stars against the SIMBAD database and excluded all
objects identified as RSGs from both the Galactic and LMC samples presented in
this work. Residual contamination by unidentified RSGs cannot be fully excluded,
but it is expected to be statistically negligible.

\subsection{PAGB stars and PNe in the Milky Way}\label{sec:pagb}

We selected 310 PAGB stars from the SIMBAD list of 311 objects. One object (IRAS
11415-6541) was excluded because it does not exist and was mistakenly included in
the SIMBAD database.

We selected 2607 PNe from the SIMBAD database, which contains a total of 12,936
objects, using five key catalogs of Galactic PNe: the Catalogue of Galactic PNe
(\citealt{kohoutek2001}; \citealt{kerber2003}), the Macquarie/AAO/Strasbourg
H$\alpha$ Catalog (MASH; \citealt{parker2006}; \citealt{miszalski2008}), the PNe
catalog by \citet{frew2013}, the Hong Kong/AAO/Strasbourg H$\alpha$ database
(HASH; \citealt{parker2016}), and the catalog of central stars of PNe compiled by
\citet{gonzales2021}.

\subsection{Catalogs of OH, H$_2$O, and SiO Maser sources} \label{sec:omaser}

Most OH, H$_2$O, and SiO maser sources are associated with OAGB stars, PAGB
stars, and PNe \citep[e.g.,][]{ks2012}. These maser sources provide valuable
probes of the dust and gas envelopes surrounding AGB stars and are particularly
useful for identifying OAGB stars, especially in the Milky Way.
Table~\ref{tab:tab2} lists the OH, H$_2$O, and SiO maser sources associated with
OAGB stars, PAGB stars, and PNe in the Milky Way compiled in this work.

For the OH masers, we use only the database of \citet{engels2015} (Version 2.5,
updated on March 19, 2023), which compiles the OH maser observations of
circumstellar envelopes published in the literature. The database lists
individual observations rather than sources: its 16,879 entries refer to 6900
distinct objects, each observed one or more times at 1612, 1665, and/or 1667 MHz,
and for each observation the database gives either the measured peak flux
densities of the blue- and red-shifted maser components or an upper limit. We
resolved the object names to SIMBAD identifiers, which merged a few objects
listed under different names; when the observations of one database object were
resolved to more than one SIMBAD identifier (typically an IRAS name and a 2MASS
name of the same star), they were merged under the identifier used in our
catalogs. This left 6874 distinct sources, of which 6598 were observed at 1612
MHz. Earlier compilations of OH maser sources \citep[e.g.,][]{ks2012} and the
OH/IR star classifications in SIMBAD are not used, so that the detection
information and the flux densities come from a single, reproducible source.

A source is regarded as an OH maser source (D in Tables~\ref{tab:tab1},
\ref{tab:tab2}, and \ref{tab:tab5}) when it was detected at 1612 MHz in at least
one observation according to the detection flags of the database, and as a
non-detection (N) when it was observed at 1612 MHz but only upper limits were
obtained; sources observed only in the main lines at 1665 and 1667 MHz are not
counted. Of the 6598 sources observed at 1612 MHz, 2730 are detected. The OH
maser peak flux density of a source is the largest peak flux density among its
1612 MHz detections.

For the H$_2$O and SiO masers, the catalogs listed in Table~\ref{tab:tab2} were
combined sequentially, starting with the compilation of \citet{ks2012} and adding
from each subsequent catalog the objects not already present. Duplicate entries
were identified by resolving the object names in each catalog to SIMBAD
identifiers; this procedure may miss a small number of duplicates listed under
names that SIMBAD does not resolve to the same object. For these masers, only the
information on whether a maser was detected or not is used, because flux
densities are not available in a homogeneous form. For each group in
Table~\ref{tab:tab2}, the numbers of observed and detected sources are given
after the duplicate entries were removed.

The maser catalogs listed in Table~\ref{tab:tab2} combine observations from many
telescopes and surveys with different sensitivities and target selections, so no
uniform detection limit can be defined and the detection statistics describe the
observed samples rather than unbiased detection rates for the OAGB population.
This applies also to the OH maser database, in which the upper limits of the
individual observations range from a few tens of mJy to a few Jy. The one effect
common to all of the surveys is that fainter maser sources are detected only at
smaller distances. The properties of the maser sources are discussed in
Section~\ref{sec:maser}.

\subsection{Infrared Photometric Data}\label{sec:photdata}

Table~\ref{tab:tab3} summarizes the visual and IR bands used to construct the
SEDs, CMDs, and 2CDs in this study. For each band, the reference wavelength
($\lambda_{ref}$) and zero-magnitude flux (ZMF) are listed. These quantities are
required for computing theoretical model colors (see Section~\ref{sec:agbmodels})
and constructing SEDs from the observed photometry. A detailed description of the
IR photometric data and ZMF values is provided in Section 2.1 of \citet{suh2021}.

This study incorporates only high-quality observational data across all
wavelength bands for photometric analysis (quality better than 1 for IRAS, AKARI,
and MSX data; quality better than U for WISE data).

\begin{table}
\setlength{\tabcolsep}{4pt}
\scriptsize
\caption{Visual and IR bands and zero magnitude flux values$^a$\label{tab:tab3}}
\centering
\begin{tabular}{lllll}
\hline \hline
Band &$\lambda_{ref}$ ($\mu$m)	&ZMF (Jy) &Telescope &Reference$^b$\\
\hline
Bp[0.5]  &0.511	&	3552	&	Gaia	& \citet{rimoldini2023}\\
G[0.6]  &0.622	&	3229	&	Gaia    & \citet{rimoldini2023}	\\
Rp[0.8] &0.777	&	2555	&	Gaia    & \citet{rimoldini2023}	\\
J[1.2]  &1.235	&	1594	&	2MASS	& \citet{cohen2003}\\
H[1.7]  &1.662	&	1024	&	2MASS	& \citet{cohen2003}\\
K[2.2]  &2.159	&	666.7	&	2MASS	& \citet{cohen2003}\\
W1[3.4]	&3.35	&	306.681	&	WISE    & \citet{jarrett2011}	\\
W2[4.6]	&4.60	&	170.663	&	WISE    & \citet{jarrett2011}	\\
S3[5.8] &5.731	&	115.0   &	Spitzer & A	\\
S4[8.0] &7.872  &	64.9	&	Spitzer & A	\\
MA[8.3]  &8.28	&	58.49	&	MSX    & \citet{egan2003}	\\
AK[9]   &9	    &	56.262	&	AKARI    & \citet{murakami2007}	\\
IA[12]	&12     &  28.3 	&	IRAS     & \citet{beichman1988}	\\
W3[12]$^c$	&12	    &  28.3 	&	WISE & \citet{jarrett2011}\\
MC[12.1]  &12.13	&	26.51	&	MSX    &  \citet{egan2003}       	\\
MD[14.7]  &14.65	&	18.29	&	MSX    & \citet{egan2003}	\\
AK[18]   &18	&	12.001	&	AKARI    & \citet{murakami2007}	\\
ME[21.3]  &21.34	&	8.8  	&	MSX    & \citet{egan2003}	\\
W4[22]	&22.08	&	8.284	&	WISE    & \citet{jarrett2011}	\\
S5[24]  &23.68  &	7.17	&	Spitzer & B	\\
IA[25]	&25 & 6.73 	&	IRAS      & \citet{beichman1988}	\\
IA[60]	&60 & 1.19 	&	IRAS      & \citet{beichman1988}	\\
AK[65]  &65	& 0.965	&	AKARI    & \citet{murakami2007}	\\
AK[90]  &90	& 0.6276	&	AKARI    & \citet{murakami2007}	\\
IA[100]	&100 & 0.43 	&	IRAS      & \citet{beichman1988}	\\
AK[140]  &140	&0.1859	&	AKARI    & \citet{murakami2007}	\\
AK[160]  &160	&0.1487	&	AKARI    & \citet{murakami2007}	\\
\hline
\end{tabular}
\begin{flushleft}
$^a$Spitzer data are used only for OAGB stars in the LMC.
$^b$A: \url{https://irsa.ipac.caltech.edu/data/SPITZER/docs/irac/iracinstrumenthandbook},
B: \url{https://irsa.ipac.caltech.edu/data/SPITZER/docs/mips/mipsinstrumenthandbook}.
$^c$See section 4.2 in \citet{suh2020}.
\end{flushleft}
\end{table}

\subsection{Cross-identification}\label{sec:cross}

This study uses the combined IRAS source catalog compiled by
\citet{abrahamyan2015} and \citet{suh2024a}, containing 345,164 sources,
including 73,769 associated sources from the PSC and FSC. Counterparts for these
IRAS sources were identified in the AllWISE, 2MASS, Akari, MSX, and Gaia DR3
catalogs, as described in Section 2.3 of \citet{suh2024a}.

For AGB stars with IRAS counterparts (OAGB\_IC and CAGB\_IC objects), we utilize
the AllWISE, 2MASS, and Gaia DR3 counterparts provided in the combined IRAS
catalog (\citealt{suh2024a}).

For AGB stars without IRAS counterparts (OAGB\_NI and CAGB\_NI objects), Gaia DR3
counterparts were identified using SIMBAD. The nearest AllWISE and 2MASS
counterparts were matched within a 3$\arcsec$ radius based on SIMBAD positions.

For PAGB stars and PNe, we use the Gaia DR3 and IRAS counterparts from SIMBAD and
related references. We identified the nearest AllWISE and 2MASS counterparts
within 3$\arcsec$ using the SIMBAD position.

\subsection{Spectroscopic data}\label{sec:specdata}

The SIMBAD astronomical database offers valuable information on the optical and
IR spectra of the sample stars. Additionally, AAVSO provides spectroscopic data
for a significant portion of the sample stars.

To identify key dust features in AGB stars, the IRAS Low Resolution Spectrograph
(LRS; $\lambda$ = 8$–$22 $\mu$m) data (\citealt{kwok1997}) are useful (see
Section~\ref{sec:seds}). Furthermore, the ISO Short Wavelength Spectrometer (SWS;
$\lambda$ = 2.4$–$45.4 $\mu$m) and Long Wavelength Spectrometer (LWS; $\lambda$ =
43$–$197 $\mu$m) data can be utilized for detailed analysis (e.g.,
\citealt{sylvester1999}).

The Gaia DR3 LPV catalog (\citealt{lebzelter2023}) provided low-resolution Red
Photometric (RP) spectra in the visual wavelength bands, enabling the
identification of carbon stars among the 1,720,558 LPV candidates.
\citet{suh2024a} utilized this data to identify CAGB stars in our Galaxy.
Accordingly, when selecting OAGB stars, objects with C-rich spectra from Gaia DR3
were excluded unless there was other evidence indicating O-rich spectra.

\subsection{Distances and Galactic extinction}\label{sec:dist-ext}

Recent parallax distance data from \citet{bailer-jones2021} (r\_med\_geo: the
median of the geometric distance) and extinction information from
\citet{lallement2022}, both derived from Gaia DR3, are crucial for determining
the absolute luminosities of many OAGB stars in our Galaxy. Gaia DR3 distances
from \citet{bailer-jones2021} are available for 17,710 of the 19,784 MW-OAGB\_IC
sources (89.5\%) and for 49,496 of the 51,093 MW-OAGB\_NI sources (96.9\%).

Large convective cells of AGB stars are known to displace the optical photocenter
(\citealt{chiavassa2018}; \citealt{beguin2024}), which degrades the astrometric
solutions of even nearby sources. Moreover, \citet{andriantsaralaza2022} found
that the parallax uncertainties of AGB stars in Gaia DR3 are underestimated, so
that distances become model-dependent when the relative parallax uncertainty
exceeds $\sim$20\%. We therefore assess the quality of the Gaia DR3 astrometry
before using these distances. Following \citet{andriantsaralaza2022}, we regard a
Gaia parallax as more reliable when the parallax is positive, the relative
parallax uncertainty is better than 20\% (\texttt{parallax\_over\_error} $> 5$),
and the renormalized unit weight error RUWE $< 1.4$. Only 1805 MW-OAGB\_IC
sources (9.1\% of the full sample) and 253 MW-OAGB\_NI sources (0.5\%) satisfy
all three conditions.

The much lower fraction for MW-OAGB\_NI is driven by distance rather than by
astrometric quality. The parallax uncertainties of the two samples are comparable
(median $0.132$ and $0.146$\,mas for MW-OAGB\_IC and MW-OAGB\_NI respectively),
but the median parallax of MW-OAGB\_NI is only $0.028$\,mas, a factor of 5.5
smaller than that of MW-OAGB\_IC ($0.153$\,mas) and well below its own
uncertainty. Only 0.7\% of the MW-OAGB\_NI sources therefore reach
\texttt{parallax\_over\_error} $> 5$, compared with 10.9\% for MW-OAGB\_IC.

Consistently, the RUWE criterion is not the main limitation for either sample: of
the sources with \texttt{parallax\_over\_error} $> 5$, 1805 of 2156 (83.7\%) in
MW-OAGB\_IC and 253 of 360 (70.3\%) in MW-OAGB\_NI also satisfy RUWE $< 1.4$. The
MW-OAGB\_IC sample as a whole nevertheless has systematically high RUWE (median
1.31), as expected for stars whose photocenter is displaced by convection. The
RUWE cut does not remove this effect but only its most extreme cases: even the
sources that pass the cut have a median astrometric excess noise of $0.61$\,mas,
far above that of ordinary field stars.

These criteria are, however, too restrictive to be applied to the sample as a
whole: they would exclude more than 90\% of the MW-OAGB\_IC sources and
essentially all of the MW-OAGB\_NI sources, and the few MW-OAGB\_NI sources that
survive may lie preferentially in crowded fields at low Galactic latitude, where
blending can distort the astrometric solution even when the formal quality
criteria are met. We therefore retain the full sample and mark the high-quality
subsample separately in the distance comparisons (Sections~\ref{sec:sed-dist} and
\ref{sec:dpmr}; Figures~\ref{f7} and \ref{f16}), quoting statistics for both.

Gaia operates primarily at visual wavelengths and therefore has limited
sensitivity to OAGB stars with very optically thick dust envelopes, leaving many
such objects without reliable Gaia measurements. Moreover, the Gaia counterparts
listed in SIMBAD are not always secure, as identifying the correct counterparts
can be challenging using available photometric data and theoretical models,
particularly in crowded fields. Together with the astrometric degradation
discussed above, these effects mean that Gaia distances are either unavailable or
of limited accuracy for a substantial fraction of our sample. Consequently,
distances estimated by fitting observed SEDs with theoretical models provide a
valuable alternative and, in many cases, may offer more robust estimates (see
Section~\ref{sec:seds}).

Using the derived distance and position of each Galactic OAGB star, the
interstellar extinction was estimated from the three-dimensional dust maps of
\citet{vergely2022}, distributed with the G-Tomo tool \citep{lallement2022}.
Beyond the boundaries of these maps, the extinction was supplemented with the
\texttt{Combined19} map of the \texttt{mwdust} package (\citealt{bovy2016}; which
merges the maps of \citealt{marshall2006}, \citealt{green2019}, and
\citealt{drimmel2003}). The resulting $A_V$ was converted to extinction in the
Gaia, 2MASS, and AllWISE (W1[3.4] and W2[4.6]) bands using the extinction law of
\citet{wang2019}, and applied to all the SEDs, CMDs, and 2CDs of Galactic OAGB
stars presented in this work, which are based on Gaia DR3, 2MASS, and AllWISE
photometry (see Table~\ref{tab:tab3}).

No correction was applied at wavelengths longer than 5~$\mu$m. At shorter
wavelengths the extinction law is smooth and well determined, whereas at longer
wavelengths it is dominated by the interstellar silicate absorption, whose
profile varies between sightlines, so that a correction based on a mean
extinction law would add uncertainty rather than reduce it. Moreover, the
extinction at these wavelengths is small: it is comparable to the extinction
neglected for the LMC sample (Section~\ref{sec:lmc-agb}) and much smaller than
the uncertainties of the model fits around the circumstellar silicate feature.
Because the extinction depends on the distance, the extinction estimate and the
SED fitting were iterated until the distances converged (see
Section~\ref{sec:seds}).

\begin{table}
\caption{Twenty five theoretical models in five groups for typical OAGB stars\label{tab:tab4}}
\centering
\begin{tabular}{lllll}
\hline \hline
Model$^a$ &Dust$^b$ &$\tau_{10}$ & $T_*$ (K) & $L_*$$^c$ \\
\hline
OM1a &	warm  	&	0.001	&	3000	&	3.6	\\
	 &	warm  	&	0.005	&	3000	&	3.6	\\
     &	warm  	&	0.01	&	3000	&	3.6	\\
	 &	warm  	&	0.02	&	3000	&	3.6	\\
\hline
OM2a &	warm  	&	0.03	&	2500	&	4.9	\\
     &	warm  	&	0.05	&	2500	&	4.9	\\
	 &	warm  	&	0.1	    &	2500	&	4.9	\\
	 &	warm  	&	0.2	    &	2500	&	4.9	\\
\hline
OM3a &	warm  	&	0.3	    &	2300	&	7.2	\\
     &	warm  	&	0.5	    &	2300	&	7.2	\\
\hline
OM4a &	cold  	&	1	&	2000	&	25.6	\\
	 &	cold  	&	2	&	2000	&	25.6	\\
     &	cold  	&	3	&	2000	&	25.6	\\
	 &	cold  	&	5	&	2000	&	25.6	\\
	 &	cold  	&	6	&	2000	&	25.6	\\
OM4f &	cold  	&	7	&	2000	&	25.6	\\
\hline
	 &	cold  	&	10	&	2000	&	30$^d$	\\
	 &	cold  	&	12	&	2000	&	30$^d$	\\
OM5c &	cold  	&	15	&	2000	&	30$^d$	\\
	 &	cold  	&	17	&	2000	&	30$^d$	\\
	 &	cold  	&	20	&	2000	&	30$^d$	\\
	 &	cold  	&	23	&	2000	&	30$^d$	\\
	 &	cold  	&	30	&	2000	&	30$^d$	\\
	 &	cold  	&	33	&	2000	&	30$^d$	\\
OM5i &	cold  	&	40	&	2000	&	30$^d$	\\
\hline
\end{tabular}
\begin{flushleft}
$^a$Seven representative OAGB models (OM1a, OM2a, OM3a, OM4a, OM4f, OM5c, and OM5i)
were used to produce CMDs and 2CDs (Figures~\ref{f2}, \ref{f3}, \ref{f7}, \ref{f8}, and \ref{f9}).
The model name OM$n$x denotes the x-th model (a, b, c, \ldots) of the $n$-th group.
$^b$Silicate dust from \citet{suh1999}.
$^c$Stellar luminosity ($10^{3} L_{\odot}$).
$^d$Because the LMC contains no OAGB stars with $\tau_{10} > 7$, the absolute
luminosity of the fifth group in the Table, which applies only to
Galactic OAGB stars, cannot be calibrated using the LMC sample. Therefore, its
adopted luminosity ($3 \times 10^{4}$~$L_{\odot}$) is more uncertain than those
of the other groups and should be interpreted with caution (see Section~\ref{sec:lum-loagb}).
For all models, $T_c$ = 1000 K.
See Section~\ref{sec:agbmodels} for details.
\end{flushleft}
\end{table}

\section{Theoretical Dust Shell Models}\label{sec:agbmodels}

To calculate theoretical model SEDs for OAGB stars, we use radiative transfer
models for spherically symmetric dust shells surrounding central stars. The
radiative transfer code RADMC-3D
(\url{http://www.ita.uni-heidelberg.de/~dullemond/software/radmc-3d/}), following
the approaches of \citet{sk2013} and \citet{suh2020}, is applied. Detailed
descriptions of the theoretical models and their limitations can be found in
\citet{suh2020}.

The dust shell is modeled with a continuous power-law density distribution ($\rho
\propto r^{-2}$) and a dust condensation temperature of $T_c=1000$ K. Its inner
and outer boundaries are set by dust temperatures of 1000 K and 30 K,
respectively. Spherical dust grains with a radius of 0.1~$\mu$m are assumed
throughout the shell. The dust optical depth, $\tau_{10}$, is defined at a
wavelength of 10~$\mu$m.

For OAGB stars, we adopt optical constants of warm and cold silicate dust from
\citet{suh1999}. Warm silicate is used for stars with thin dust shells
($\tau_{10} < 1$), while cold silicate is applied for stars with thick dust
shells ($\tau_{10} \geq 1$). The model parameters for OAGB stars are listed in
Table~\ref{tab:tab4}, with additional details provided in
Section~\ref{sec:lum-loagb}.

For CAGB stars, we apply the optical constants of amorphous carbon (AMC) from
\citet{suh2000}. The model parameters for seven representative CAGB models are
listed in Table 3 of \citet{suh2026}.

\begin{figure}
\centering
\smallplotthree{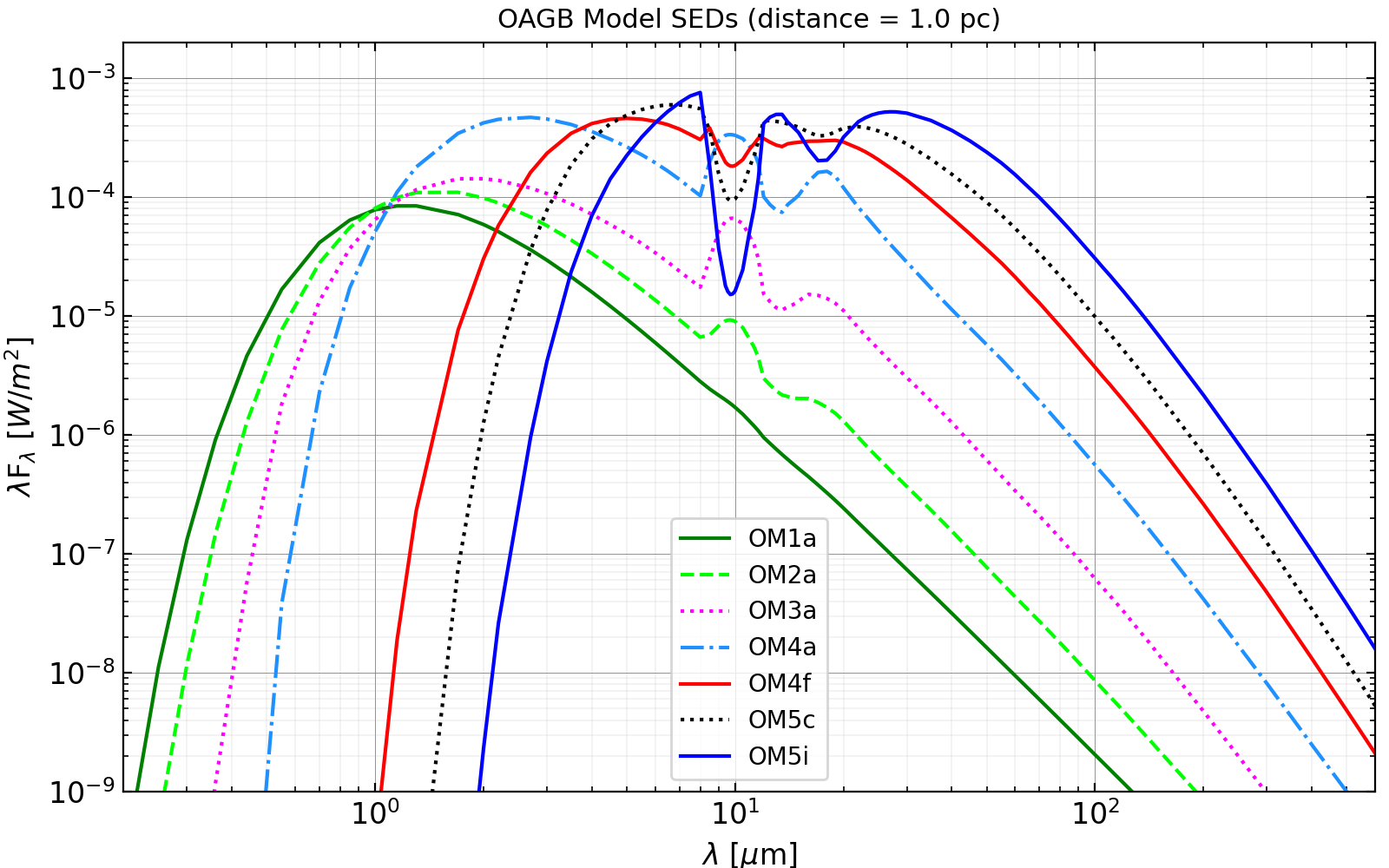}{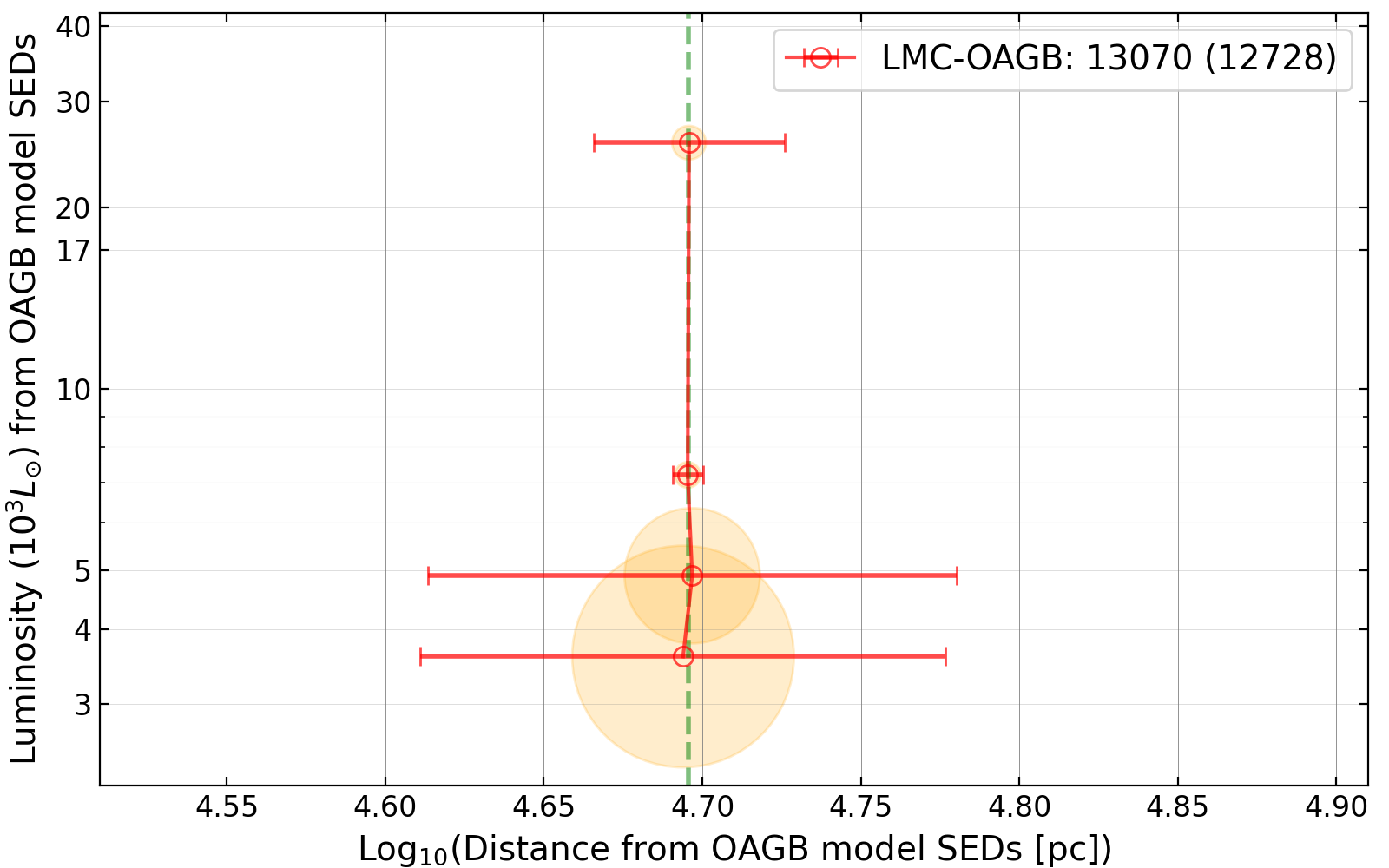}{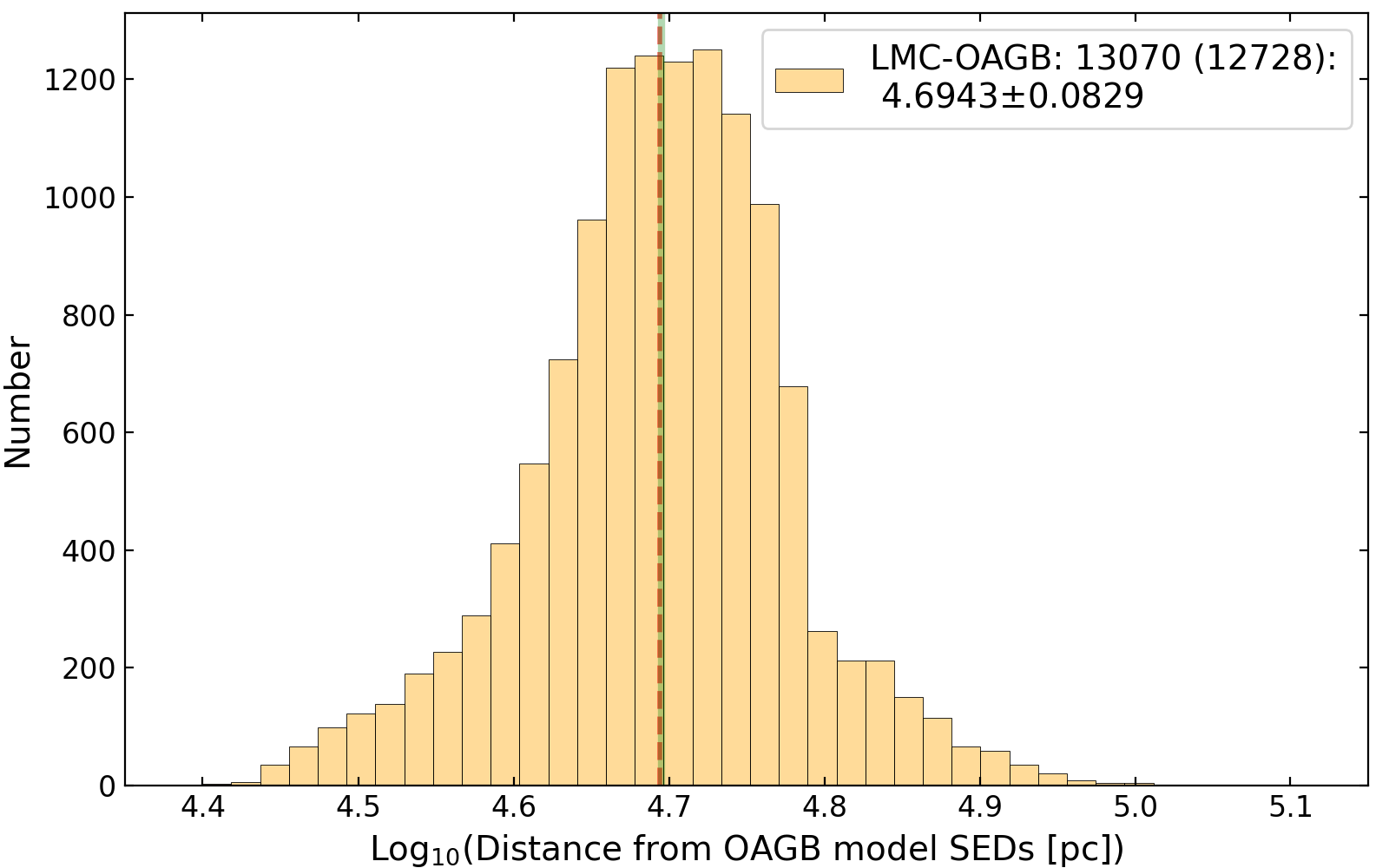}
\caption{The top panel shows OAGB model SEDs for seven representative models (see Table~\ref{tab:tab4}).
For LMC-OAGB stars, they show the distance distribution (middle) and
the relations between dust optical depth and luminosity (bottom)
derived from OAGB model SEDs (Table~\ref{tab:tab4}).
See Section~\ref{sec:lum-loagb} for details.
} \label{f1}
\end{figure}

\begin{figure*}
\centering
\smallplotfour{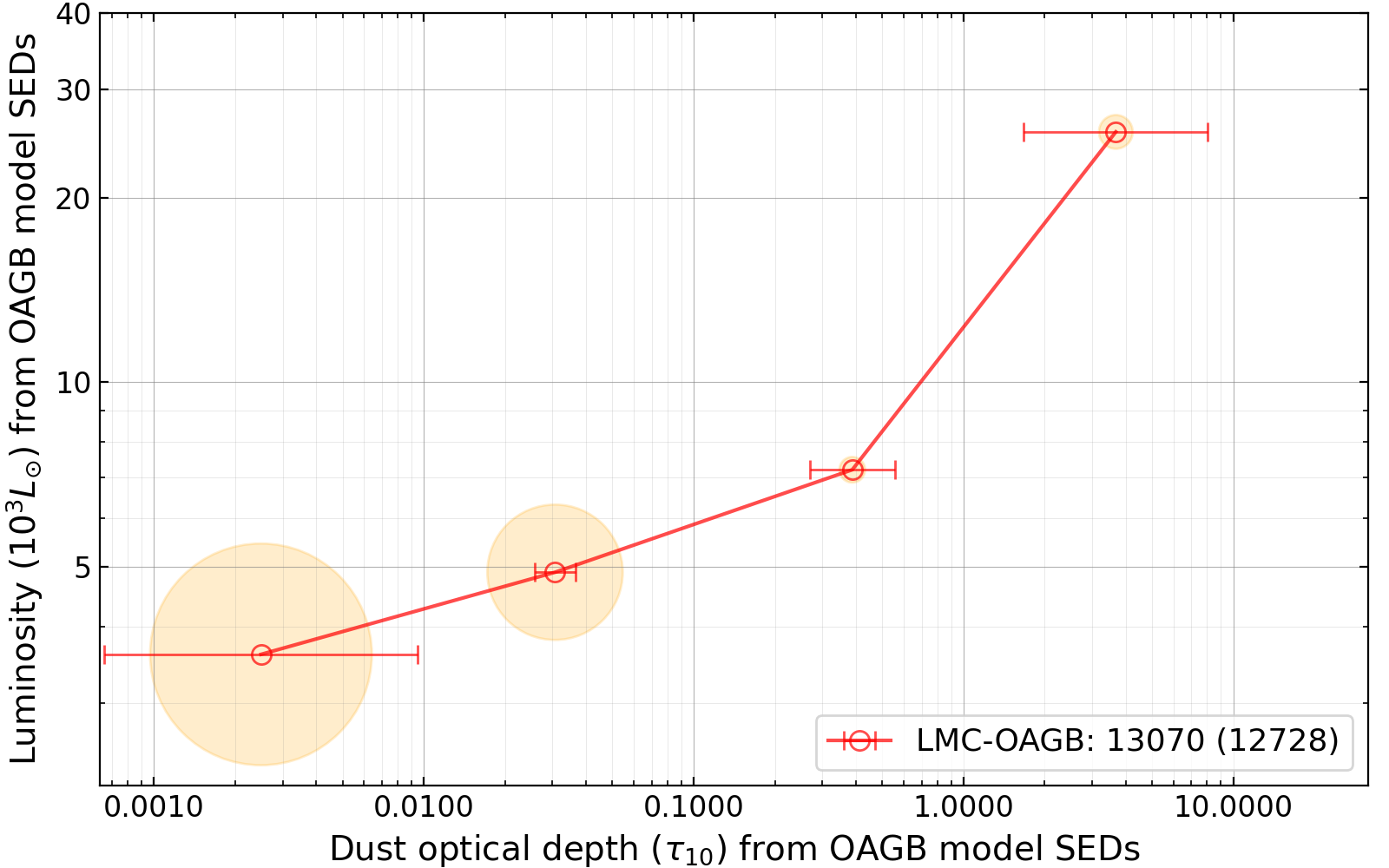}{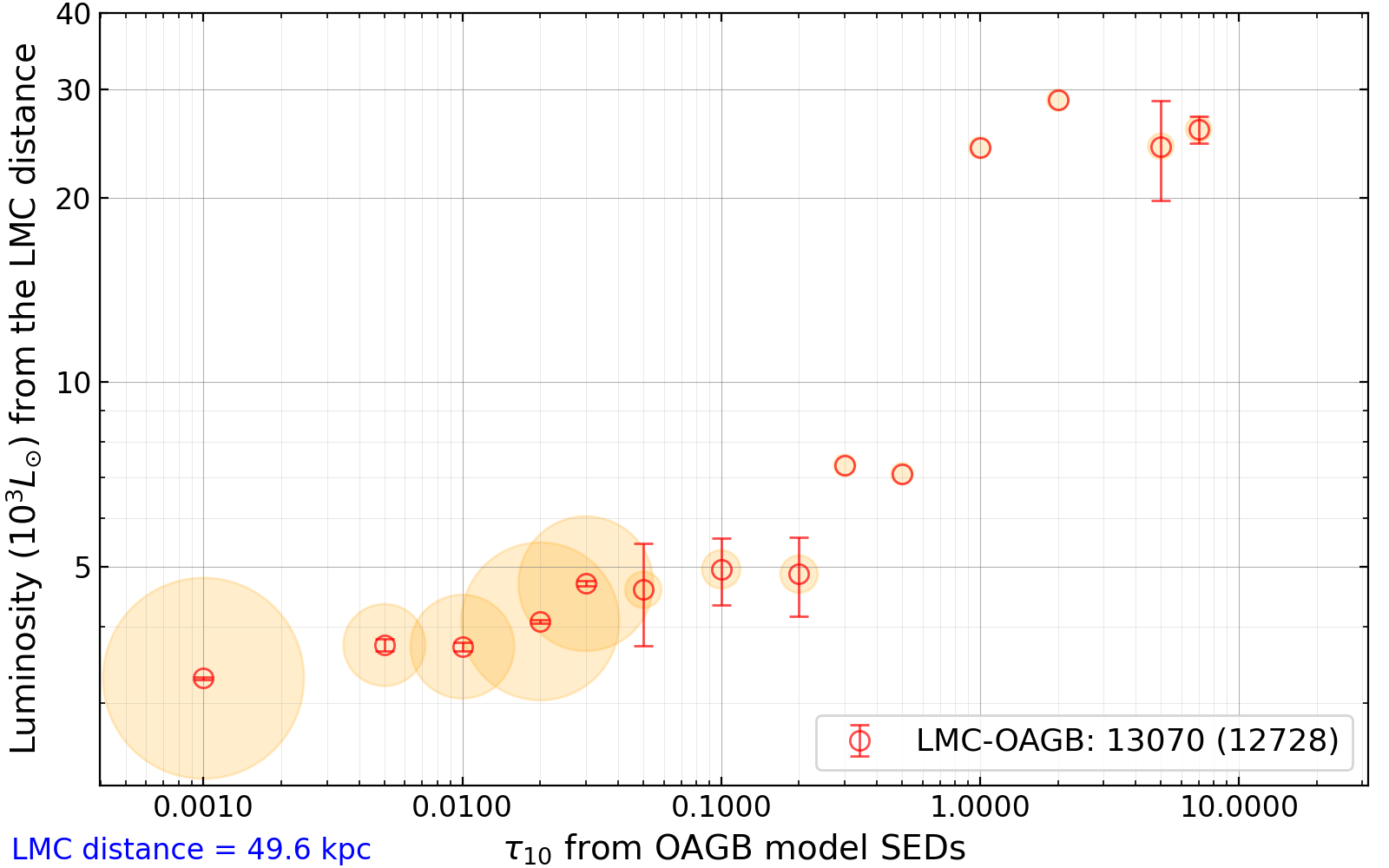}{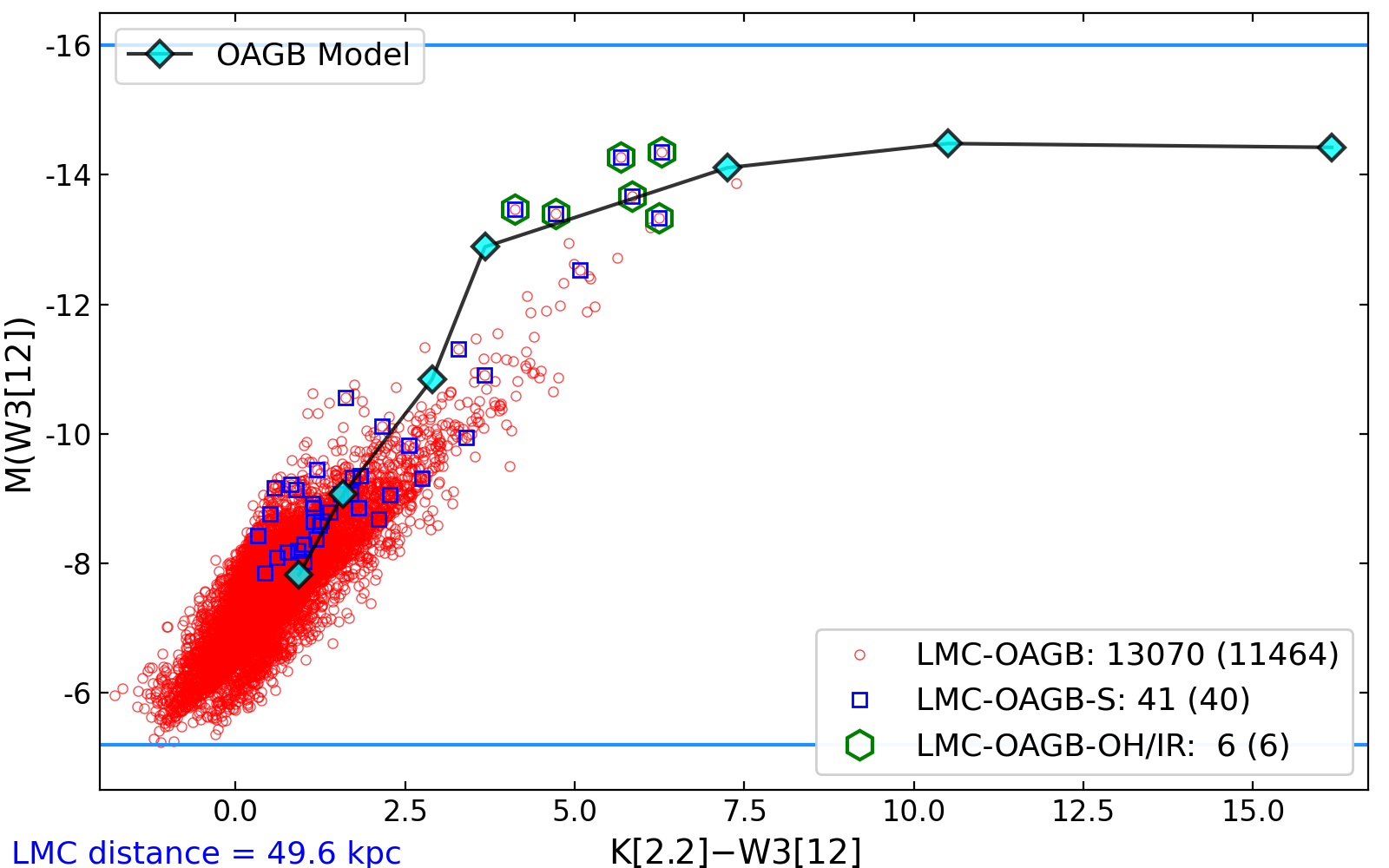}{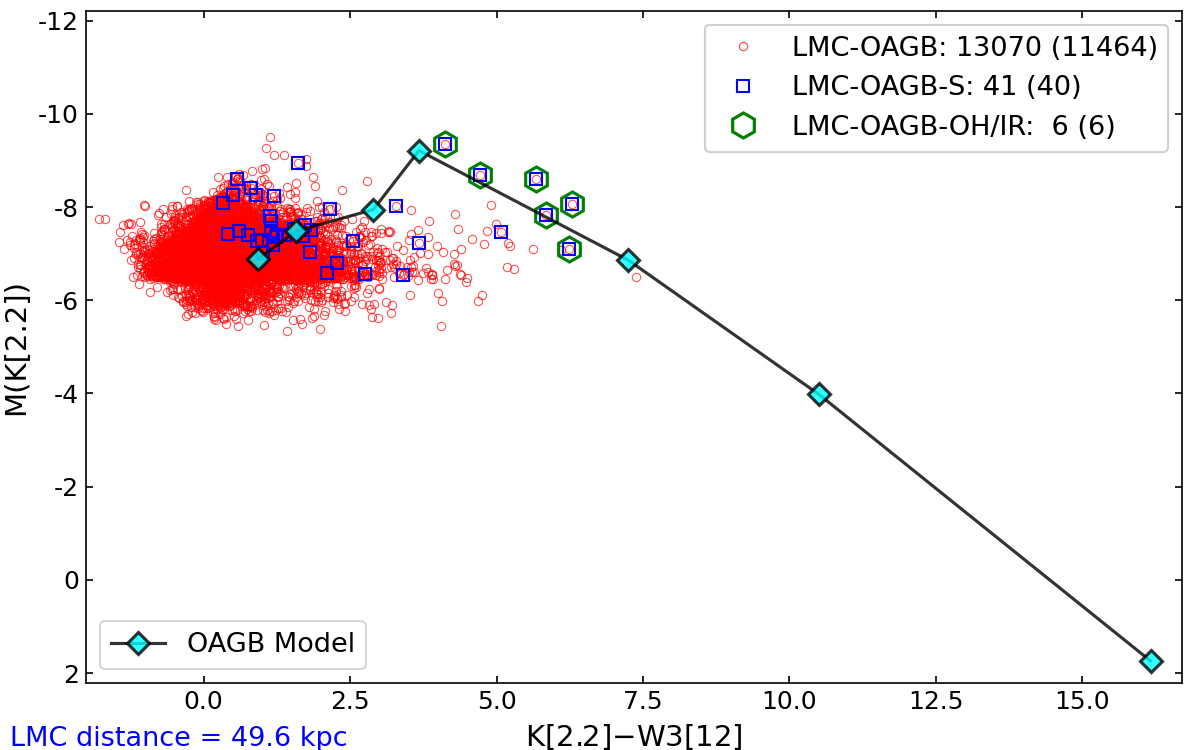}\caption{
For LMC-OAGB stars, the top panels show the relations between optical depth and luminosity
derived using two different methods based on OAGB models.
In the top panels, the sizes of the filled circular markers scale with the square root of the number of
objects represented at each location.
The bottom panels present IR CMDs for LMC-OAGB stars.
For OAGB models: $\tau_{10}$ = 0.001, 0.03, 0.3, 1, 7, 15, and 40 from left to right (see Table~\ref{tab:tab4}).
The total number of objects in each subgroup is listed, with the numbers in parentheses
representing the subset for which the required data are available and therefore plotted.
The blue horizontal lines in the lower-left panel mark the upper and lower limits of the magnitude
range of the LMC-OAGB and MW-OAGB stars.
See Section~\ref{sec:lum-loagb} for details.}\label{f2}
\end{figure*}

\subsection{Modeling LMC-OAGB stars}\label{sec:lum-loagb}

The intrinsic luminosities of Galactic OAGB stars remain uncertain because of the
large uncertainties in their distance estimates. By contrast, the
well-established distance to the LMC makes it an ideal laboratory for determining
the intrinsic luminosities of OAGB stars.

For the full sample of 13,070 LMC-OAGB stars (Table~\ref{tab:tab1}), we tested an
extensive grid of radiative transfer models covering a broad range of parameter
combinations to identify the most suitable set of OAGB models. The observed SEDs
were fitted using a least-squares optimization method (see
Sections~\ref{sec:seds} and \ref{sec:mseds}).

We began the procedure with the initial set of OAGB models for Galactic OAGB
stars derived by \citet{suh2004} and \citet{suh2021} and applied them to the
LMC-OAGB stars, refining and adjusting the model parameters.

To investigate the dependence on dust optical depth, the sample was divided into
four $\tau_{10}$ groups. Within each group, we determined the stellar temperature
in the range 2000-3000 K that best reproduced the observed SED shapes of the
LMC-OAGB stars, yielding four representative temperatures of 3000, 2500, 2300,
and 2000 K. Using these $\tau_{10}$ groups, we then derived SED-based scaling
distance distributions by scaling the model SEDs over a stellar luminosity range
of $3\times10^{3}$ to $3\times10^{4}$~$L_{\odot}$.

Four representative luminosity groups (3.6, 4.9, 7.2, and
25.6~$\times10^{3}$~$L_{\odot}$) were adopted by selecting the values that
yielded the smallest discrepancies between the derived scaling distances and the
known distance to the LMC. Each luminosity group corresponds to the same
temperature group (see Table~\ref{tab:tab4} and the middle panel of
Figure~\ref{f1}).

The LMC-OAGB sample contains no stars with very large dust optical depths
($\tau_{10} > 7$), whereas such heavily obscured objects are common among
Galactic OAGB stars (e.g., \citealt{suh2004}; \citealt{suh2021}), whose observed
SEDs are well reproduced by models with a stellar temperature of 2000 K. We
therefore introduced a fifth model group for Galactic OAGB stars with $\tau_{10}
> 7$ by extrapolating the relation established at lower optical depths in the
LMC. This group adopts a stellar temperature of 2000 K and a luminosity of
$3.0\times10^{4}$~$L_{\odot}$ (see Table~\ref{tab:tab4}).

Table~\ref{tab:tab4} summarizes the final grid of 25 OAGB models, organized into
the five groups. These models were selected to maximize the agreement with the
observed SEDs of LMC-OAGB stars while minimizing the differences between the
scaling distances and the known distances. The models are arranged in order of
increasing $\tau_{10}$ and include the corresponding dust opacity and stellar
parameters ($T_*$ and $L_*$). The top panel of Figure~\ref{f1} displays the SEDs
of the seven representative models.

The middle panel of Figure~\ref{f1} illustrates the relationship between distance
and luminosity derived from the OAGB model SEDs (Table~\ref{tab:tab4}). For all
four luminosity groups, the model-derived distances agree well with the accepted
distance to the LMC. In this panel, the sizes of the filled circular symbols
scale with the square root of the number of objects represented by each point.

The optimal OAGB model (listed in Table~\ref{tab:tab4}) together with its
associated scaling distance was determined for each LMC-OAGB star. Reliable fits
were achieved for 12,728 out of the 13,070 sources (see Section~\ref{sec:seds}).

The bottom panel of Figure~\ref{f1} illustrates the logarithmic distribution of
the derived distances for LMC-OAGB stars. The $\log_{10} d$ distribution is well
described by a Gaussian with a dispersion of 0.0829 dex. Both the median and mean
distances (49.46 and 50.37 kpc, respectively) are in good agreement with the
canonical LMC distance of 49.6 kpc. The small dispersion in $\log_{10} d$ (0.0829
dex) further highlights the internal consistency and reliability of the OAGB
models (Table~\ref{tab:tab4}). While the large sample size reveals a slight
asymmetry in the $\log_{10} d$ distribution, the difference between the left and
right dispersions is small ($<$ 10 \% of the total width), and all results remain
unchanged when asymmetric dispersions are used.

The upper panels of Figure~\ref{f2} compare the relationships between optical
depth and luminosity for LMC-OAGB stars using two approaches derived from the
OAGB model SEDs. The upper-left panel displays the relation obtained using the
four luminosity groups computed by the method described above (see
Table~\ref{tab:tab4}; Figure~\ref{f1}). In contrast, the upper-right panel shows
an alternative estimate in which luminosities were calculated after rescaling
each derived distance to the adopted LMC distance of 49.6 kpc. This method
reveals a more detailed luminosity distribution at each $\tau_{10}$ but results
in noticeably greater scatter. When the four luminosity groups defined by
$\tau_{10}$ (see Table~\ref{tab:tab4}) are combined, however, the two approaches
produce broadly consistent results. In these plots, the areas of the filled
circles reflect the square root of the number of stars represented at each
location.

The bottom panels of Figure~\ref{f2} show the WISE–2MASS CMDs for the LMC-OAGB
sample, plotting absolute magnitudes M(W3[12]) and M(K[2.2]) versus the
K[2.2]$-$W3[12] color. These observed absolute magnitudes, calculated under the
assumption of a 49.6 kpc distance, are generally consistent with the predictions
of the OAGB models (see Table~\ref{tab:tab4}). We find that M(W3[12]) traces
absolute luminosity reasonably well and thus serves as a practical proxy for the
luminosity distribution. The maximum and minimum values of M(W3[12]) for the
LMC-OAGB stars and Galactic OAGB stars fall within the range of $-5.2$ and $-16$
mag (see Section~\ref{sec:lum-goagb}), indicated by blue horizontal lines.

Conversely, M(K[2.2]) becomes systematically fainter for intrinsically brighter
OAGB stars with higher dust optical depths, as circumstellar dust attenuates
shorter wavelengths and redistributes the energy to longer wavelengths. As a
result, absolute magnitudes in the NIR or at still shorter wavelengths are not
reliable indicators of luminosity for OAGB stars.

The scatter in $\log_{10} d$ (0.0829 dex, corresponding to approximately 20 \% in
distance; bottom panel of Figure~\ref{f1}) for LMC-OAGB stars is likely
attributable to three effects: (1) the finite spatial extent of the LMC, whose
radius is about 10 \% of its mean distance, resulting in an expected distance
scatter of $\sim$5 \%; (2) the degeneracy between dust optical depth and stellar
luminosity, whereby a given $\tau_{10}$ can correspond to multiple luminosities
because stars with different initial masses evolve to different luminosities; and
(3) intrinsic luminosity variations caused by stellar pulsations. Because the
first contribution is relatively small, the observed scatter in $\log_{10} d$
primarily reflects the uncertainty in the adopted luminosities. The luminosity
range implied by this distance scatter spans approximately a factor of 2.1.

\begin{figure*}
\centering
\smallplotsix{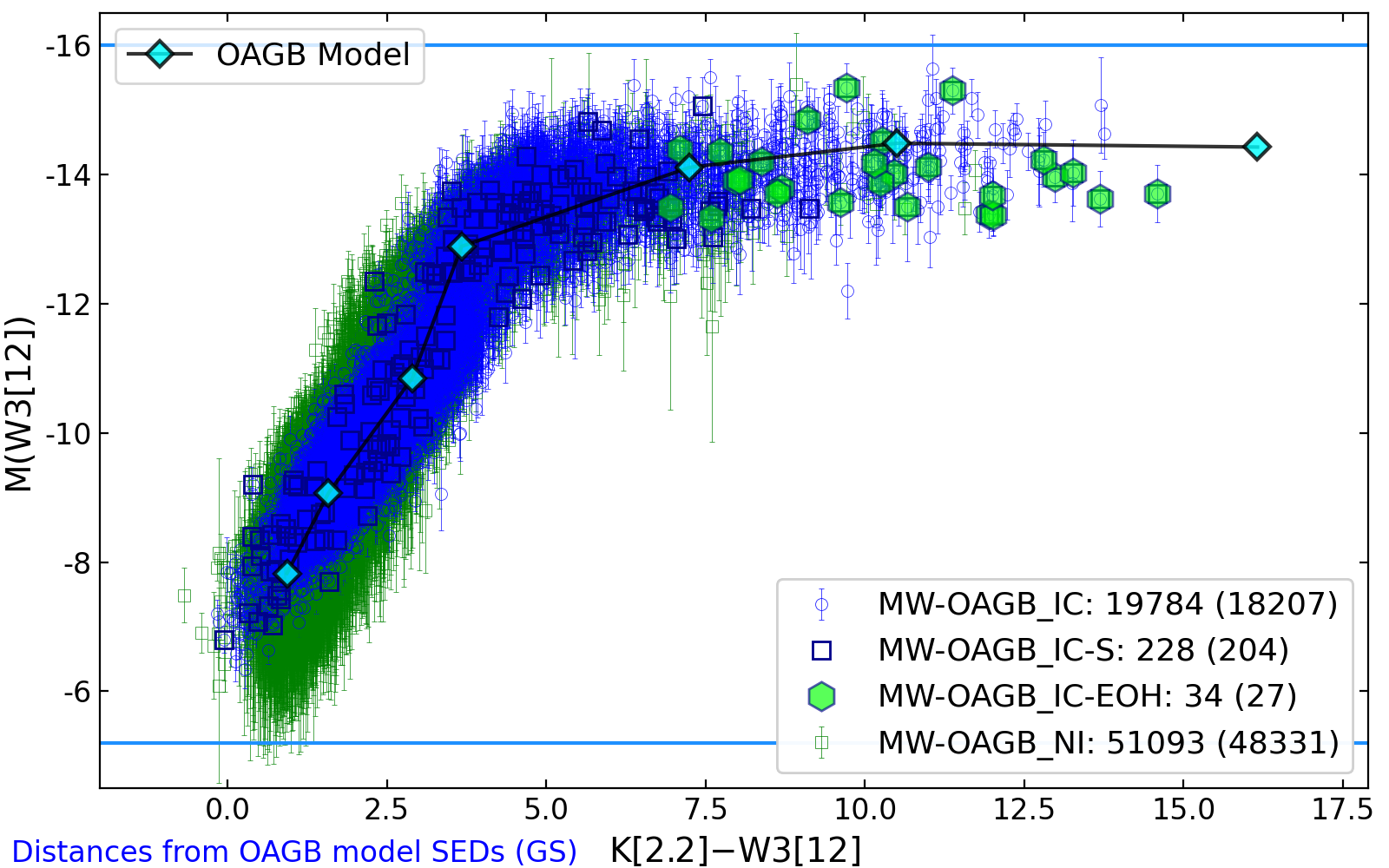}{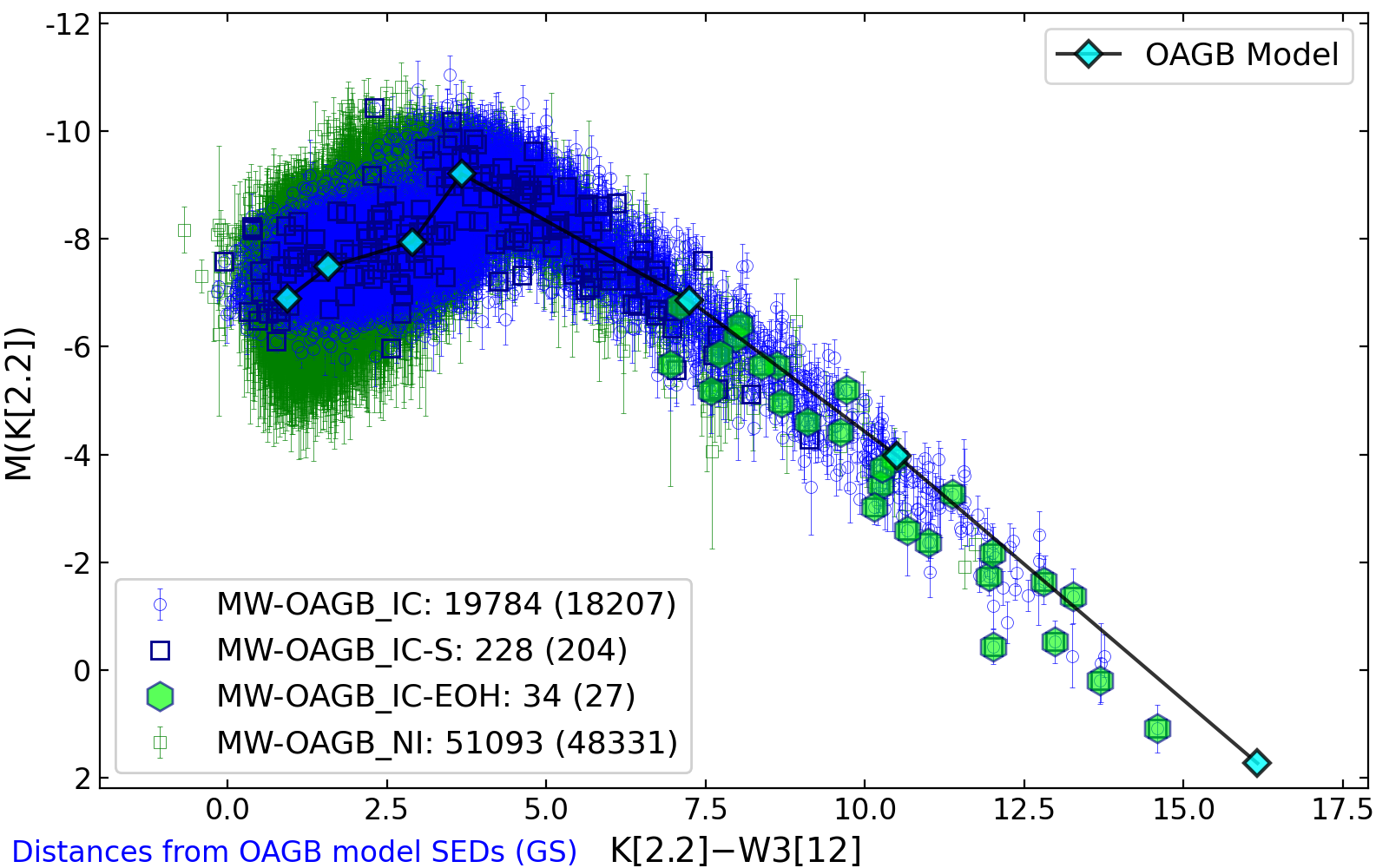}{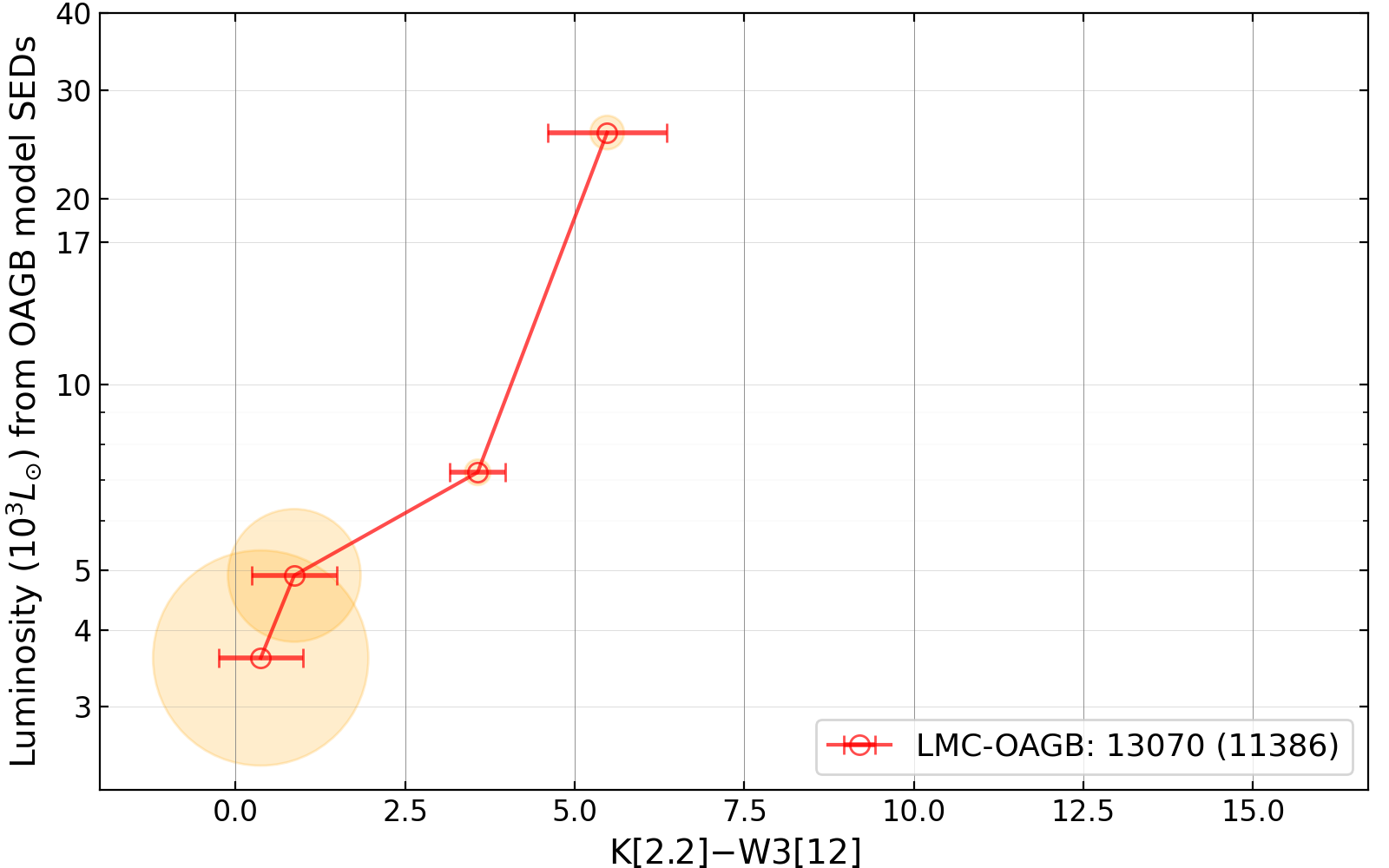}{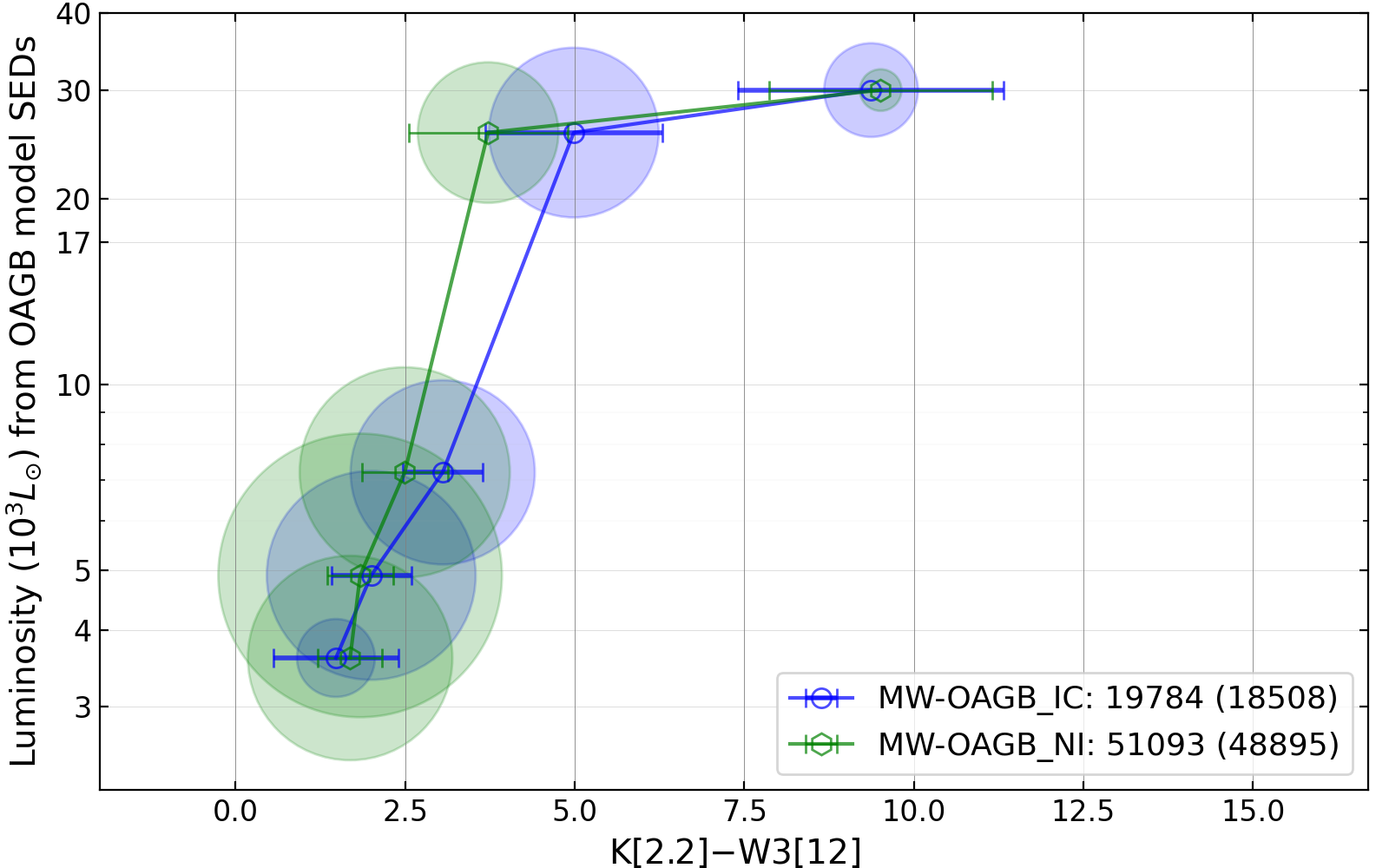}{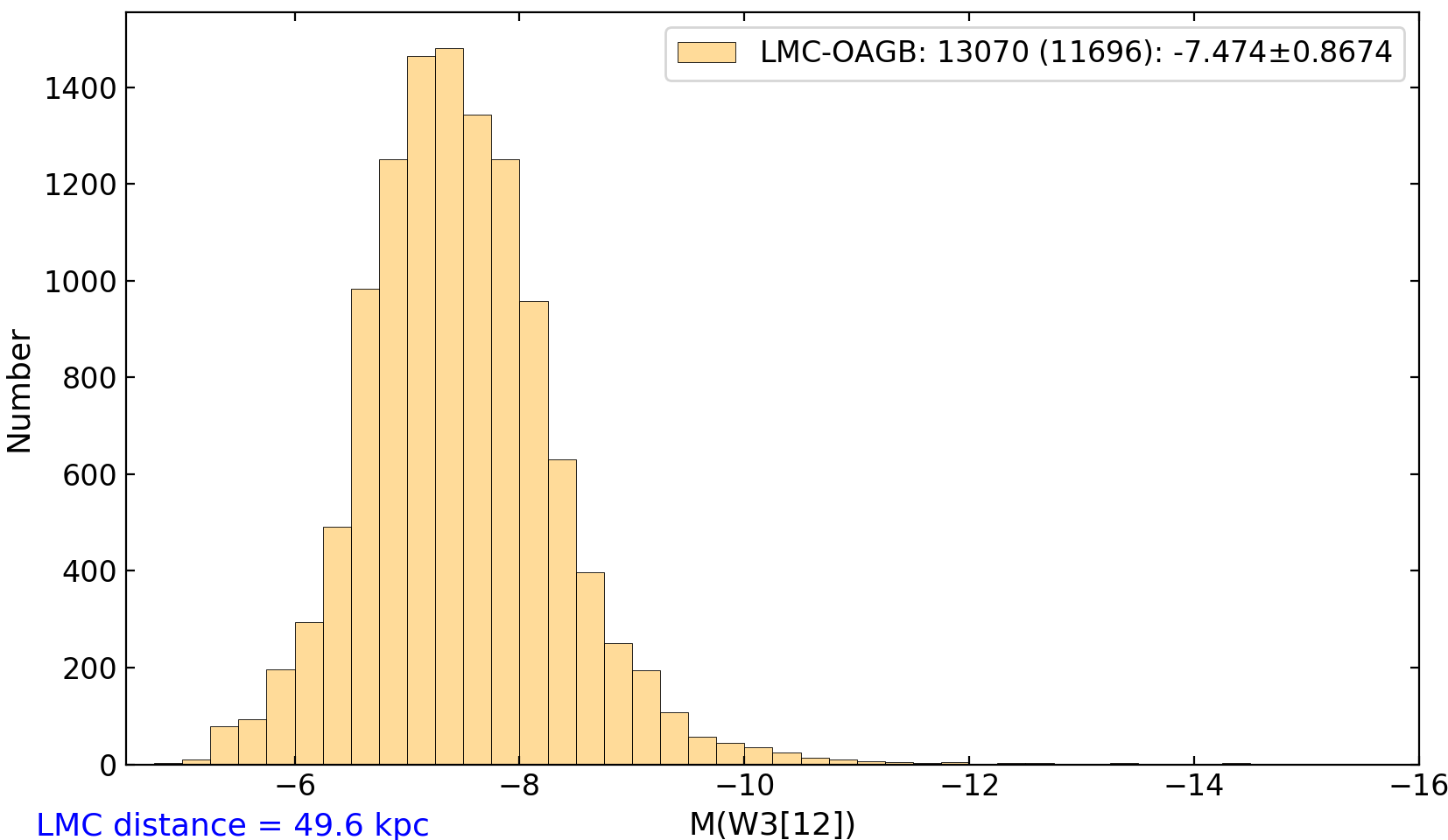}{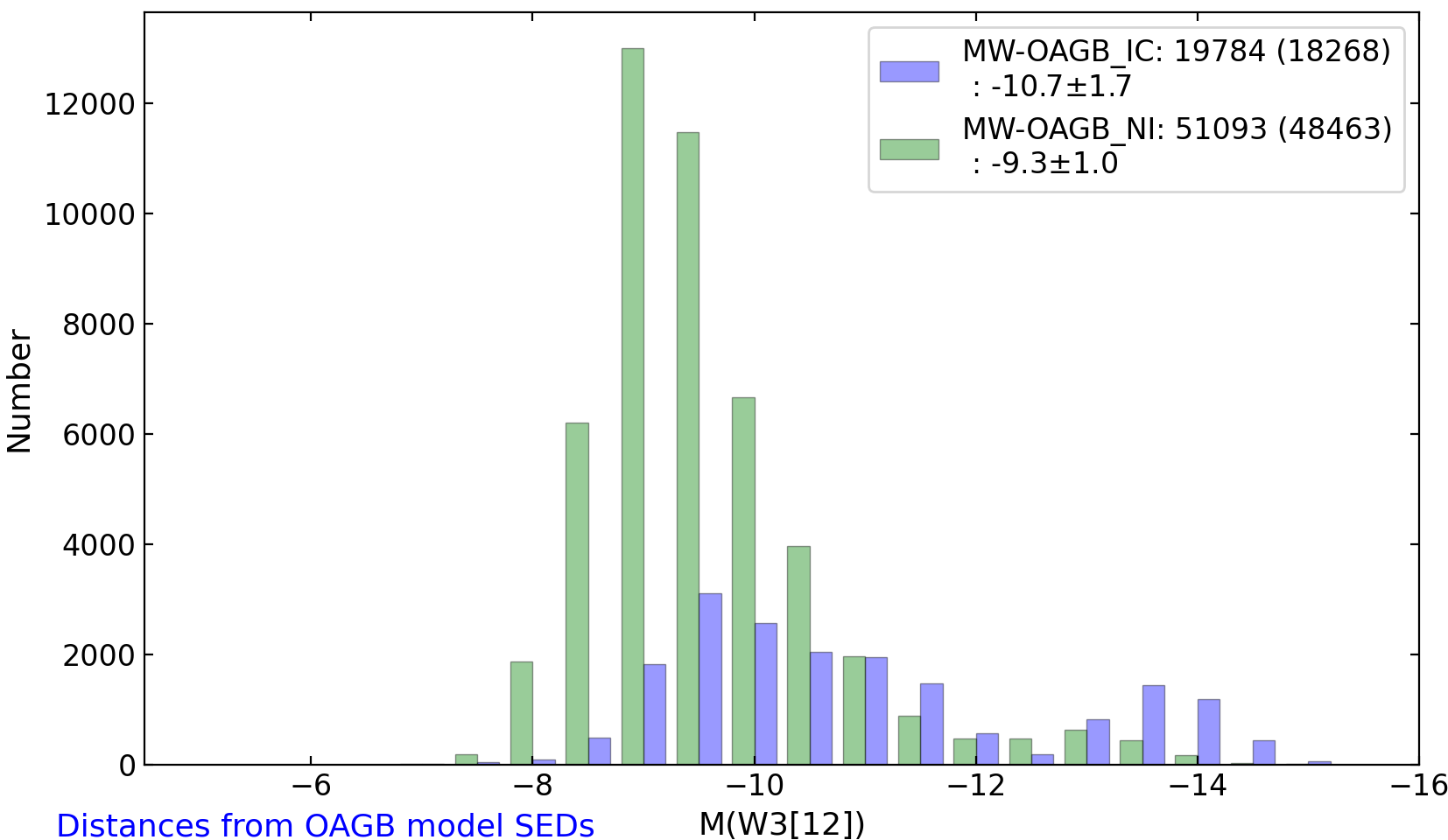}
\caption{The top panels show IR CMDs for Galactic OAGB stars.
For Galactic OAGB stars, distances were derived from OAGB model SEDs, with Gaussian Monte Carlo scatter (GS)
in $\log_{10} d$ ($\sigma$=$\pm$0.0829 dex; obtained from LMC-OAGB stars; see Figure~\ref{f1}).
For OAGB stars in the LMC and the Milky Way, the middle panels illustrate the
relationship between dust optical depth and luminosity, while the bottom panels
present the corresponding luminosity distributions. In the middle panels, the sizes
of the filled circular markers scale with the square root of the number of
objects represented at each location.
For OAGB models: $\tau_{10}$ = 0.001, 0.03, 0.3, 1, 7, 15, and 40 from left to right (see Table~\ref{tab:tab4}).
The total number of objects in each subgroup is listed, with the numbers in parentheses
representing the subset for which the required data are available and therefore plotted.
The blue horizontal lines in the upper-left panel mark the upper and lower limits of the magnitude range of the LMC-OAGB and MW-OAGB stars
(See Section~\ref{sec:lum-loagb}).
See Section~\ref{sec:lum-goagb} for details.
} \label{f3}
\end{figure*}

\subsection{Application to Galactic OAGB stars}\label{sec:lum-goagb}

Having established the reliability of the OAGB models and quantified their
uncertainties using the LMC-OAGB sample (Section~\ref{sec:lum-loagb}), we applied
the same set of 25 OAGB models (Table~\ref{tab:tab4}) to the Galactic sample
(MW-OAGB objects; Table~\ref{tab:tab1}). The observed SEDs of the Galactic OAGB
stars were fitted using a least-squares optimization method
(Sections~\ref{sec:seds} and \ref{sec:mseds}).

The top panels of Figure~\ref{f3} show WISE-2MASS CMDs for Galactic OAGB stars,
with M(W3[12]) and M(K[2.2]) plotted as a function of the K[2.2]$-$W3[12] color.
Distances were derived by fitting the observed SEDs with the OAGB models (see
Table~\ref{tab:tab4}). To account for uncertainties in the luminosity of the OAGB
models for Galactic OAGB stars, we applied Gaussian Monte Carlo scatter (GS) in
distance ($\log_{10} d$) with $\sigma = \pm$0.0829 dex, consistent with the
dispersion measured for LMC-OAGB stars (see the bottom panel of Figure~\ref{f1}).
OAGB stars in the LMC and the Milky Way exhibit similar distributions on the CMDs
(see Figures~\ref{f2} and \ref{f3}) and show good overall agreement with the OAGB
models. However, the LMC lacks objects with very red K[2.2]$-$W3[12] colors
corresponding to large dust optical depths ($\tau_{10}
> 7$).

The middle panels of Figure~\ref{f3} show the observed K[2.2]$-$W3[12] color as a
function of luminosity derived from the best-fitting OAGB model SEDs
(Table~\ref{tab:tab4}) for stars in the LMC and the Milky Way. More luminous
objects tend to exhibit redder IR colors, reflecting the behavior of the OAGB
models, in which larger dust optical depths correspond to both redder colors and
higher luminosities. Although the overall distributions are similar in the two
galaxies, Galactic OAGB stars span a wider range of colors. Most notably,
extremely red and luminous sources associated with very large dust optical depths
($\tau_{10} > 7$) are exceedingly rare in the LMC sample, whereas they are
readily found among Galactic OAGB stars.

The bottom panels of Figure~\ref{f3} show the W3[12]-band luminosity
distributions for OAGB stars in the two galaxies. While both distributions have
roughly Gaussian profiles, they differ substantially in detail. The LMC sample is
dominated by a single luminosity component with $-5 \lesssim M(W3[12]) \lesssim
-10$ mag. By contrast, the Galactic sample contains not only a similar main
component between $-8$ and $-12$ mag, but also a distinct population of
exceptionally luminous stars extending to $M(W3[12]) \simeq -15$ mag. This
luminous population is largely absent from the LMC sample.

Among the Galactic OAGB stars, MW-OAGB\_IC objects span a wide range of
K[2.2]$-$W3[12] colors and extend to the bright end of the luminosity
distribution, whereas MW-OAGB\_NI objects are concentrated at bluer colors and
are rare at the bright end (see Figure~\ref{f3}). Generally, MW-OAGB\_IC objects
are brighter and/or more evolved than MW-OAGB\_NI objects (see
Section~\ref{sec:gagb}).

In Figures~\ref{f2} and \ref{f3}, blue horizontal lines mark the maximum and
minimum M(W3[12]) values of the OAGB sample, $-5.2$ and $-16$ mag, respectively.
Although the CMDs and luminosity distributions of OAGB stars in the LMC and the
Milky Way share many common features, the LMC exhibits a pronounced deficiency of
luminous OAGB stars (M(W3[12]) $< -10$ mag). These high-luminosity objects are
typically heavily obscured by optically thick dust envelopes and are commonly
identified as OH/IR stars. They are thought to represent either the latest stages
of OAGB evolution or the descendants of relatively massive progenitors.

\subsection{Limitations of the theoretical models}\label{sec:limit}

The relation between SED shape (dust optical depth) and luminosity used in this
work is an empirical calibration based on the LMC sample, which contains no stars
with very thick dust envelopes. As discussed in Section~\ref{sec:comoc}, this is
likely due to a combination of a relative deficiency of high-mass progenitors and
the metallicity-dependent effects specific to O-rich chemistry, and the two
cannot be separated with the present data. To the extent that the first factor
contributes, the LMC calibration is weighted toward lower-mass stars. In
addition, theoretical models (e.g., \citealt{dicriscienzo2016};
\citealt{ventura2018}) predict that OAGB stars of all masses pass through an
optically thin phase, so that the luminosity adopted for each optically thin
group in Table 4 is set mainly by the lowest-mass stars reaching that luminosity,
since these are far more numerous. The SED distances of higher-mass Galactic OAGB
stars in optically thin phases may therefore be underestimated.

The opposite bias is expected for optically thick stars. In the LMC, only six
OAGB stars with available model parameters, all of them OH/IR stars, reach
$\tau_{10} \ge 1$, and these luminous objects alone set the luminosity of the
fourth group in Table~\ref{tab:tab4} ($25.6 \times 10^{3}~L_\odot$). In the Milky
Way, by contrast, 3958 of the 19,126 MW-OAGB\_IC objects with model parameters
(20.7\%) fall in the fourth group. Because silicate dust forms more efficiently
in the Milky Way (Section~\ref{sec:comoc}), stars of considerably lower
luminosity may also reach these optical depths, and applying the luminosity
calibrated with the six LMC stars to them may overestimate their distances (see
Section~\ref{sec:sed-dist} for details).

Because the LMC contains no OAGB stars with very thick dust envelopes ($\tau_{10}
> 7$), the absolute luminosity of the fifth group in Table~\ref{tab:tab4}, which
applies only to Galactic OAGB stars, cannot be calibrated using the LMC sample
(see Section~\ref{sec:lum-loagb}). Therefore, its adopted luminosity ($3 \times
10^{4}$~$L_{\odot}$) is more uncertain than those of the other groups and should
be interpreted with caution.

AGB stars can exhibit luminosity variations by a factor of up to 3–4 (e.g.,
\citealt{suh2004}), which implies that distances inferred from OAGB model SEDs
may vary by up to a factor of 2.

While the theoretical dust shell models employed in this study are valuable for
replicating various observations of AGB stars in IR bands, they do not consider
gas-phase radiation processes or non-spherical dust envelopes. AGB stars often
exhibit a range of gas-phase emission and absorption features in the visual and
NIR bands (e.g., \citealt{lancon2000}; \citealt{LeBertre2005}). As a result,
discrepancies between theoretical predictions and observations are expected to be
more pronounced in the visual and NIR bands, where gas-phase radiation effects
dominate.

Additionally, for AGB stars with non-spherical dust envelopes, the observed
colors and magnitudes in both visual and IR bands can deviate substantially from
theoretical model predictions (e.g., \citealt{suh2016}).

\begin{figure*}
\centering
\smallploteight{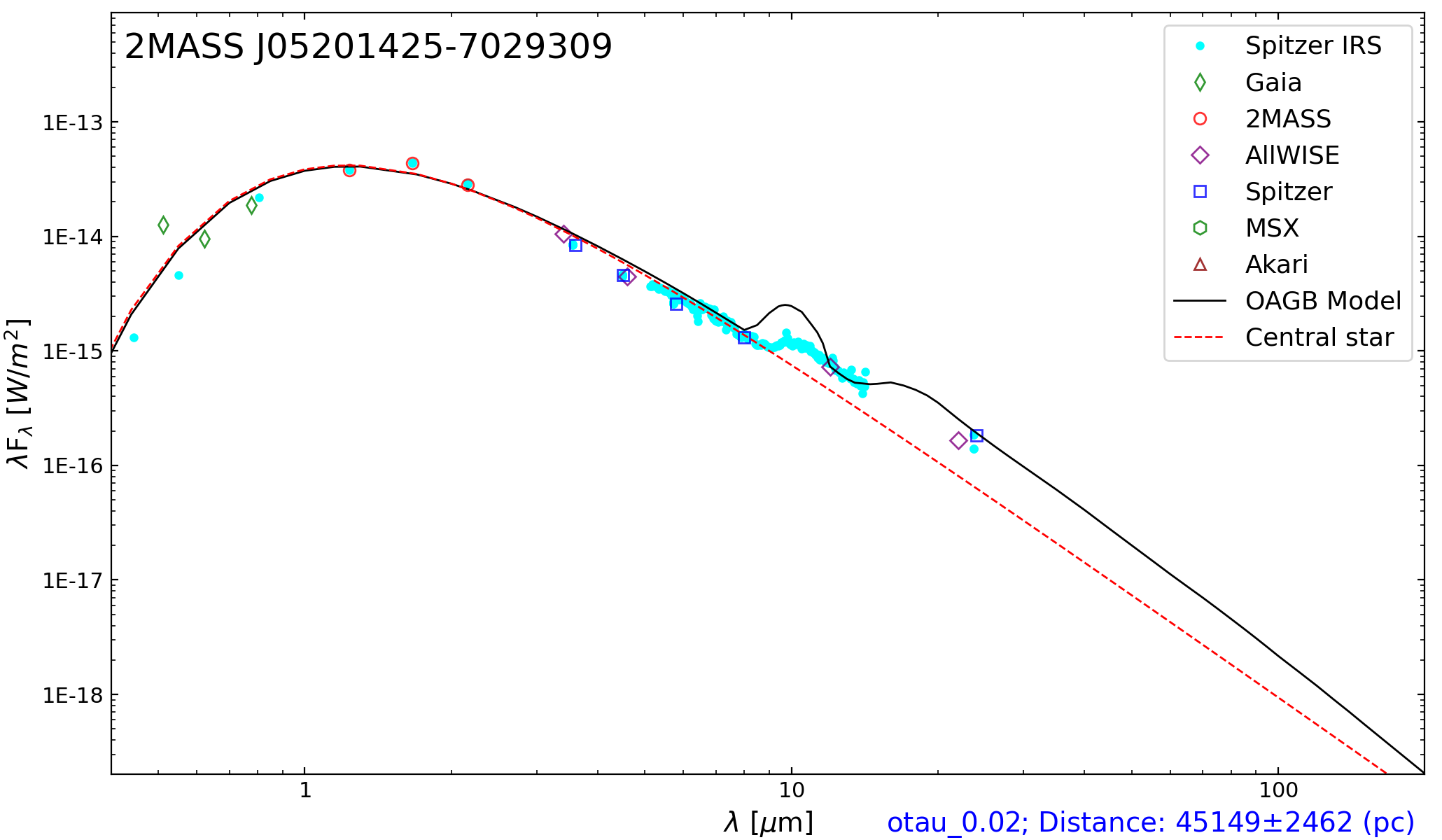}{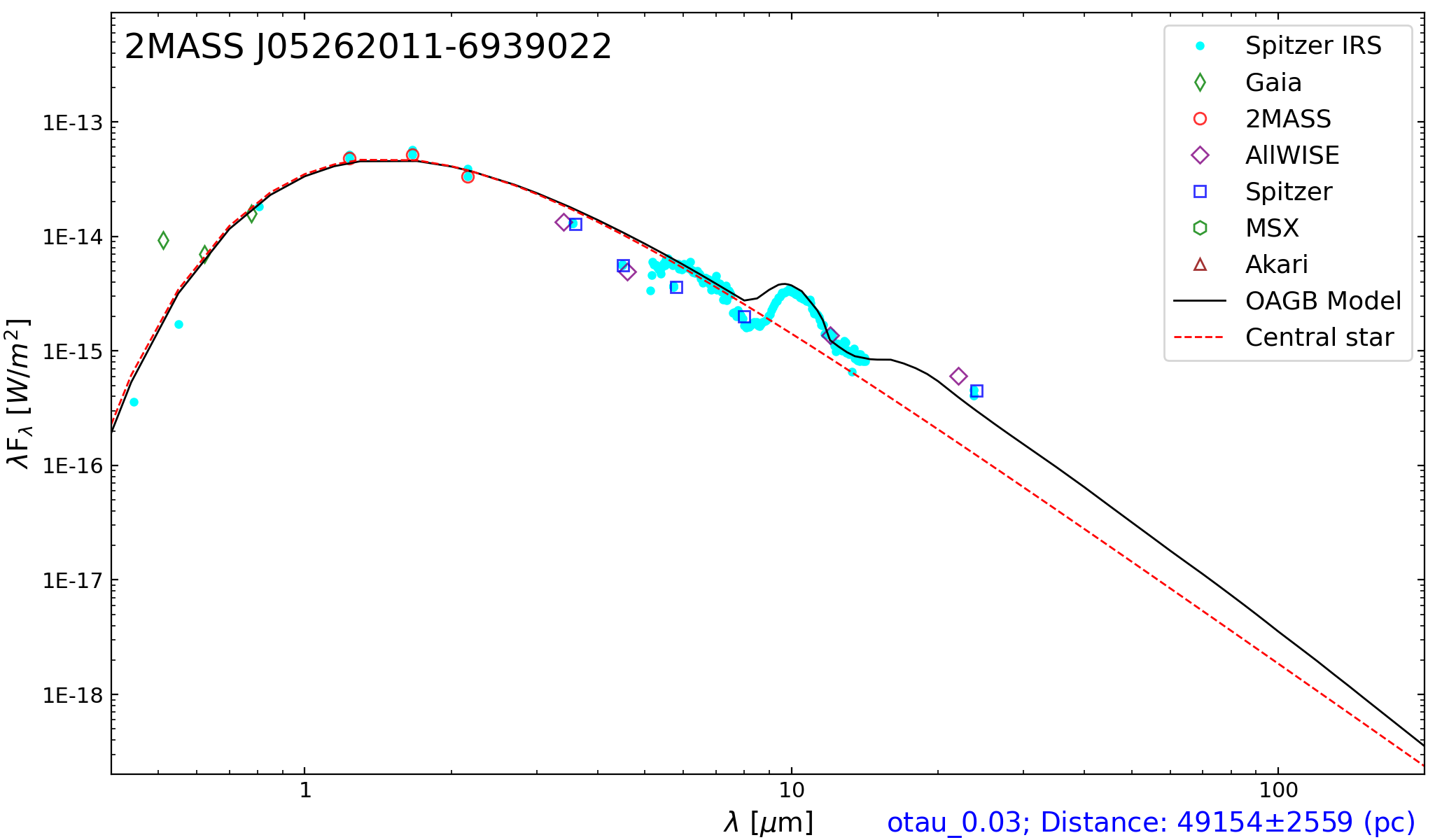}{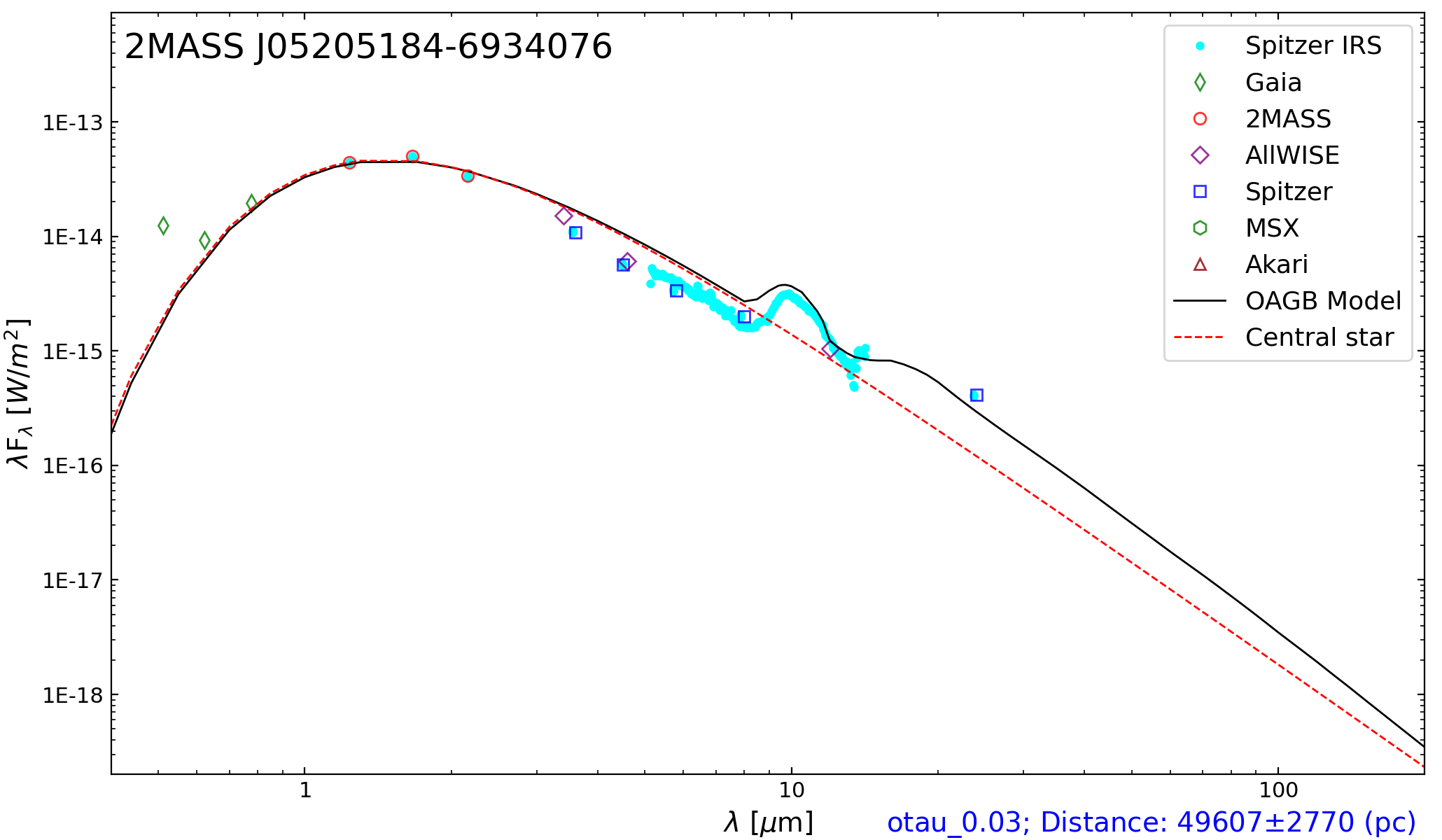}{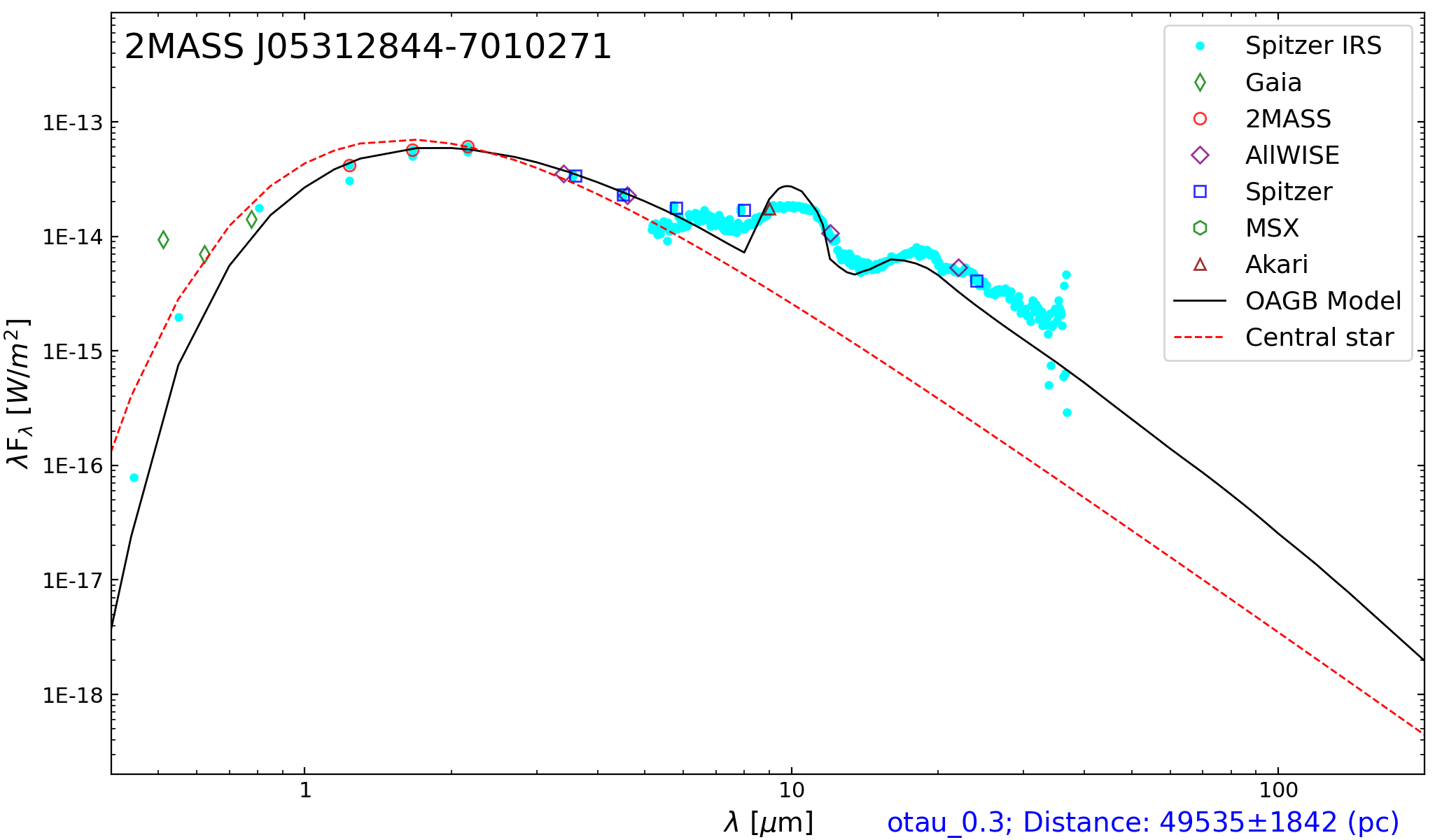}{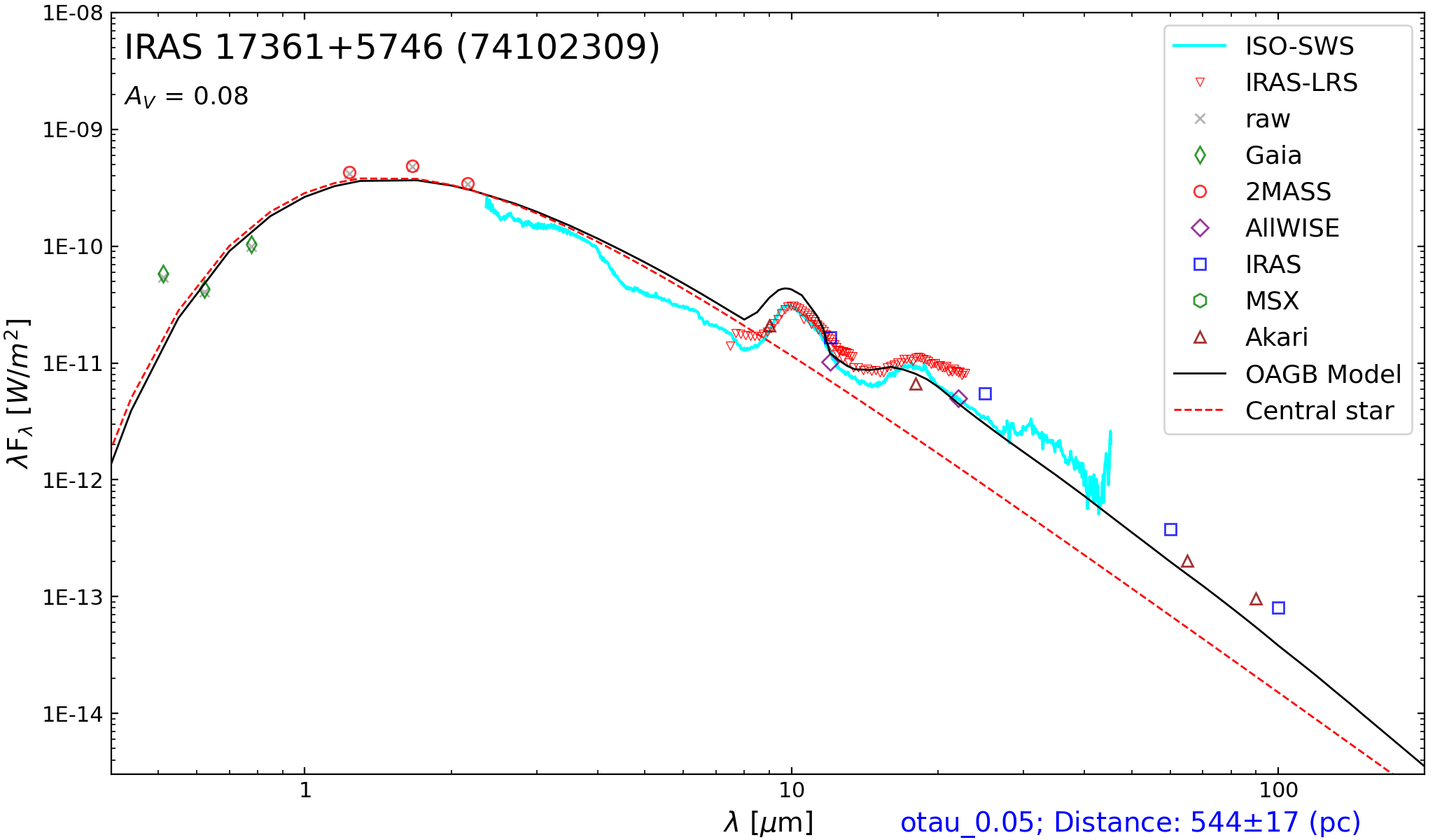}{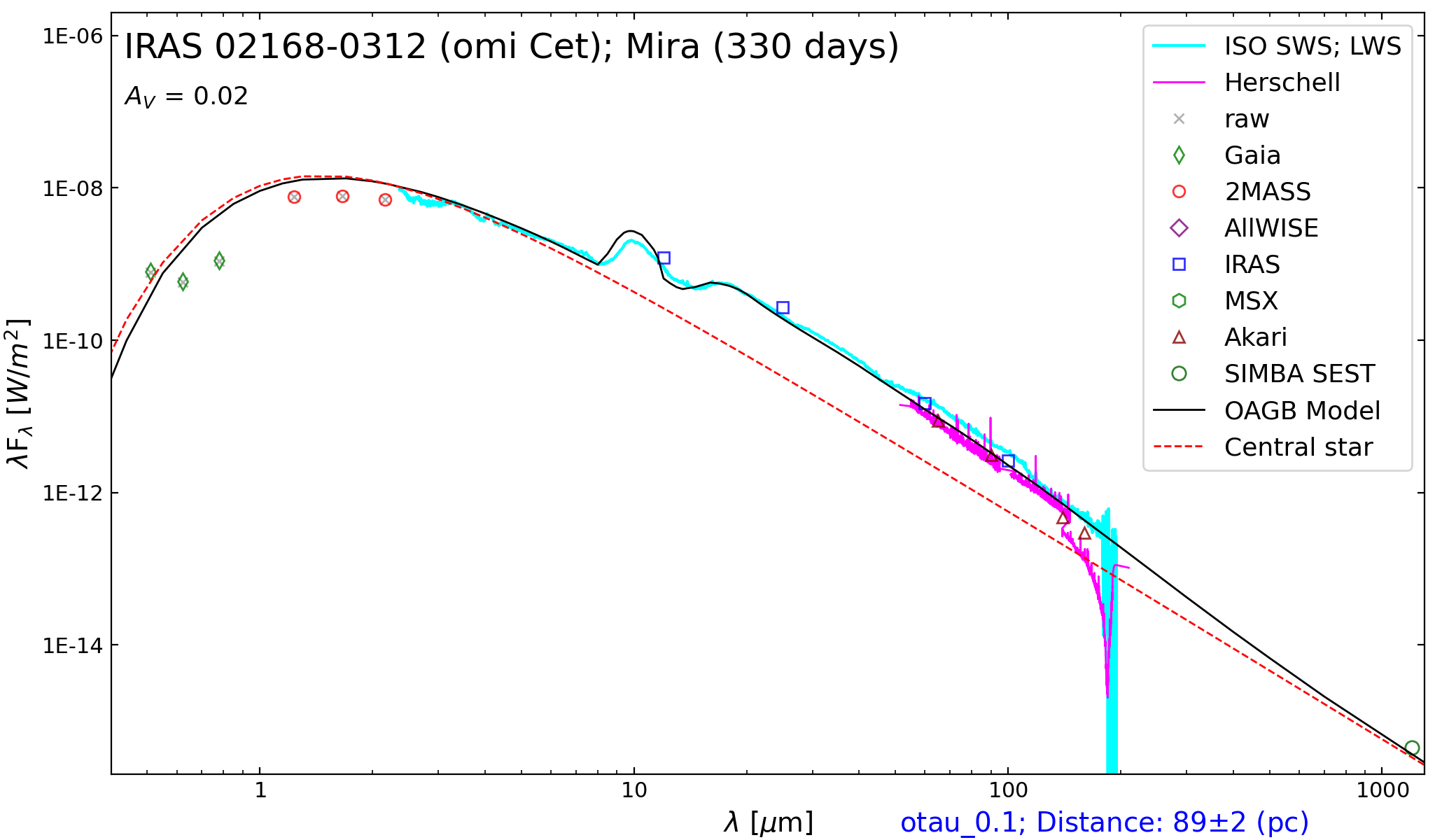}{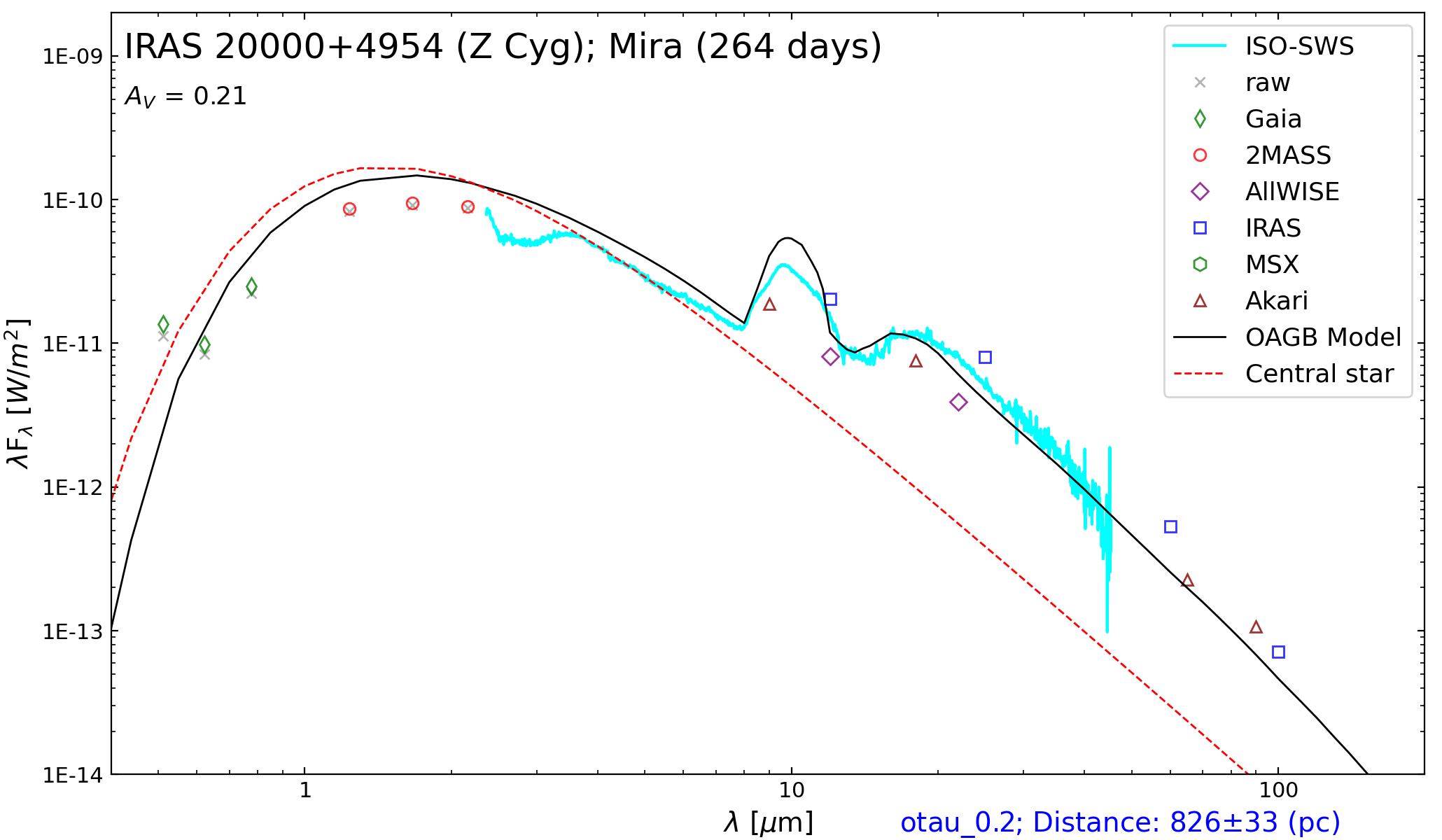}{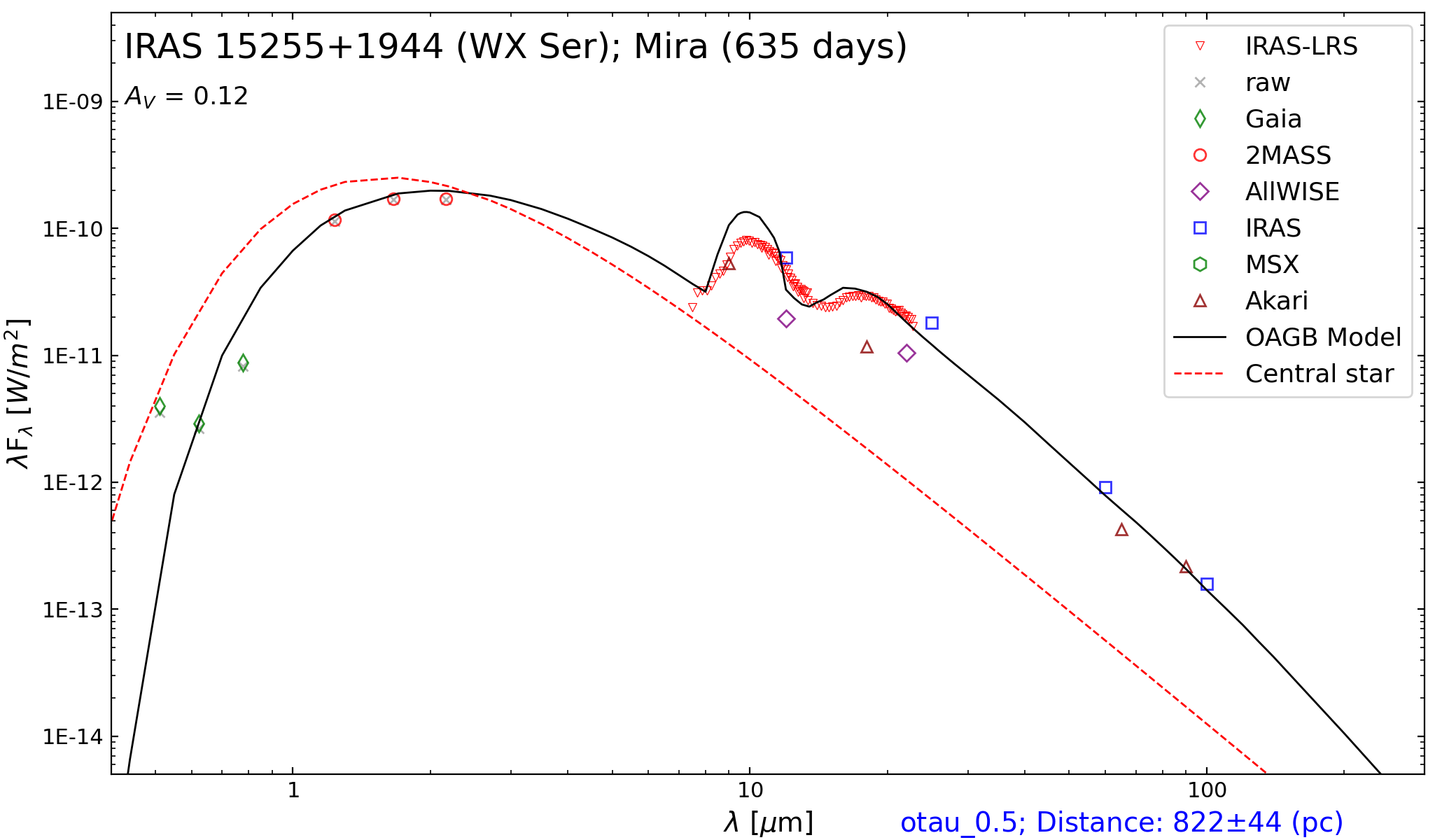}
\caption{Observed and model SEDs for OAGB stars with thin dust envelopes
($0.001 \leq \tau_{10} < 1$).
The top four panels show LMC stars, while the bottom four panels show Milky Way stars.
Each panel indicates the $\tau_{10}$ value and the corresponding scaling distance (bottom right).
See Section~\ref{sec:seds}.} \label{f4}
\end{figure*}

\begin{figure*}
\centering
\smallploteight{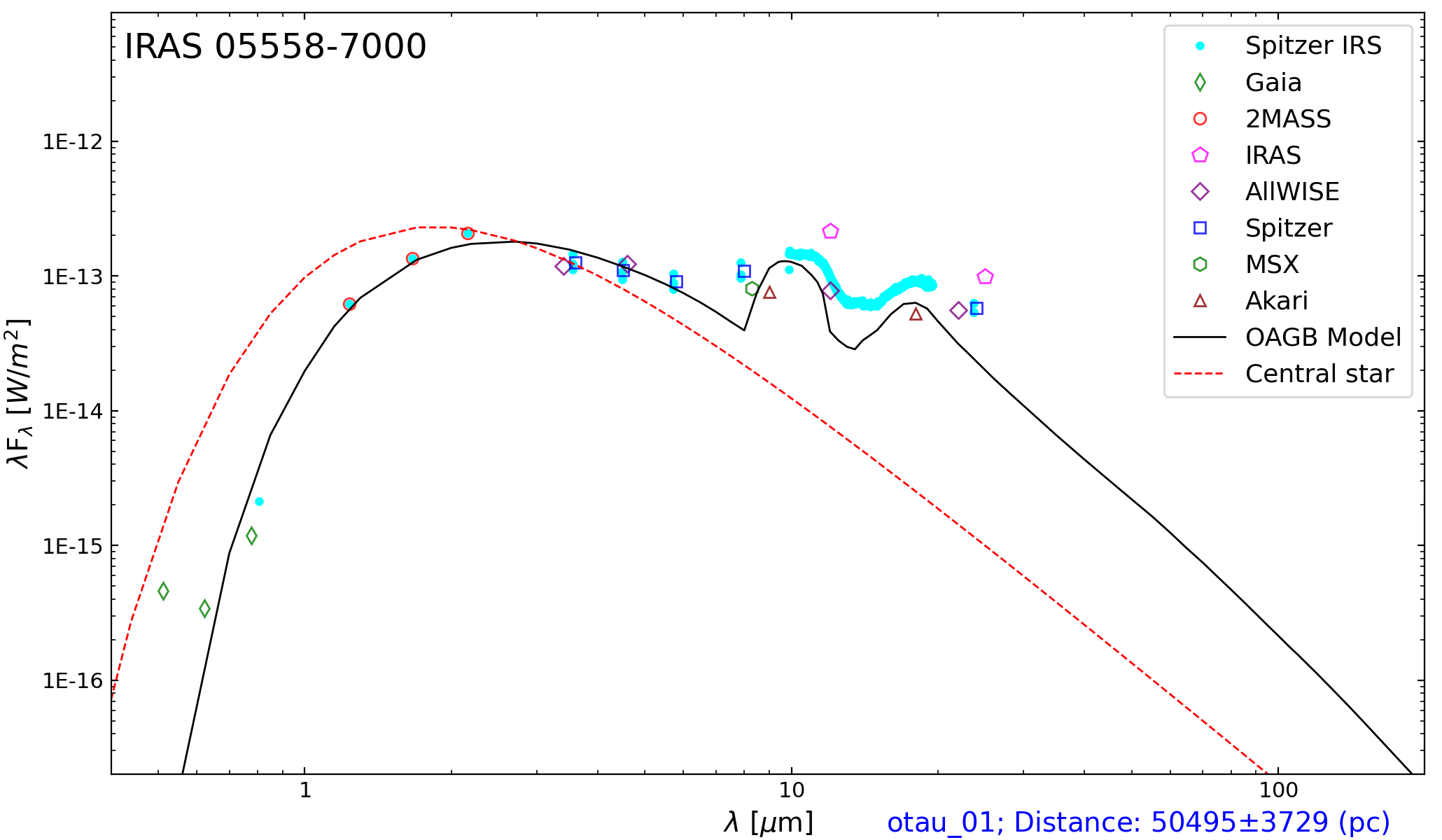}{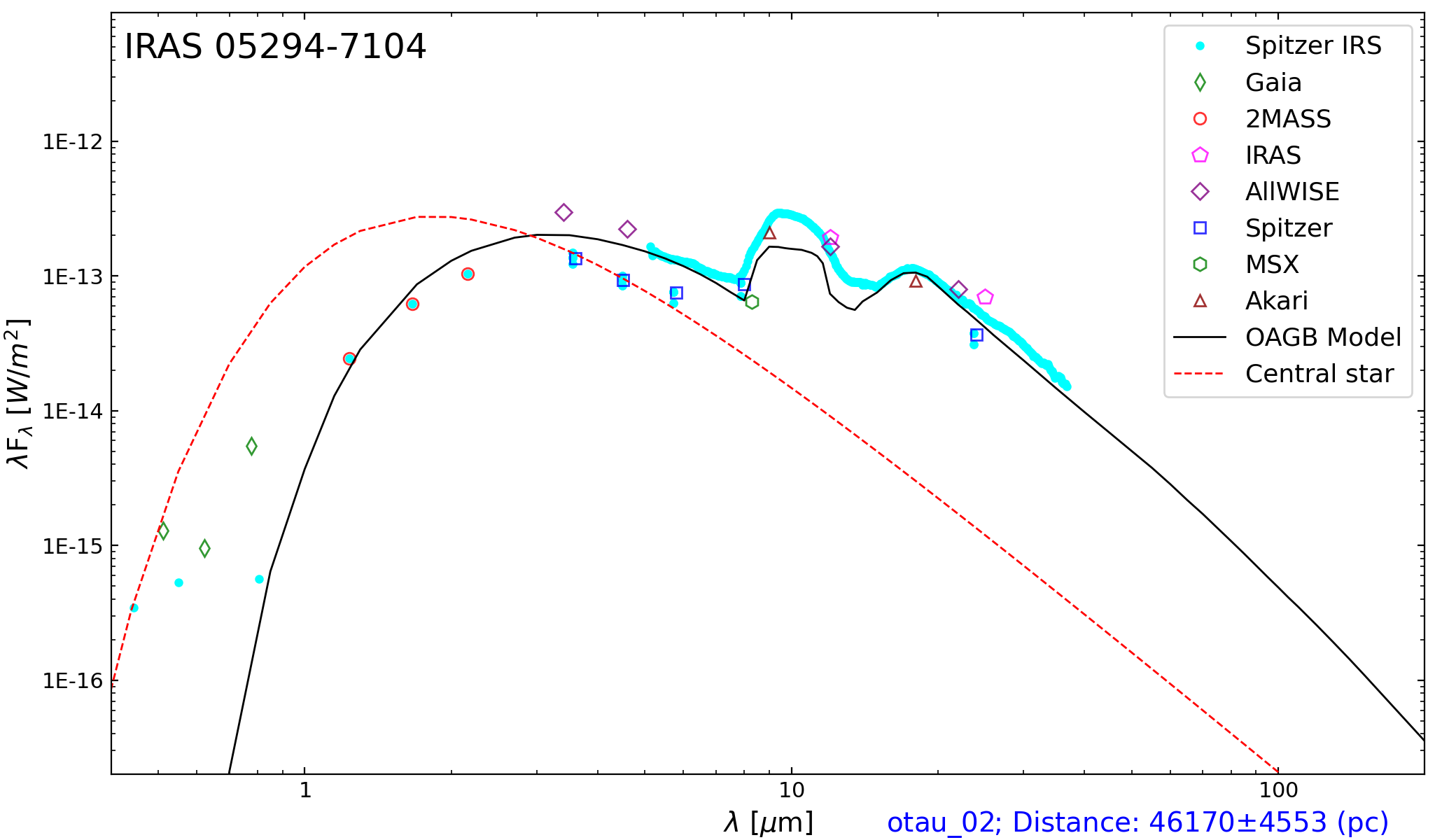}{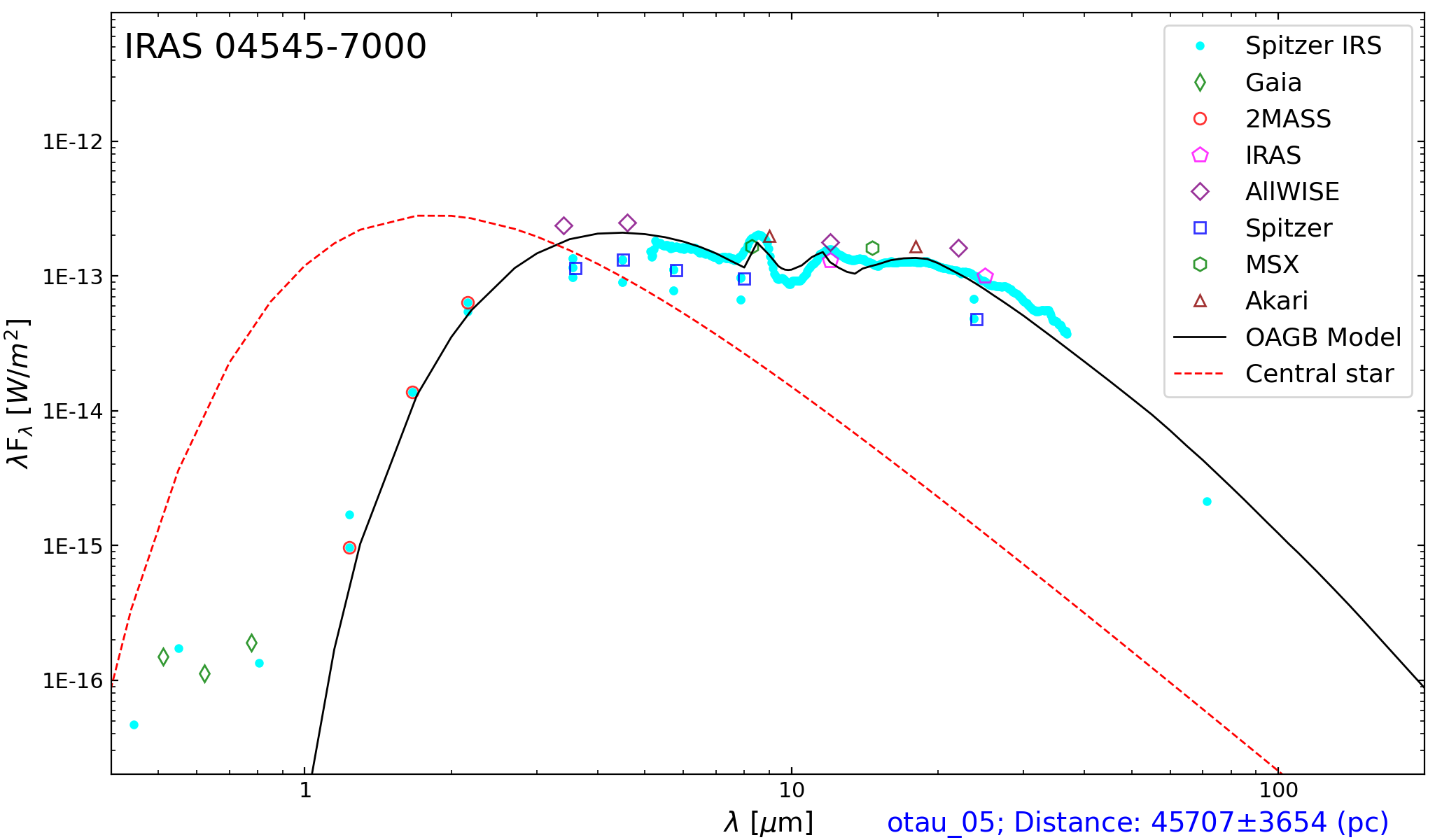}{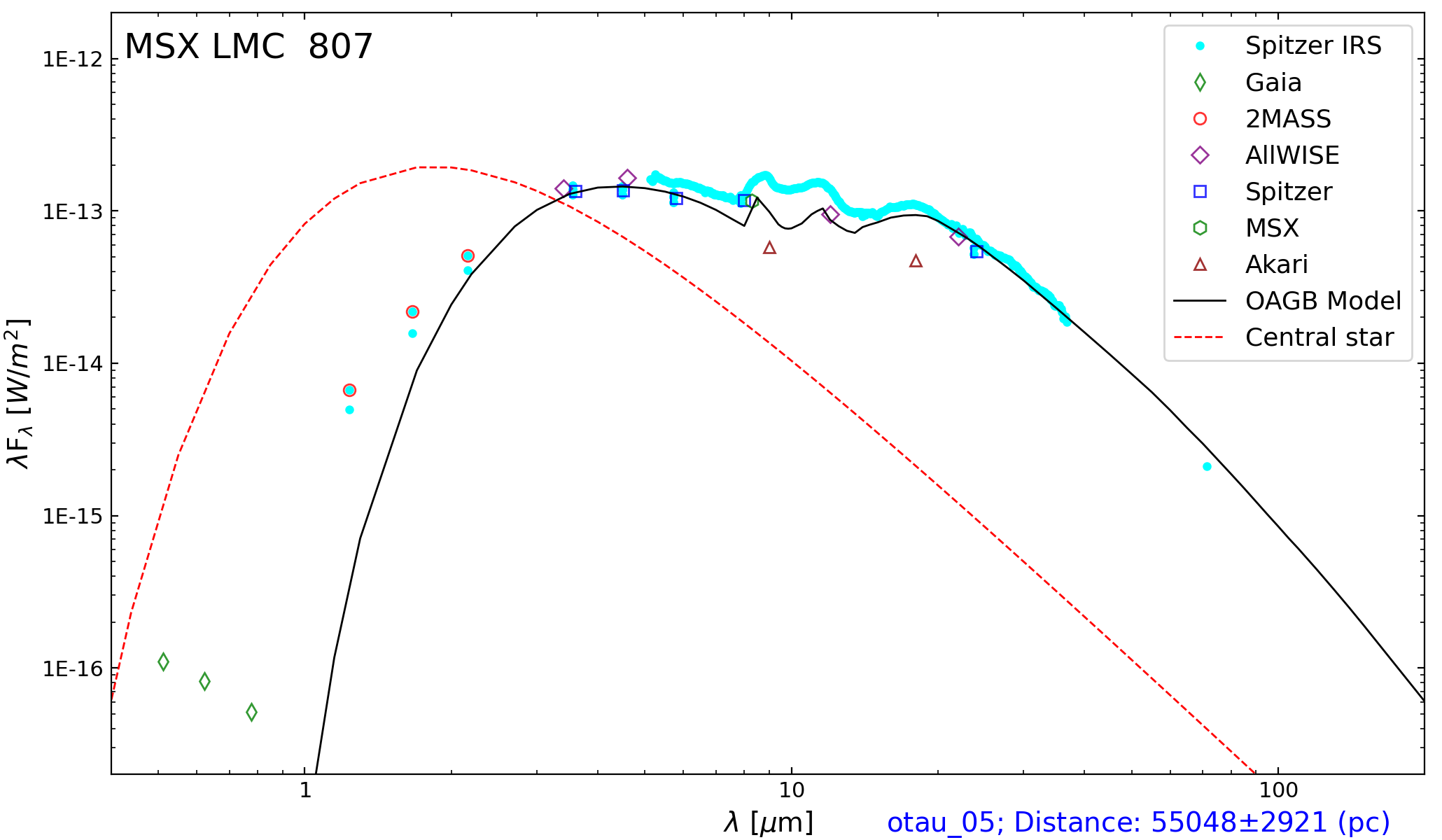}{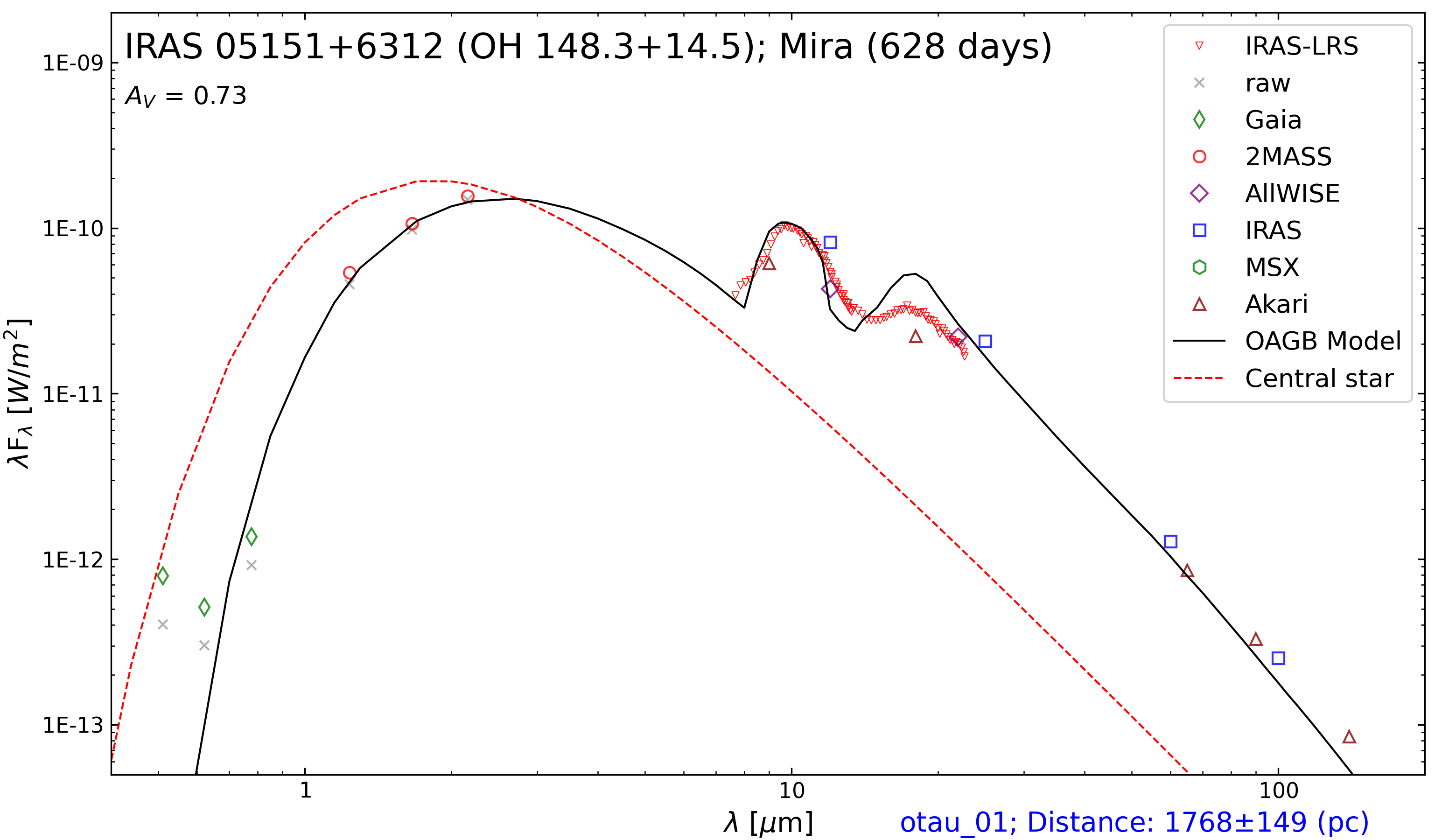}{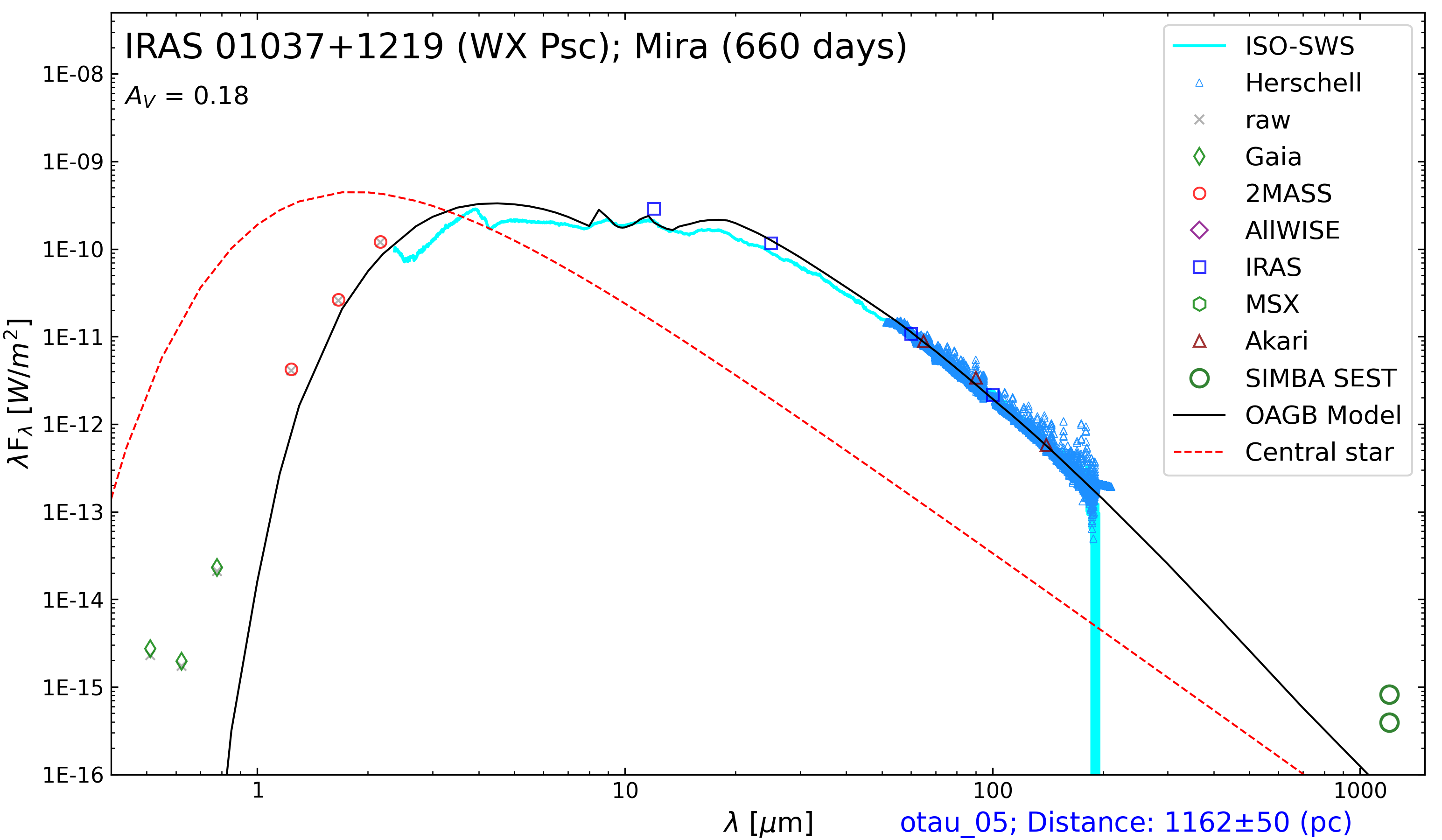}{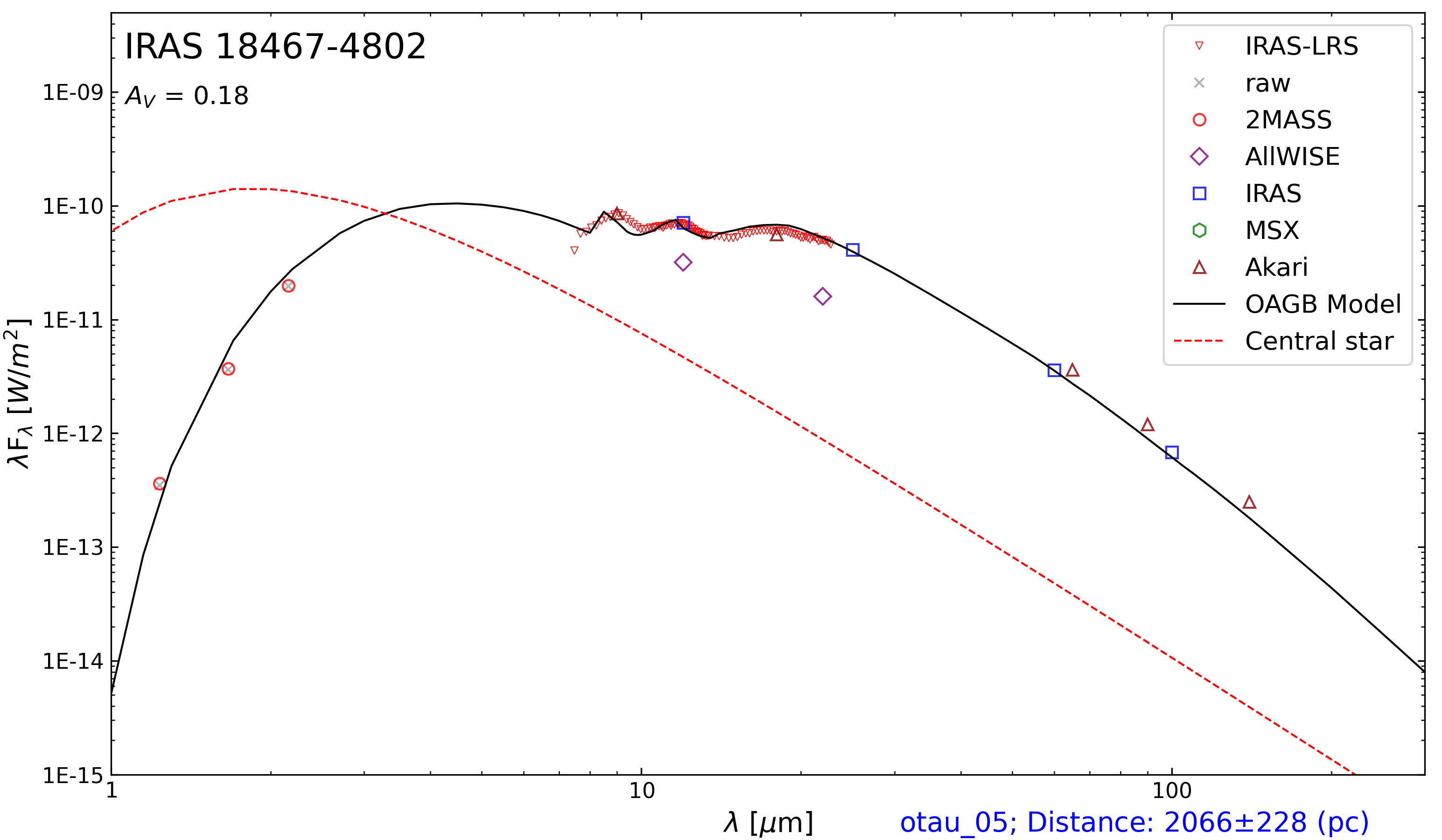}{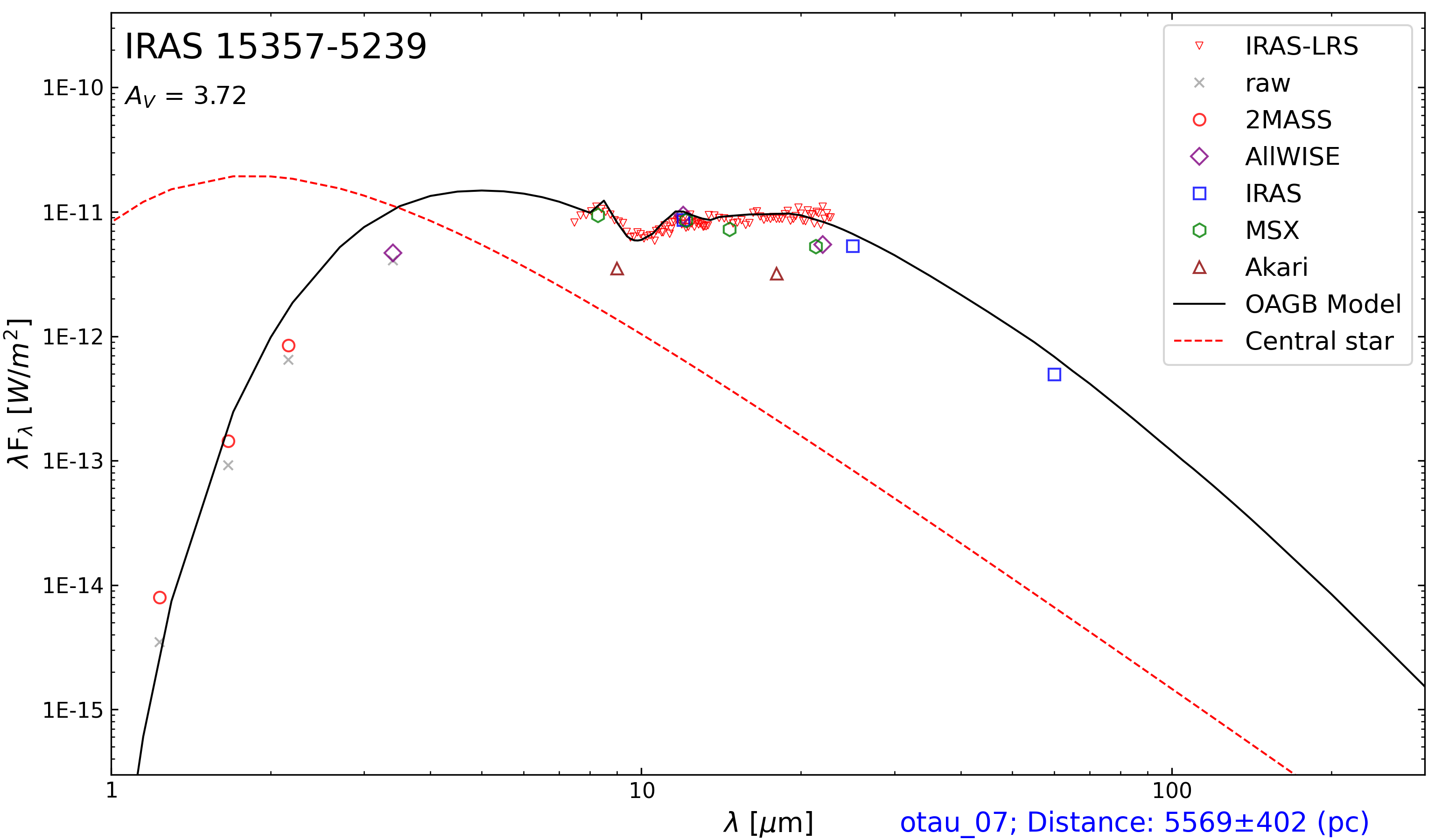}
\caption{Observed and model SEDs for OAGB stars with thick dust envelopes
($1 \leq \tau_{10} \leq 7$).
The top four panels show LMC stars, while the bottom four panels show Milky Way stars.
See Section~\ref{sec:seds}.} \label{f5}
\end{figure*}

\begin{figure*}
\centering
\smallplotfour{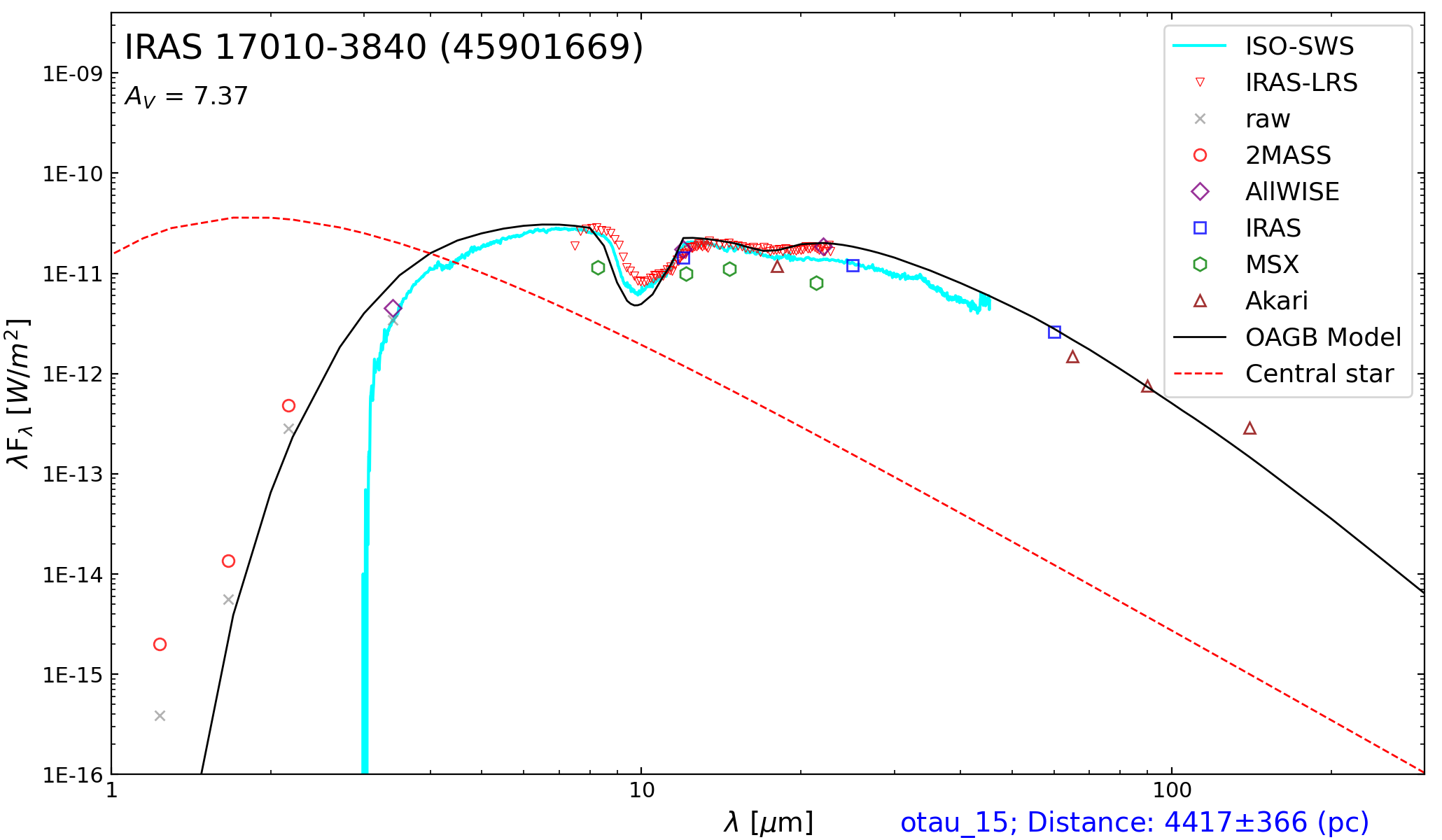}{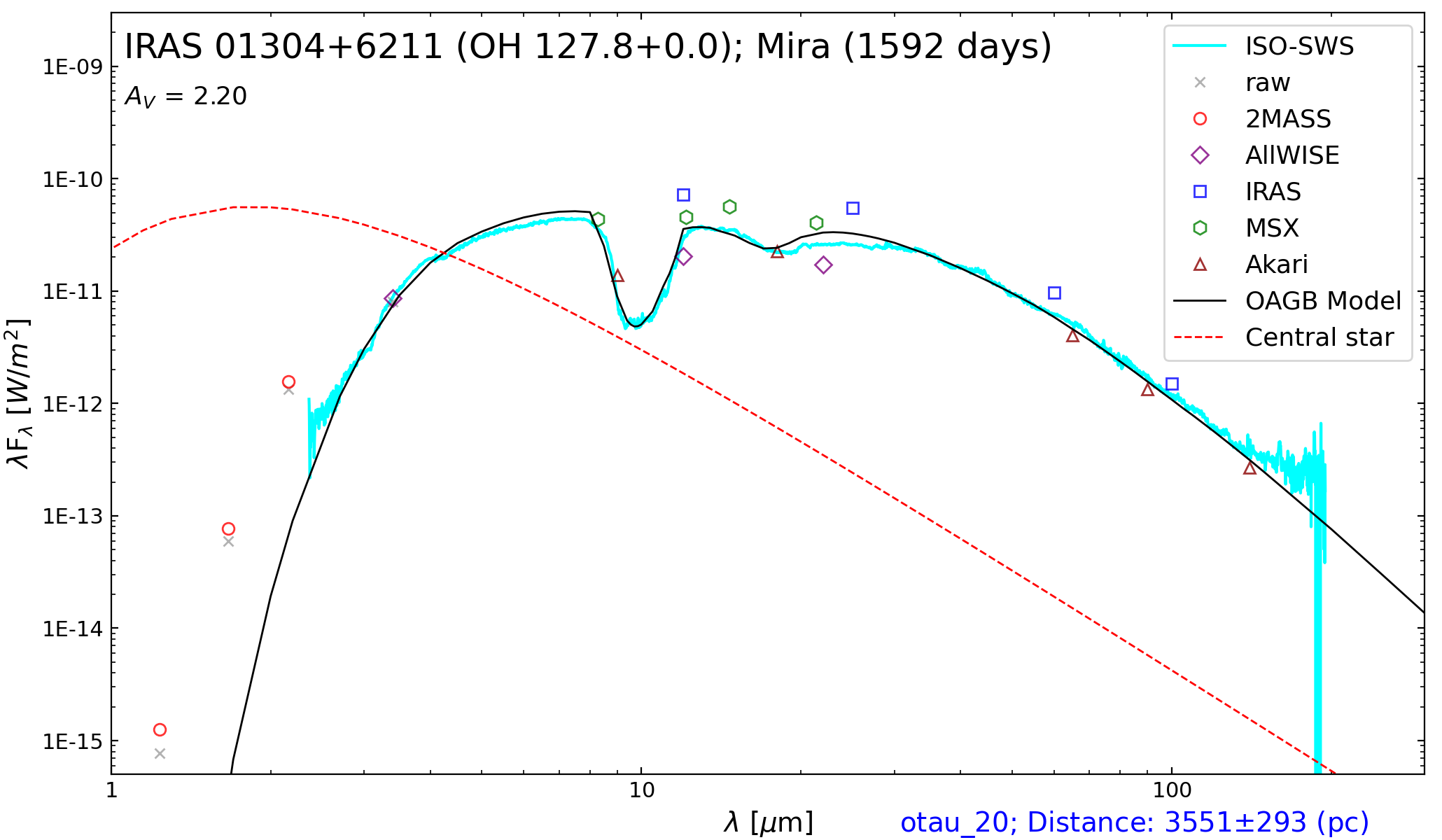}{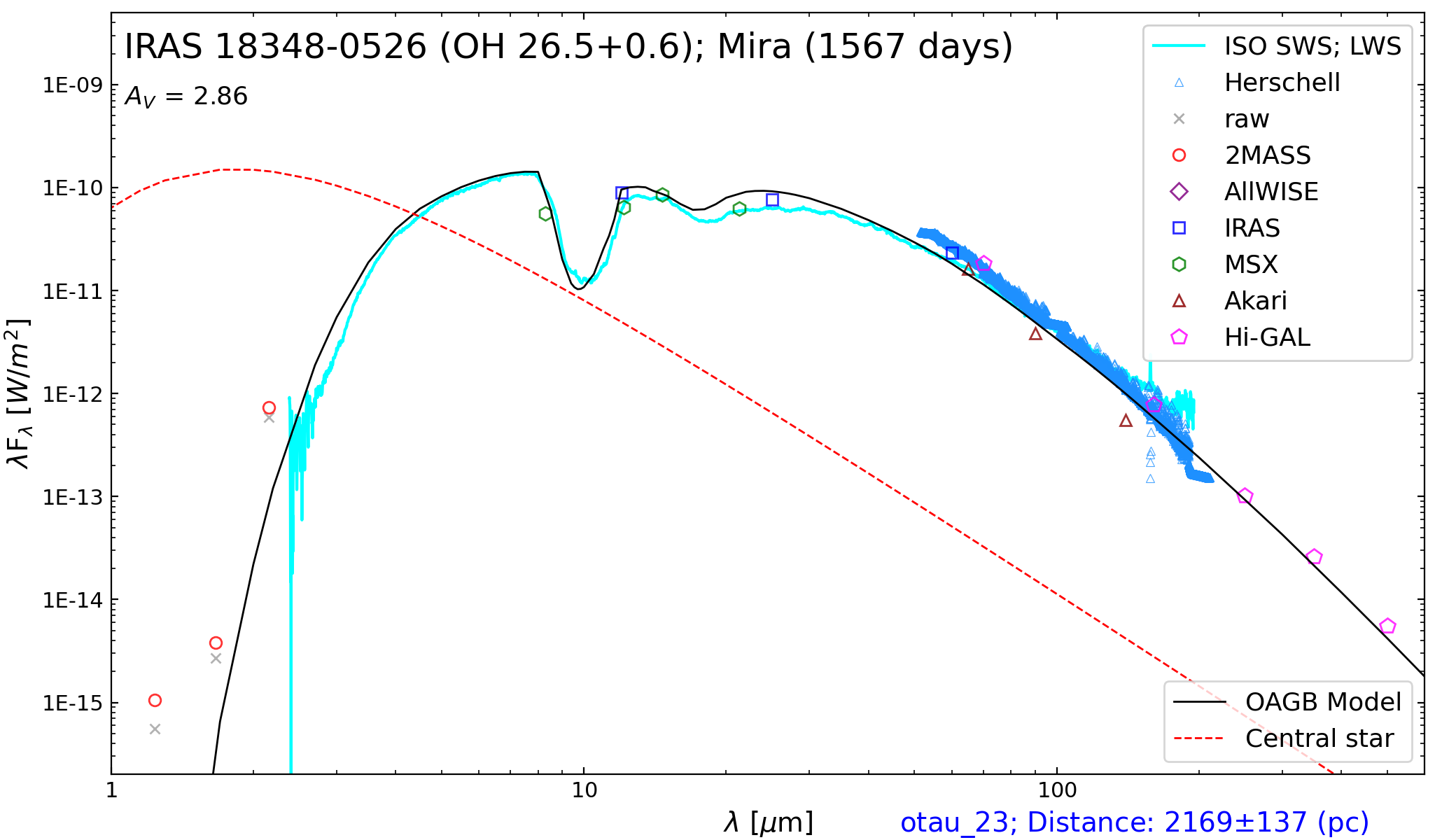}{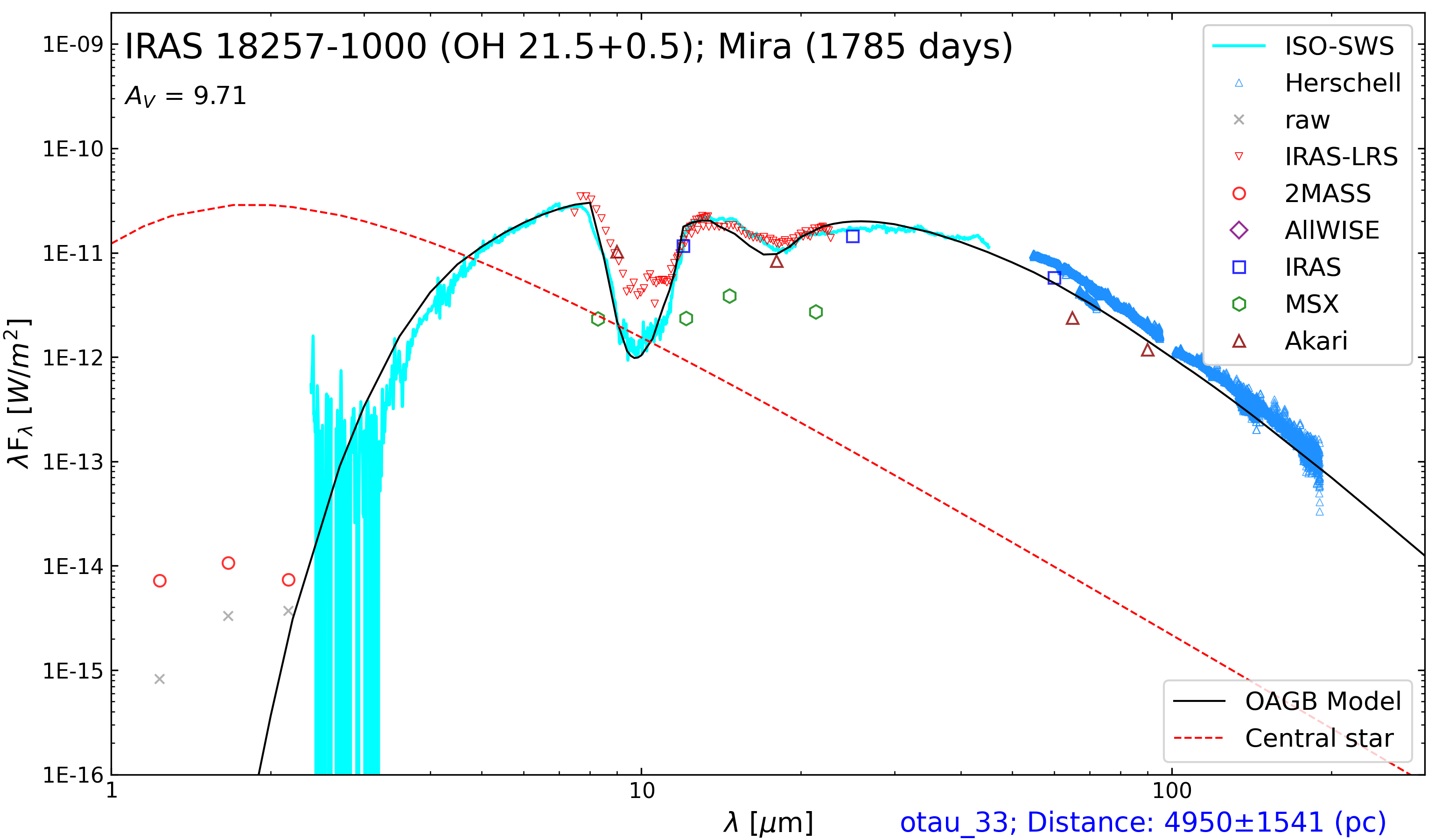}
\caption{Observed and model SEDs for Galactic OAGB stars with very thick dust envelopes ($\tau_{10} > 7$).
Such heavily obscured objects are not seen among LMC-OAGB stars (see Section~\ref{sec:seds}).} \label{f6}
\end{figure*}

\section{SEDs of OAGB stars}\label{sec:seds}

In this section, we compare observed SEDs of the OAGB stars in the LMC and the
Milky Way with theoretical models (see Section~\ref{sec:agbmodels}), exploring
the properties of OAGB stars in depth. For each object, theoretical model
parameters for the dust shell and central star can be derived by fitting the
observed SED. This also allows us to determine a theoretical distance that
closely matches the overall level of the observed SED.

The observed SEDs were constructed using photometric measurements from Gaia DR3,
2MASS, IRAS, MSX, AKARI, Spitzer, and AllWISE. Sub-millimeter (1.2 mm)
observational data from SIMBA detector installed at Swedish-ESO Submillimetre
Telescope (SEST; \citealt{dehaes2007}) and photometric data from Herschel
InfraRed Galactic Plane Survey (Hi-GAL; \citealt{elia2017}) in five bands between
70 and 500 $\mu$m are used when accessible.

For LMC-OAGB stars, Spitzer IRS (\citealt{jones2017}) spectral data were included
when available. For Galactic OAGB stars, IR spectra from the IRAS Low Resolution
Spectrograph (LRS; \citealt{kwok1997}), ISO Spectrometer (SWS and LWS;
\citealt{sylvester1999}), and THROES (caTalogue of HeRschel Observations of
Evolved Stars; \citealt{ramos-medina2018}) are used when accessible.

From the sample of Galactic OAGB stars with IRAS counterparts (MW-OAGB\_IC
objects; see Table~\ref{tab:tab1}), we selected a subset of 228 well-studied OAGB
stars, designated as MW-OAGB\_IC-S, for which additional observational data
(e.g., ISO spectra or IRAS LRS spectra) are available (see Section
\ref{sec:gagb}). Of these 228 MW-OAGB\_IC-S objects, 162 have distance
measurements based on Gaia DR3 parallaxes.

The observed SEDs of LMC-OAGB stars were fitted using a least-squares
optimization method, resulting in the 25 OAGB models (Table~\ref{tab:tab4}; see
Section~\ref{sec:lum-loagb} and Figure~\ref{f1}) that best reproduced the
observed SED shapes and minimized the differences between the scaling and true
distances of LMC-OAGB stars. These LMC-validated OAGB models were then applied to
the full sample of Galactic OAGB stars.

For the complete samples of OAGB stars in the LMC and the Milky Way, the
best-fitting OAGB model ($\tau_{10}$), along with the corresponding scaling
distance and its uncertainty, was determined by least-squares fitting of the
observed SEDs to the grid of the 25 OAGB models listed in Table~\ref{tab:tab4}.

To ensure reliable results, only SED fits with a coefficient of determination
$R^2 > 0.4$ and a relative uncertainty of the scaling distance $\sigma_d/d < 0.4$
are retained for all LMC and Galactic sample stars. $R^2$ is computed for
$\lambda F_\lambda$ in linear scale using the photometric data at $\lambda \ge
2~\mu$m, which determine the scaling distance. These criteria remove only the
objects for which no model reproduces the overall level of the SED or the scaling
distance is poorly constrained. Under this criterion, reliable best-fitting
models and scaling distances were obtained for 12,728 of 13,070 LMC-OAGB stars
(97.4 \%), 19,126 of 19,784 MW-OAGB\_IC stars (96.7 \%), and 48,971 of 51,093
MW-OAGB\_NI stars (95.8 \%).

For the LMC-OAGB stars, no interstellar extinction correction was applied (see
Section~\ref{sec:lmc-agb}). For the Galactic OAGB stars, the Gaia, 2MASS, and
AllWISE W1[3.4] and W2[4.6] fluxes were corrected for interstellar extinction
before the SED fitting (see Section~\ref{sec:dist-ext}). The SEDs of MW-OAGB
stars shown in Figures~\ref{f4}--\ref{f6} are the extinction-corrected SEDs used
in the fitting.

\subsection{SEDs for OAGB stars in both galaxies}\label{sec:mseds}

Figure~\ref{f4} compares the observed SEDs with OAGB models (see
Table~\ref{tab:tab4}) for eight representative OAGB stars with thin dust
envelopes ($0.001 \leq \tau_{10} < 1$) in the LMC (top four panels) and the Milky
Way (bottom four panels), arranged in order of increasing dust optical depth
($\tau_{10}$). The observed SEDs of OAGB stars in the two galaxies are broadly
similar. For each panel, the adopted $\tau_{10}$ value and the corresponding
scaling distance are indicated in the lower-right corner.

Figure~\ref{f5} compares the observed SEDs with OAGB models (see
Table~\ref{tab:tab4}) for eight representative OAGB stars with thick dust
envelopes ($1 \leq \tau_{10} \leq 7$, corresponding to the fourth model group in
Table~\ref{tab:tab4} with $L_* = 25.6 \times 10^{3}~L_\odot$) in the LMC (top
four panels) and the Milky Way (bottom four panels), arranged in order of
increasing dust optical depth. As in Figure~\ref{f4}, the observed SEDs of OAGB
stars in the LMC and the Milky Way exhibit broadly similar characteristics.

Figure~\ref{f6} compares the observed SEDs with OAGB models (see
Table~\ref{tab:tab4}) for four representative Galactic OAGB stars with very thick
dust envelopes ($\tau_{10} > 7$, corresponding to the fifth model group in
Table~\ref{tab:tab4} with $L_* = 30 \times 10^{3}~L_\odot$), arranged in order of
increasing dust optical depth. No counterparts with such high dust optical depths
are found among the LMC-OAGB sample stars.

Overall, a simple single-shell dust model incorporating silicate grains
(Table~\ref{tab:tab4}) provides a reasonable representation of the observed SEDs.
However, the theoretical dust shell model adopted in this study does not include
gas-phase radiative processes (see Section~\ref{sec:limit}) and therefore cannot
fully reproduce the observed visual and near-IR fluxes.

\begin{figure*}
\centering
\smallplotsix{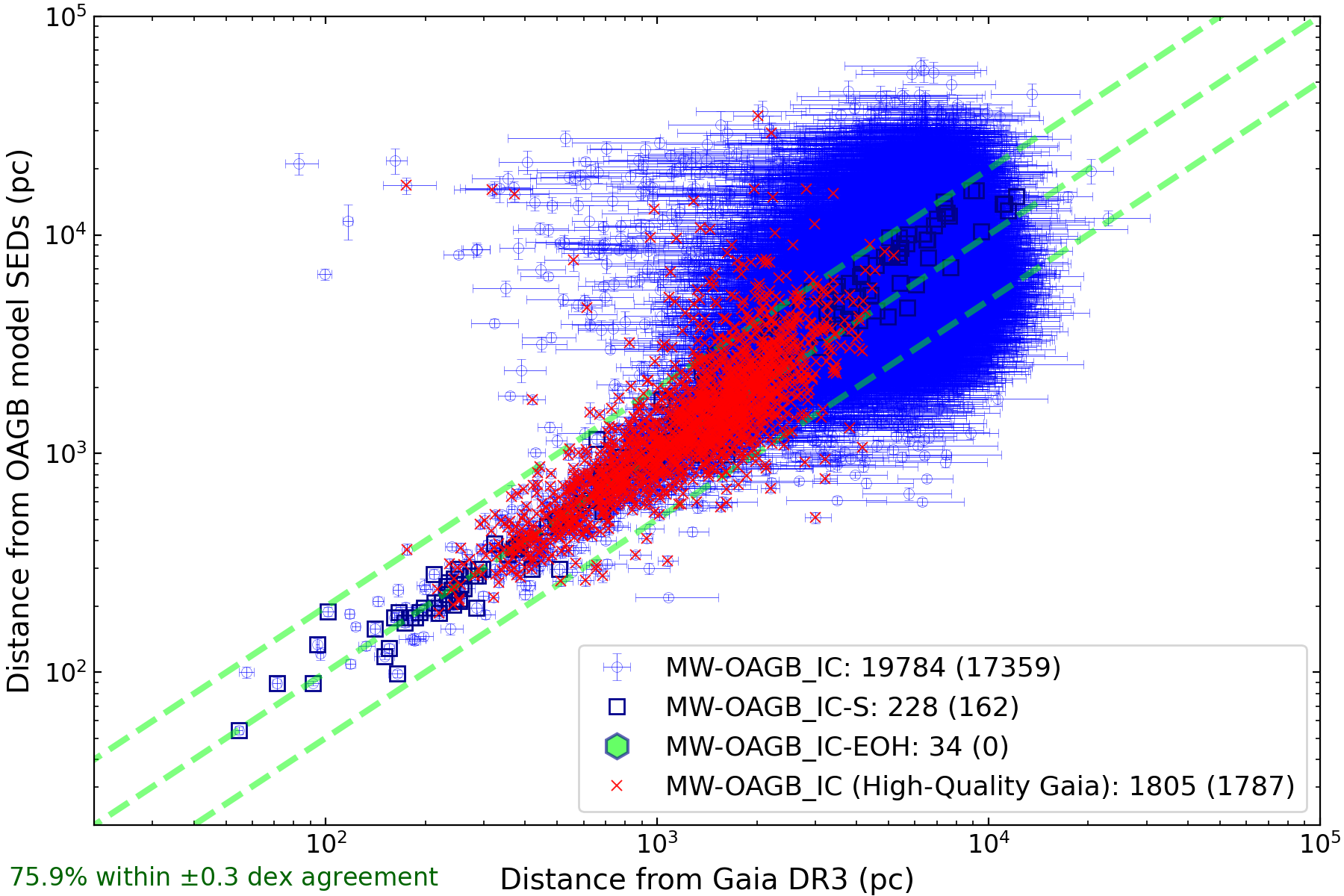}{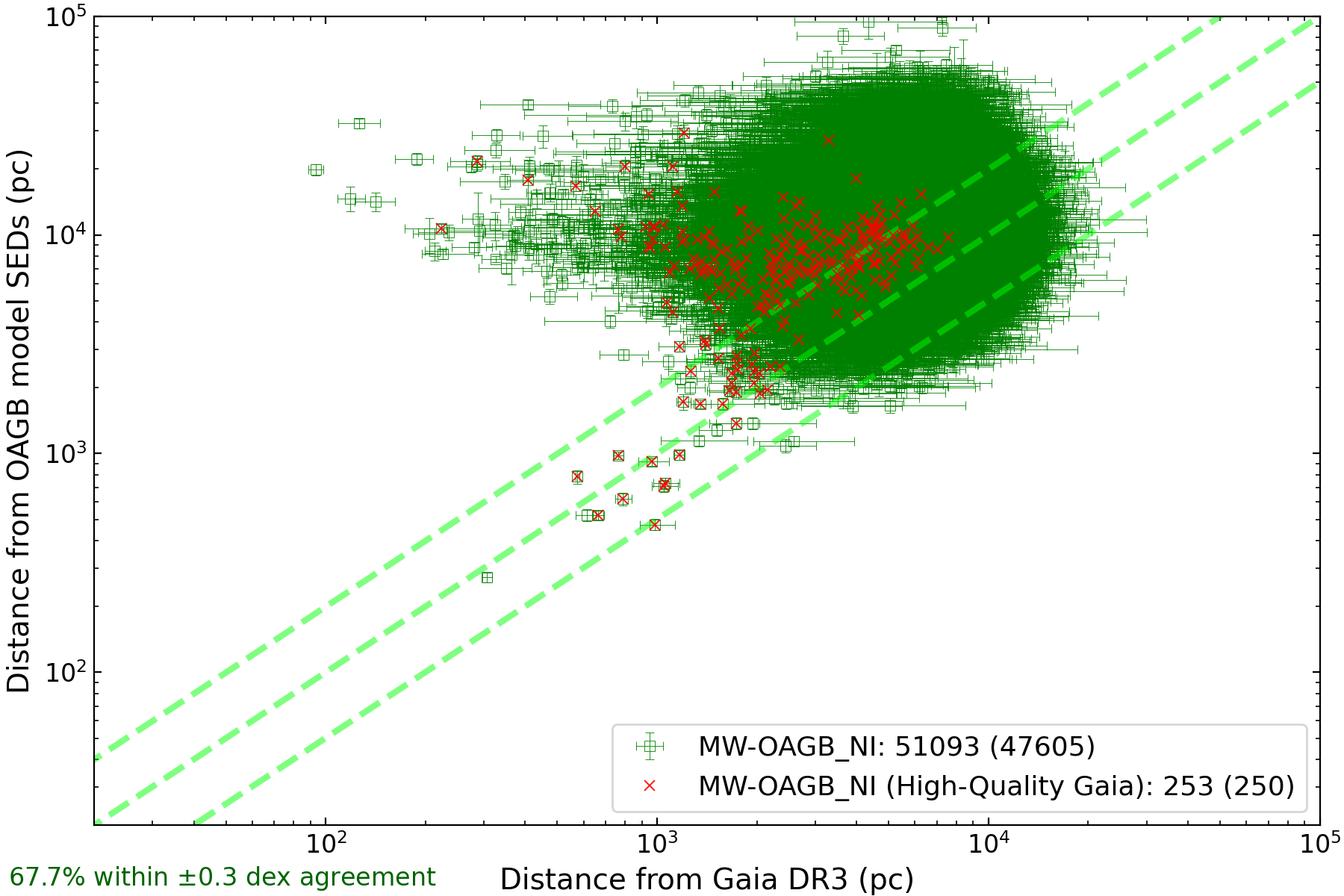}{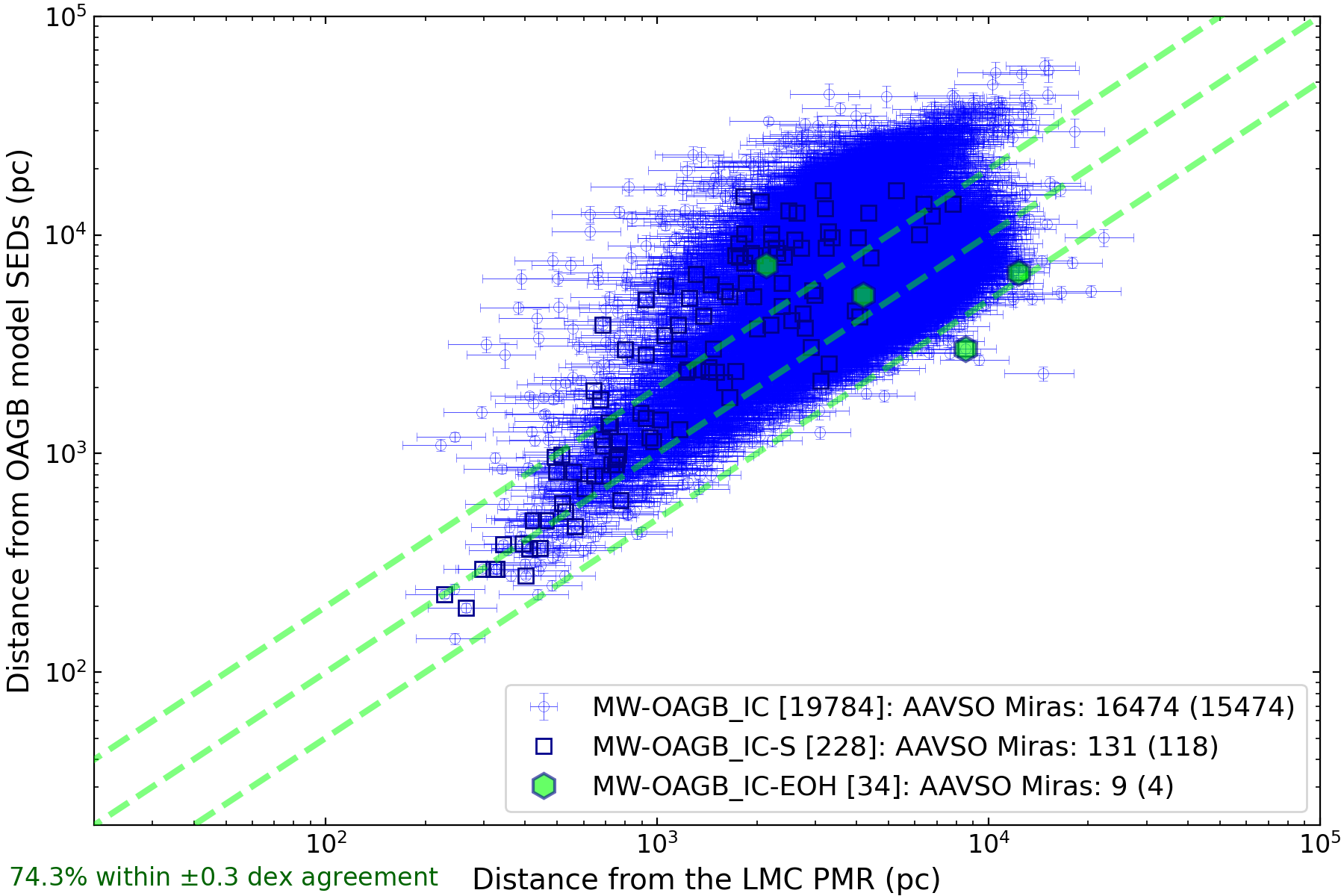}{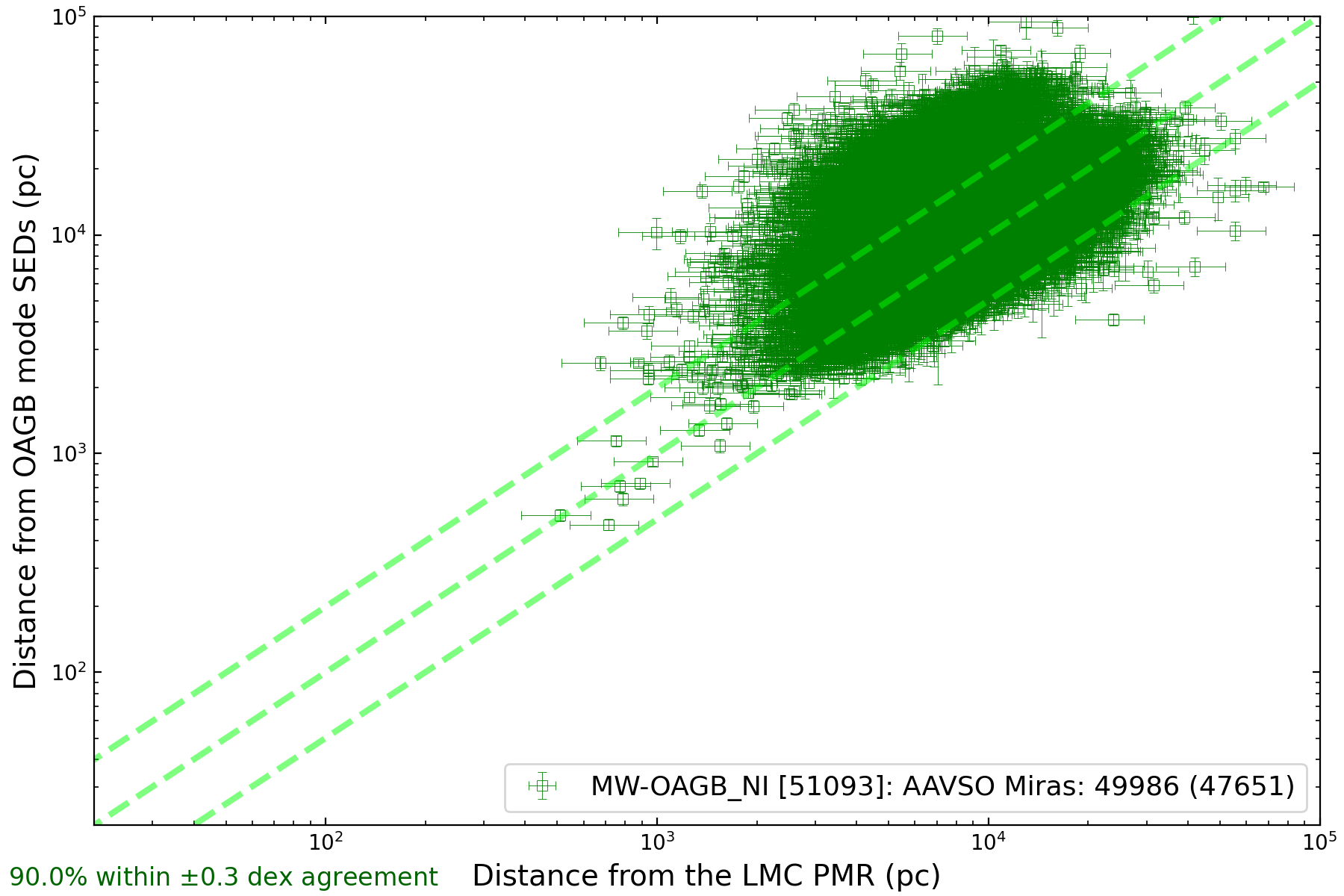}{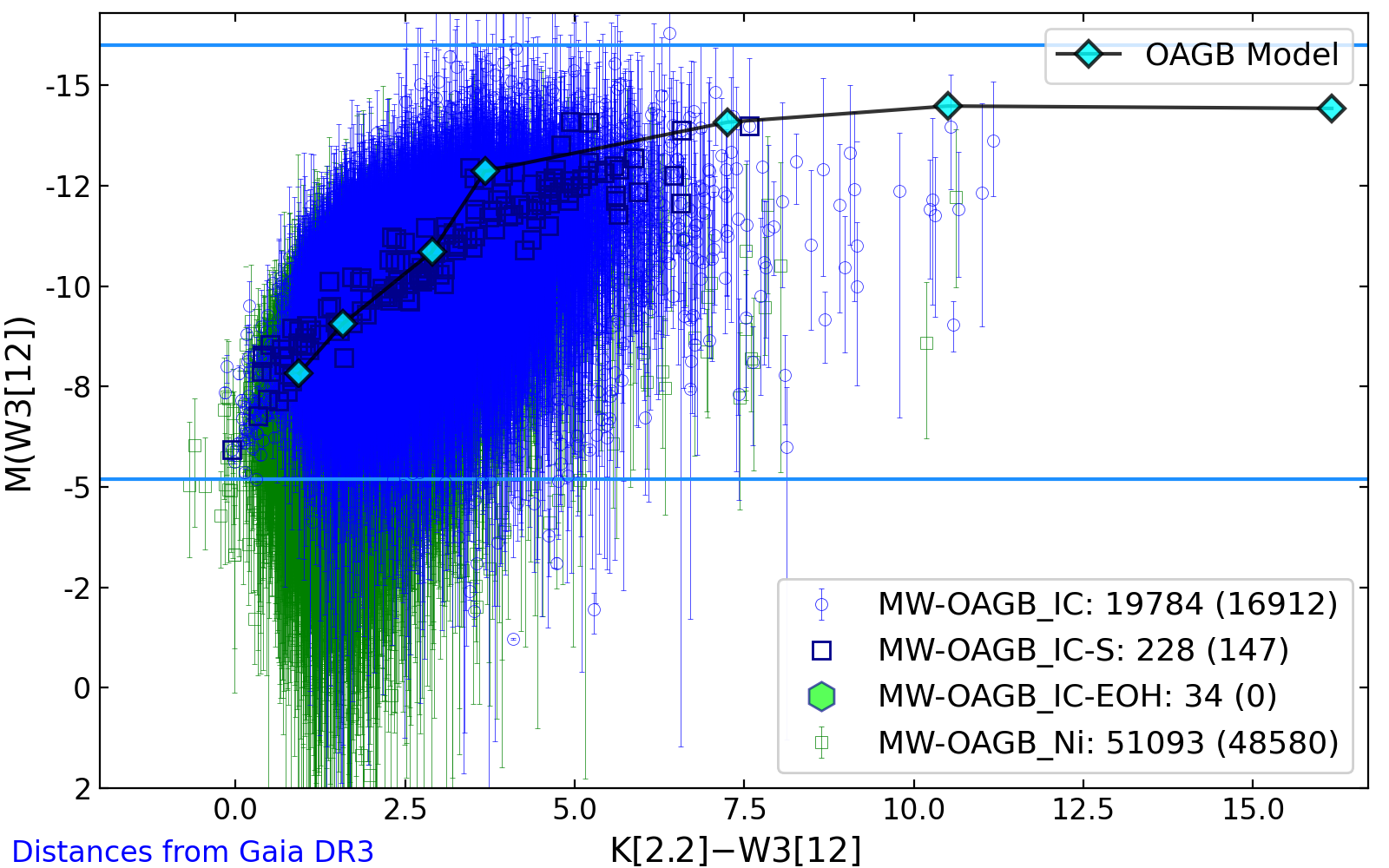}{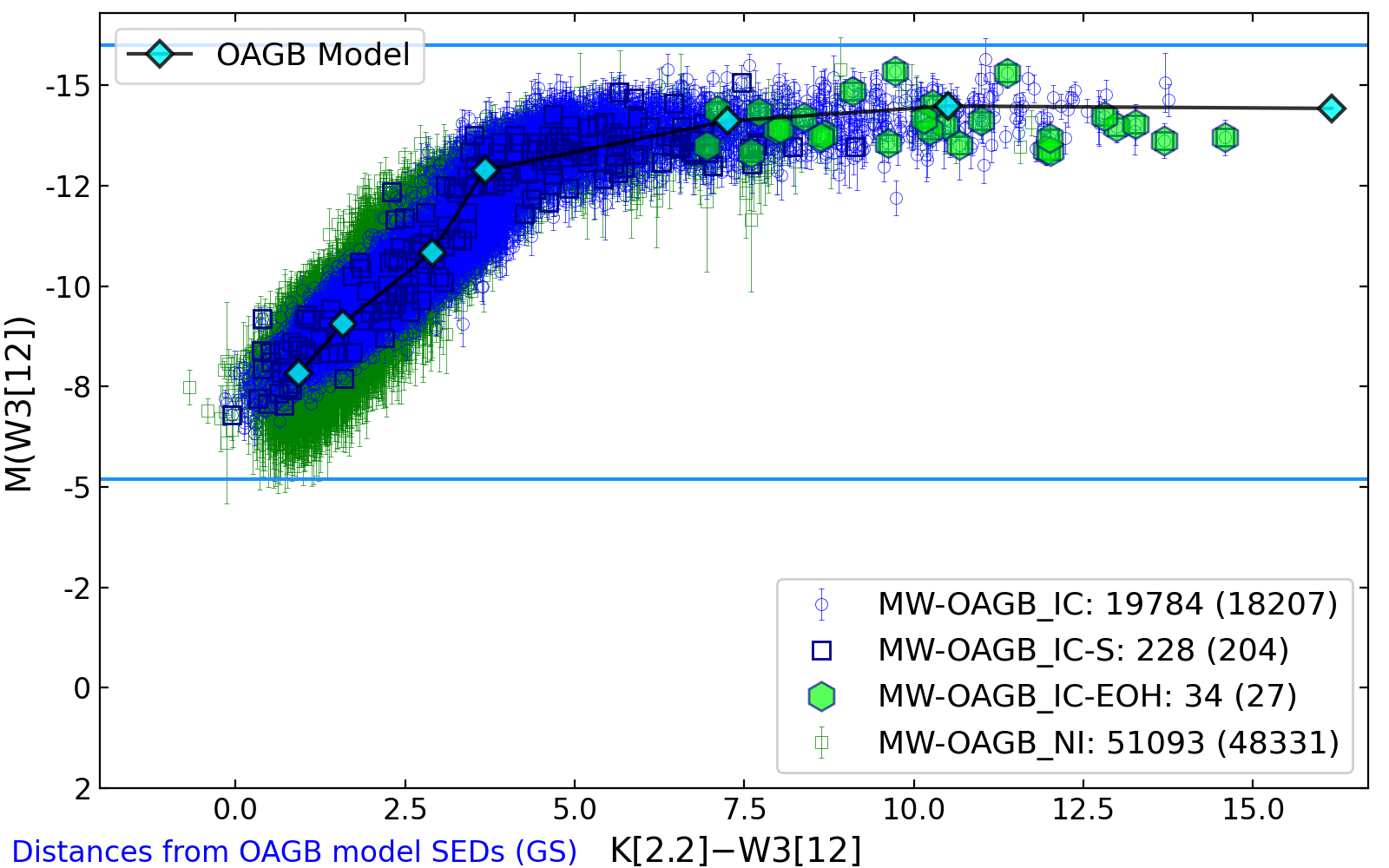}
\caption{For Galactic OAGB stars, the top four panels present a comparison of distance
estimates obtained from theoretical OAGB model SEDs, Gaia DR3 parallaxes, and the
PMR calibrated with LMC Mira variables (See Section~\ref{sec:sed-dist}).
The bottom panels show IR CMDs for Galactic OAGB stars using distances from Gaia DR3 (left)
and OAGB model SEDs (right; the same as the upper left panel of Figure~\ref{f3} with a different scale).
For OAGB models: $\tau_{10}$ = 0.001, 0.03, 0.3, 1, 7, 15, and 40 from left to right (see Table~\ref{tab:tab4}).
The error bars represent the uncertainties in the derived distances.
Red crosses in the top panels mark sources with high-quality Gaia astrometry (see Section~\ref{sec:dist-ext}).
The blue horizontal lines in the lower-left panel mark the upper and lower limits of the magnitude range of
the LMC-OAGB and MW-OAGB stars (See Section~\ref{sec:lum-loagb}).
The total number of objects in each subgroup is listed, with the numbers in parentheses
representing the subset for which the required data are available and therefore plotted.
See Section~\ref{sec:sed-dist} for details.}
\label{f7}
\end{figure*}

\subsection{SEDs and distances for Galactic OAGB stars}\label{sec:sed-dist}

To evaluate the reliability of distances derived from theoretical OAGB model
SEDs, a benchmark is necessary. A useful approach is to compare these distances
with those obtained through other methods: specifically, distances based on Gaia
parallaxes (\citealt{bailer-jones2021}) and those derived from the
period–magnitude relation (PMR) of Mira variables, which constitute the vast
majority of OAGB stars. The top four panels of Figure~\ref{f7} compare the
distances derived from the OAGB model SEDs, Gaia parallax measurements, and the
PMR for Miras in the LMC (see Section~\ref{sec:pmr}; Equation~\ref{eq:2}).

The top panels of Figure~\ref{f7} compare distances derived from parallaxes with
those obtained from the theoretical OAGB model SEDs, for both MW-OAGB\_IC and
MW-OAGB\_NI stars. 1805 MW-OAGB\_IC and 253 MW-OAGB\_NI objects have high-quality
Gaia astrometry (Section~\ref{sec:dist-ext}) and are marked separately in the
figure. Red crosses in the top panels mark sources with high-quality Gaia
astrometry. Despite considerable scatter, the MW-OAGB\_IC objects show better
overall agreement, and the 162 MW-OAGB\_IC-S objects agree very well.

Of the 17,359 MW-OAGB\_IC objects (upper-left panel of Figure~\ref{f7}), 75.9\%
fall within $\pm$0.3 dex agreement between the two distance estimates, as
indicated by the lime dotted lines. This corresponds to agreement within a factor
of 2.0 (approximately $-50$\% to $+100$\%). Restricting the comparison to the
stars with high-quality Gaia astrometry (Section~\ref{sec:dist-ext}; 1787 of the
1805 stars have both SED and Gaia distances) raises this to 91.4\%. Part of this
improvement reflects the fact that the astrometrically selected sample is
preferentially nearby, but the improvement persists when the samples are compared
over the same distance range, here defined by the SED distance: within 2 kpc the
agreement improves from 86.5\% to 97.0\%, and beyond 2 kpc from 74.6\% to 77.9\%.
The SED distance scale is therefore consistent with Gaia parallaxes wherever the
latter are reliable, and the residual scatter at larger distances is likely to be
dominated by parallax uncertainties, although the present comparison cannot
separate this from the contribution of the SED models. Among the 162
MW-OAGB\_IC-S objects, the agreement reaches 100\%.

Of the 47,605 MW-OAGB\_NI objects (upper-right panel of Figure~\ref{f7}), 67.7\%
fall within $\pm$0.3 dex. Here the astrometric selection does not improve the
overall agreement: for the stars with high-quality Gaia astrometry
(Section~\ref{sec:dist-ext}; 250 of the 253 stars have both SED and Gaia
distances) it falls to 33.6\%. The behavior differs with distance, however:
within 2~kpc the agreement improves from 77.1\% to 94.1\% (16 of the 17
high-quality stars), as for MW-OAGB\_IC, whereas beyond 2~kpc it falls from
67.7\% to 29.2\%. The overall decrease arises because most of the high-quality
MW-OAGB\_NI stars lie beyond $\sim$2~kpc, where their parallaxes are comparable
to their uncertainties. Selecting such stars by
\texttt{parallax\_over\_error}~$>$~5 preferentially retains those whose measured
parallaxes are scattered upward, i.e., whose Gaia distances are underestimated,
and this bias is enhanced when the parallax uncertainties are underestimated
\citep{andriantsaralaza2022}. Consistent with this, both the SED and the PMR
distances of these stars are systematically larger than their Gaia distances
(upper-right panel of Figure~\ref{f7} and middle panel of Figure~\ref{f16}). In
addition, many of them may lie in crowded fields at low Galactic latitude, where
blending can yield formally precise but unreliable parallaxes. For MW-OAGB\_NI
objects, therefore, the comparison with Gaia distances is not informative beyond
$\sim$2~kpc, because the Gaia parallaxes of these distant stars are too uncertain
to provide a reliable reference.

In the upper-left regions of both upper panels in Figure~\ref{f7}, groups of
objects show Gaia DR3 distances significantly smaller than the SED-based
distances (beyond $\sim$2 kpc). The discrepancy is more pronounced for
MW-OAGB\_NI objects, which are generally more distant than MW-OAGB\_IC objects.
Most of these are faint Mira variables, which are luminous and distant, yet Gaia
DR3 reports much smaller distances for them. A likely explanation is that either
the Gaia counterpart identification is inaccurate, or the Gaia astrometry itself
is unreliable for these sources; blending in crowded fields can yield formally
precise but incorrect parallaxes. A systematic offset in the same sense has been
reported by several authors. \citet{bhattacharya2024}, who also derive distances
from SED fits, find that their SED distances are consistently larger than the
Gaia parallax distances once the former exceed 2 kpc. \citet{mcdonald2025} and
\citet{lopezmarti2025} similarly conclude that Gaia distances become unreliable
beyond 2-3 kpc, and optical and radio VLBI parallaxes of LPVs can currently be
compared only out to $\sim$2.5 kpc (\citealt{nakagawa2026}).

Similar discrepancies were observed when using alternative distances derived from
the period–magnitude relation for Miras in the LMC (see Section~\ref{sec:dpmr};
upper two panels of Figure~\ref{f16}), further supporting this interpretation.
Since most nearby OAGB stars (where Gaia parallax-based distances are more
reliable) show much better agreement, the SED-based distances appear to be
generally trustworthy for OAGB stars.

The middle panels of Figure~\ref{f7} show a comparison between distances derived
using the PMR obtained from Miras in the LMC (see Section~\ref{sec:pmr};
Equation~\ref{eq:2}) and those obtained from OAGB model SEDs for both OAGB\_IC
and OAGB\_NI stars that are Miras according to AAVSO. The results demonstrate a
high level of consistency between the two distance estimates. Among the 15,474
MW-OAGB\_IC objects (see the left panel), 74.3\% fall within a $\pm$0.3 dex
agreement between the two distance estimates. Of the 47,651 MW-OAGB\_NI objects
(see the right panel), 90.0\% lie within this range of agreement. While both
distance estimation methods have inherent limitations and uncertainties, they
currently represent the only practical approaches for deriving reliable distances
from observed photometric data, such as SEDs and variability, for a large sample
of OAGB stars.

The agreement between the three distance estimates depends on the dust optical
depth. For Galactic OAGB stars with $\tau_{10} < 1$, which constitute 90\% of the
sample, the SED distances agree with the PMR distances to within 0.1 dex in the
median at all distances, and with the Gaia distances of stars with high-quality
astrometry to within 0.02 dex within 2 kpc (97\% within $\pm$0.3 dex); the larger
offsets beyond 2 kpc reflect the bias of the Gaia distances discussed above. For
$\tau_{10} \ge 1$, the SED distances are larger than the PMR distances by 0.6 dex
in the median, in both subgroups and at all distances. Almost all of these stars
lie beyond 2 kpc and only 26 have high-quality Gaia astrometry, so the comparison
with Gaia is not informative for them. For the 18 stars with $\tau_{10} \ge 1$
within 2 kpc, the SED distances agree with the Gaia distances in the median
($-0.01$ dex) and exceed the PMR distances by 0.4 dex, which suggests that the
LMC calibration holds for the luminous nearby OH/IR stars, although the number is
too small for a firm conclusion.

The luminosity of the fourth model group in Table~\ref{tab:tab4} ($25.6 \times
10^{3}~L_\odot$) is set by the few LMC-OAGB stars reaching these optical depths,
whereas in the Milky Way, where silicate dust forms more efficiently (see
Sections~\ref{sec:limit} and \ref{sec:comoc}), stars of lower luminosity may also
develop such thick envelopes, so that the SED distances of Galactic stars with
$\tau_{10} \ge 1$ may be overestimated. The PMR distances of these stars are
biased in the opposite sense, because the W3[12] period--magnitude relation is
calibrated with LMC Miras with thin dust shells and does not account for the dust
emission of thicker envelopes. The 0.6 dex offset is therefore an upper limit to
the overestimate of the SED distances, which should be regarded as less reliable
than those of stars with thinner dust shells.

The lower-left panel of Figure~\ref{f7} displays the WISE-2MASS CMD for Galactic
OAGB stars, using distances obtained from parallax observations (Gaia DR3). This
panel exhibits significant scatter, with many derived absolute magnitudes falling
outside the upper or lower limits of the OAGB star luminosity distribution
determined in Section~\ref{sec:lum-loagb}. Specifically, 30 MW-OAGB\_IC objects
and 662 MW-OAGB\_NI objects (all of these are Mira variables) exhibit much
fainter M(W3[12]) magnitudes than -5.2, which is the faintest value observed
among OAGB stars in the LMC (see Section~\ref{sec:lum-loagb}). For these objects,
the Gaia DR3 counterparts may be inaccurate, or the Gaia astrometry itself is
unreliable (Section~\ref{sec:dist-ext}), resulting in underestimated distances.

In contrast, the lower-right panel of Figure~\ref{f7} shows the WISE-2MASS CMD
for Galactic OAGB stars, based on distances estimated from theoretical OAGB model
SEDs. Compared with the Gaia DR3-based CMD, this diagram displays substantially
reduced scatter, with the inferred absolute magnitudes falling predominantly
within the expected luminosity range. These results suggest that, for a large
sample of Galactic OAGB stars, distances estimated from model SED fitting are
generally more robust than those derived from Gaia DR3 parallaxes. The plotted
distance error bars are based on the assumption that the theoretical OAGB models
are accurate. However, as these models rely on various approximations (see
Section~\ref{sec:limit}), they may introduce additional sources of uncertainty.
Nevertheless, the derived absolute magnitudes for each color index (or
$\tau_{10}$) are still expected to represent reasonable average values for a
large sample of Galactic OAGB stars. This comparison is not a fully independent
test of the SED distances, since the expected luminosity range is itself defined
by the OAGB models; its force lies in the fact that the Gaia-based magnitudes
fall well outside the range spanned by the LMC-OAGB stars, whose distance is
independently known.

\begin{figure*}
\centering
\smallplottwo{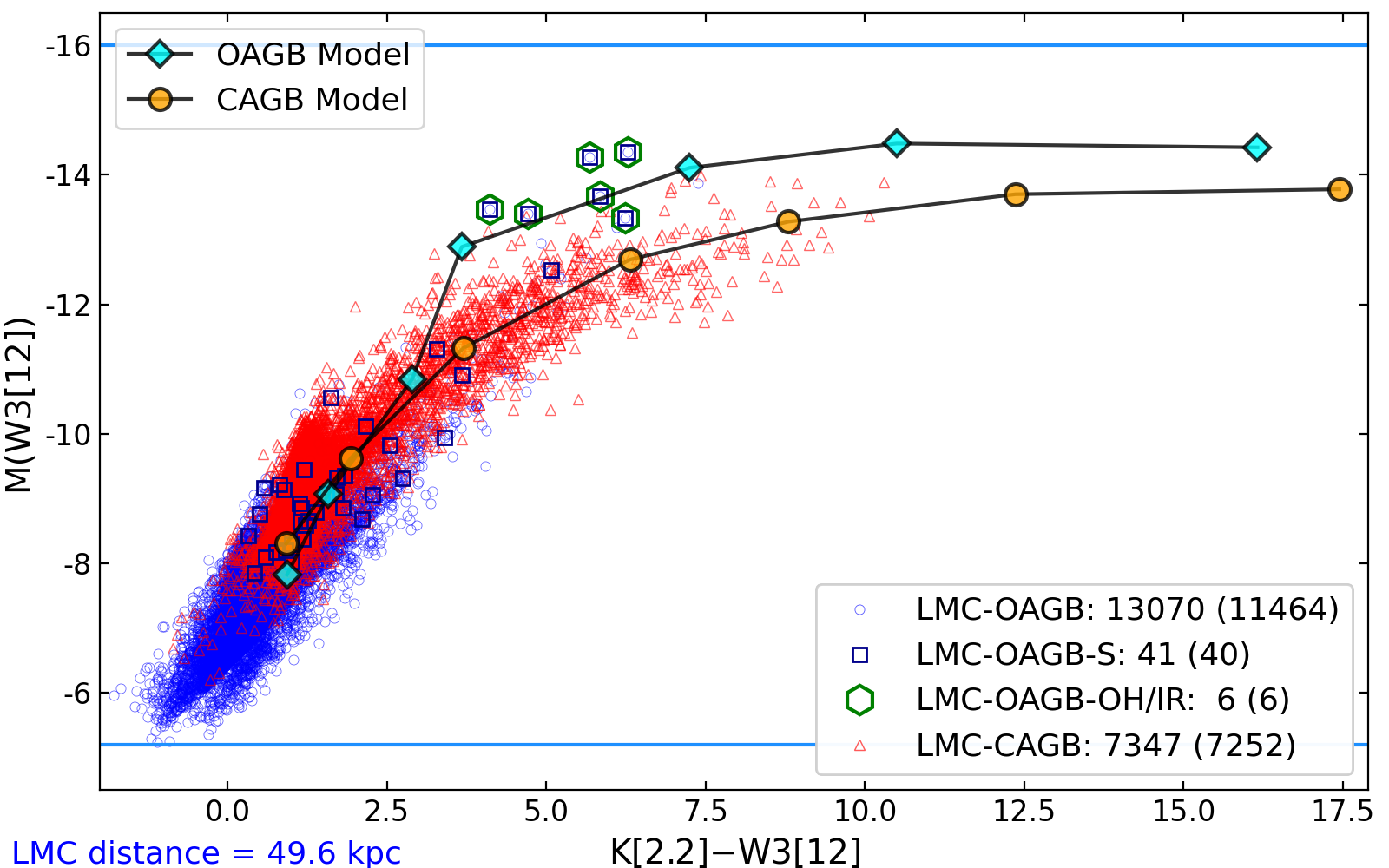}{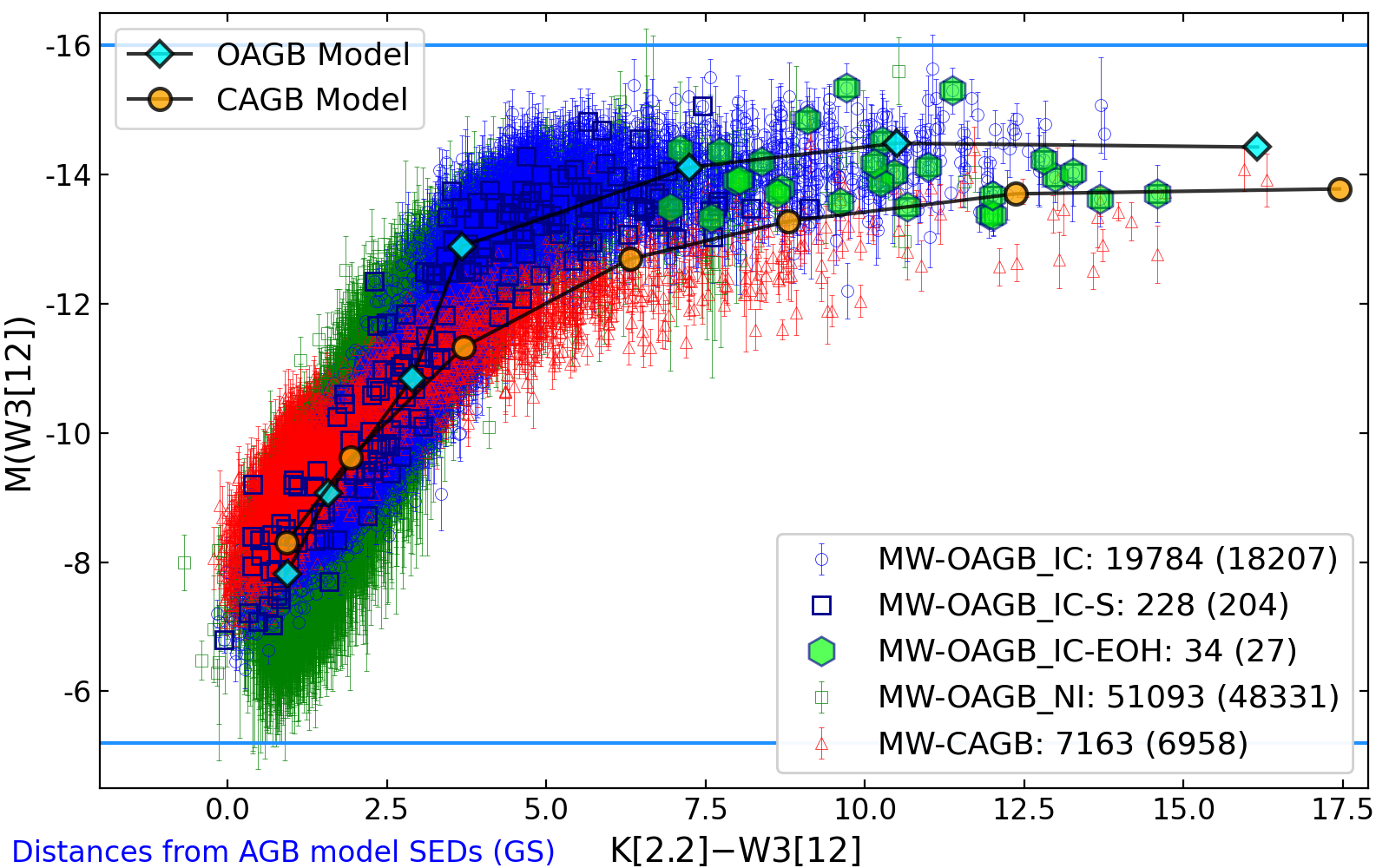}\caption{
WISE–2MASS CMDs for AGB stars in the LMC and the Milky Way (see
Table~\ref{tab:tab1}), compared with the theoretical models. The OAGB models are
listed in Table~\ref{tab:tab4}, while the CAGB models are given in Table 3 of
\citealt{suh2026}. For Galactic AGB stars, distances were derived from the AGB model SEDs, incorporating
Gaussian Monte Carlo scatter (GS) in $\log_{10} d$, with $\sigma=\pm0.0829$ dex
for OAGB stars (see Section~\ref{sec:lum-loagb}) and $\sigma=\pm0.0798$ dex for CAGB stars
(see \citealt{suh2026}).
The total number of objects in each subgroup is listed, with the numbers in parentheses
representing the subset for which the required data are available and therefore plotted.
Vertical error bars for AGB stars represent the uncertainties in the distances derived from the AGB model SEDs.
The blue horizontal lines in the left panel mark the upper and lower limits of the magnitude range of
the LMC-OAGB and MW-OAGB stars (See Section~\ref{sec:lum-loagb}).
See Section~\ref{sec:ircmd} for details.} \label{f8}
\end{figure*}

\begin{figure*}
\centering
\smallplotfour{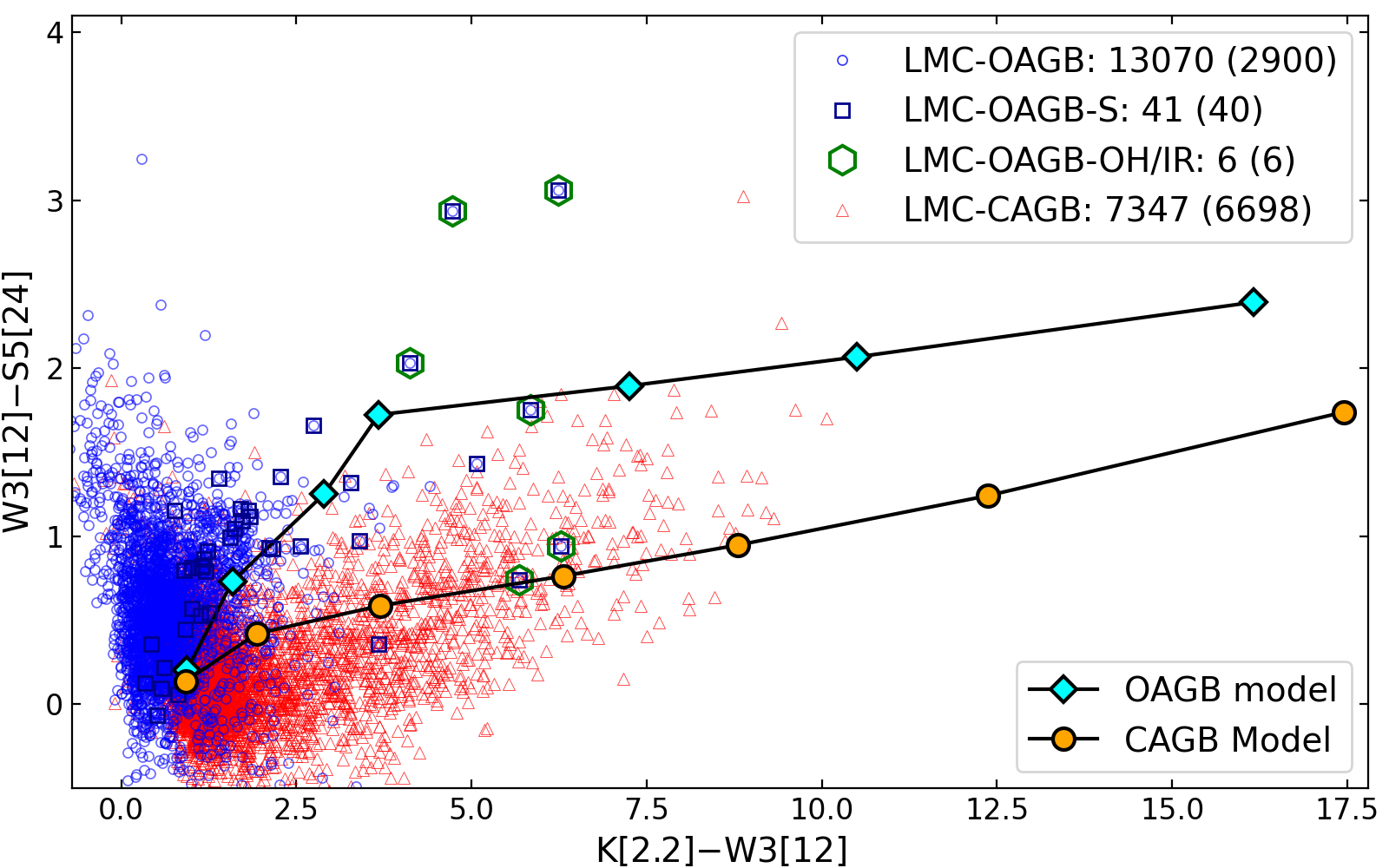}{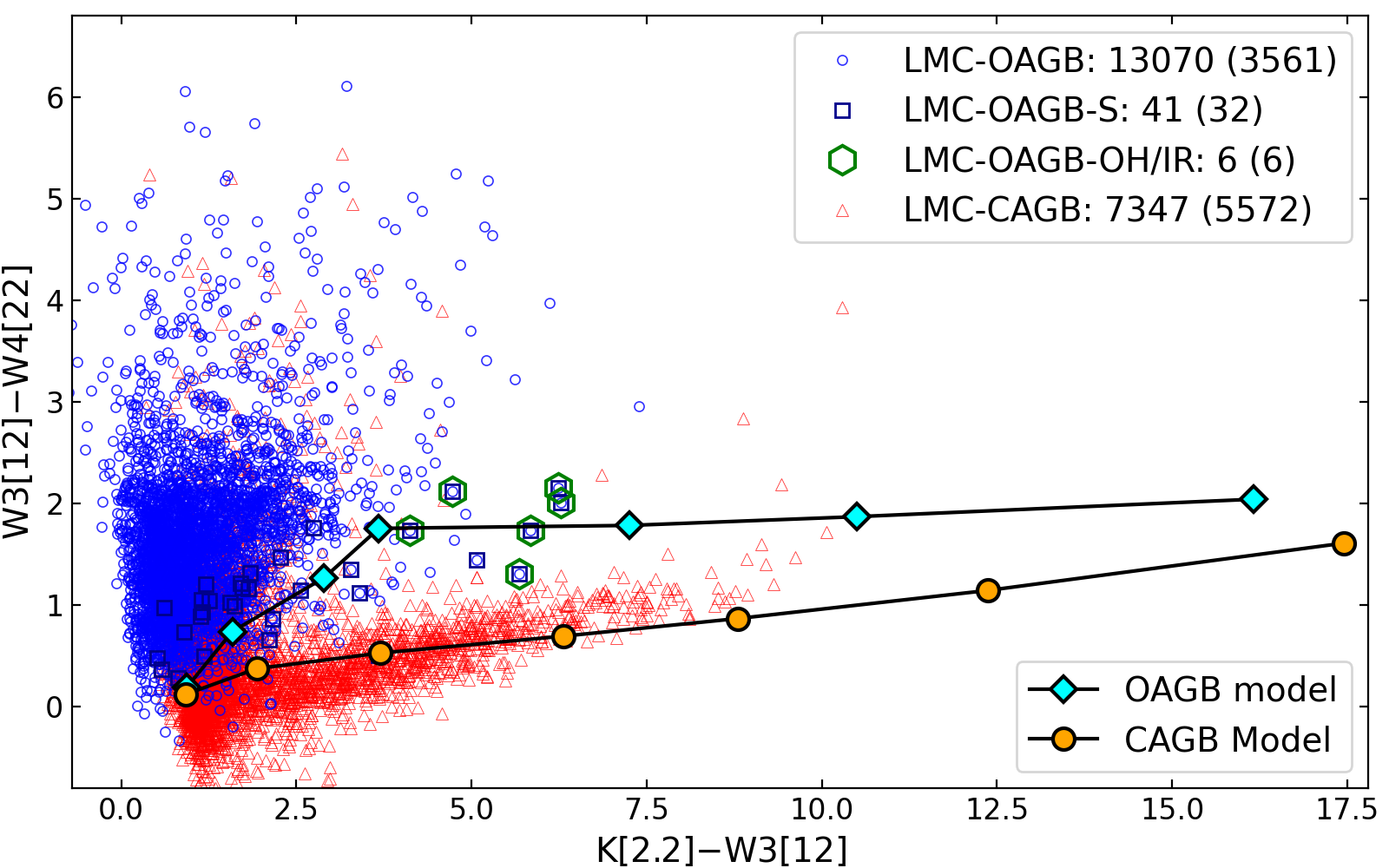}{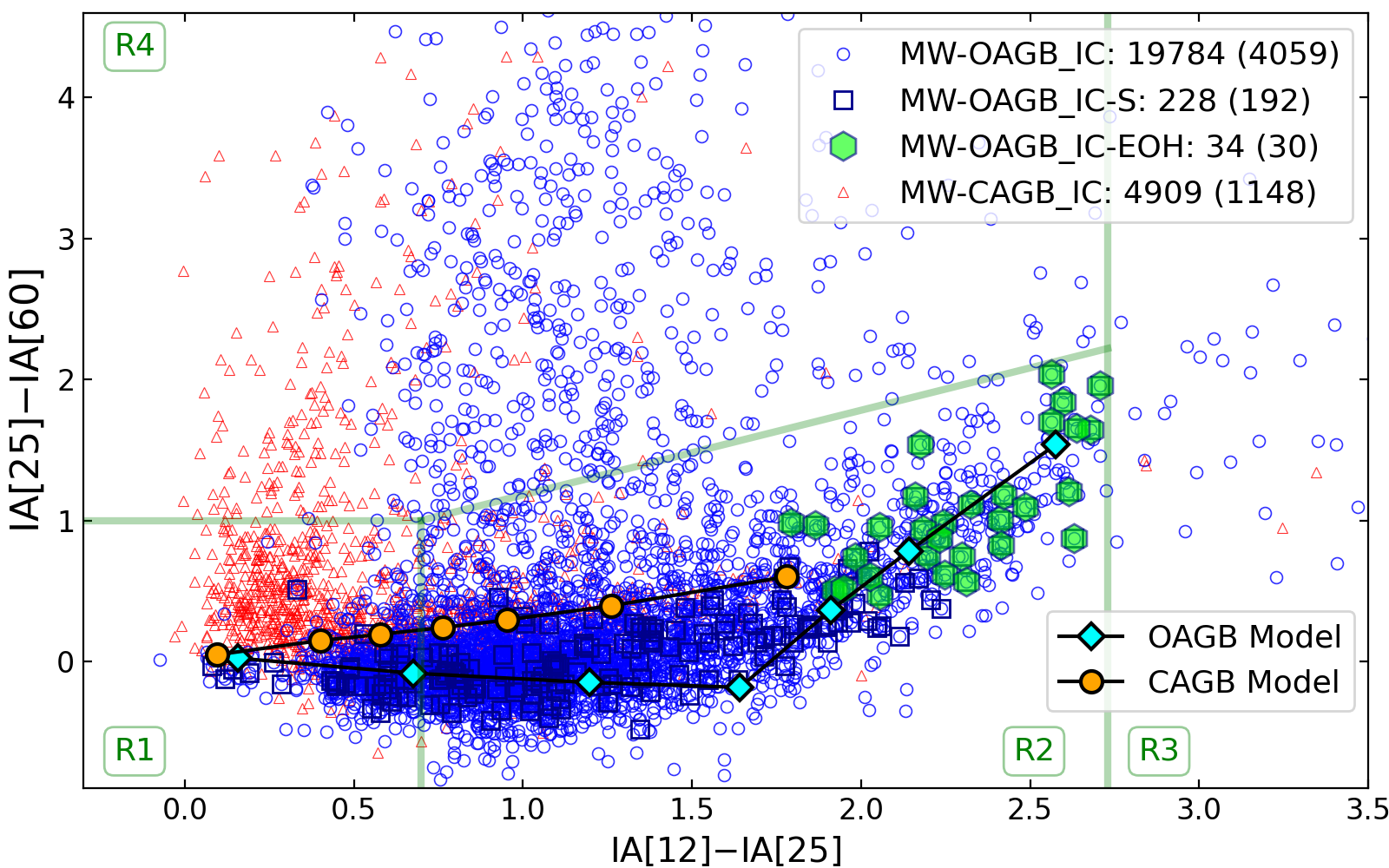}{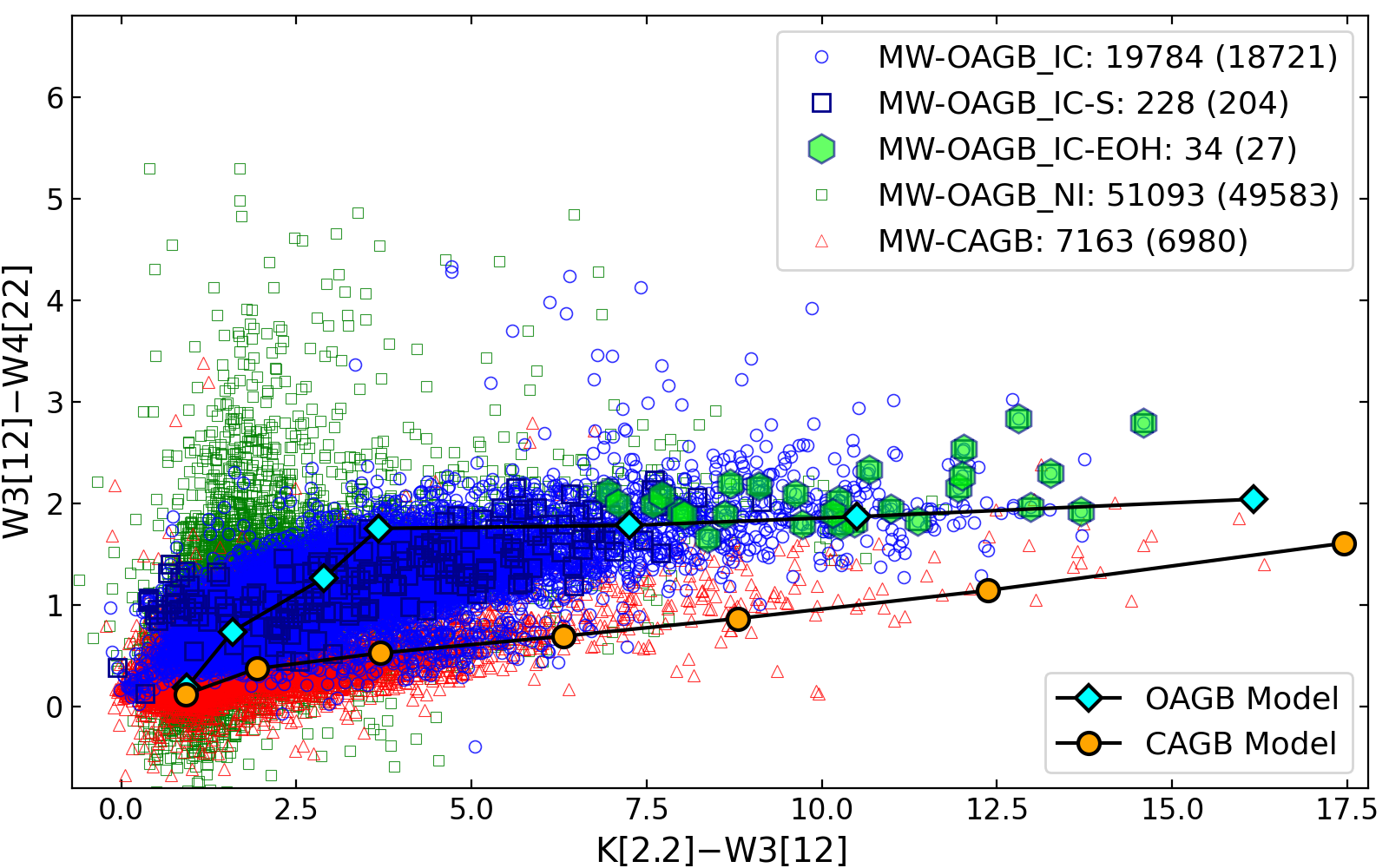}\caption{IR 2CDs for sample OAGB stars (see Table~\ref{tab:tab1}) in the LMC (upper panels)
and the Milky Way (lower panels) compared with theoretical models (Section~\ref{sec:agbmodels}).
For OAGB models: $\tau_{10}$ = 0.001, 0.03, 0.3, 1, 7, 15, and 40 from left to right (see Table~\ref{tab:tab4}).
For CAGB models: $\tau_{10}$ = 0.0001, 0.01, 0.1, 0.5, 1, 2, and 4 from left to right (see Table 3 in \citealt{suh2026}).
The total number of objects in each subgroup is listed, with the numbers in parentheses
representing the subset for which the required data are available and therefore plotted.
See Section~\ref{sec:ir2cd}.} \label{f9}
\end{figure*}

\section{CMDs and 2CDs in IR bands}\label{sec:irpro}

\citet{srinivasan2009} investigated OAGB stars in the LMC using a variety of
modeling approaches, presenting several CMDs and 2CDs together with comparisons
between observations and theoretical predictions. In the present study, we extend
that study by incorporating additional IR wavelength bands and a larger sample of
OAGB stars in both the LMC and the Milky Way, allowing a more detailed comparison
with theoretical models constructed using a more sophisticated modeling
framework.

In this section, we investigate properties of OAGB stars in the LMC and the Milky
Way (see Section~\ref{sec:sample}) in IR bands using CMDs and 2CDs for the large
sample of the AGB stars compared with the theoretical models for AGB stars (see
Section~\ref{sec:agbmodels}). For comparison, we also consider CAGB stars in both
galaxies (see \citealt{suh2026}). Table~\ref{tab:tab3} lists the visual and IR
bands used for the CMDs and IR 2CDs presented in this work (see
Section~\ref{sec:photdata}).

For Galactic AGB stars, we use distance estimates derived from model SEDs for
Galactic OAGB stars (see Section~\ref{sec:seds}) and Galactic CAGB stars (see
\citealt{suh2026}) for the CMDs and 2CDs. For the Galactic sample stars, using
the distance and position of each object, we obtained the corresponding
extinction value as explained in Section~\ref{sec:dist-ext}. We then calculated
absolute magnitudes and intrinsic (unreddened) colors across a broad range of
wavelengths (see Section~\ref{sec:dist-ext}). Utilizing these parameters, we
present various CMDs and 2CDs.

A distinct subgroup of 228 well-known MW-OAGB\_IC-S stars, including 34
MW-OAGB\_IC-EOH stars (extreme OH/IR stars, a subset of MW-OAGB\_IC-S stars), is
prominently highlighted in all the CMDs and 2CDs. Detailed SEDs for these stars
were compared with theoretical models (see Section~\ref{sec:mseds}).

\subsection{IR CMDs}\label{sec:ircmd}

In this Subsection, we compare IR CMDs for both OAGB and CAGB stars in the LMC
and the Milky way. Figure~\ref{f8} presents IR CMDs featuring absolute magnitudes
in W3[12] plotted against K[2.2]$-$W3[12] colors. Seven representative OAGB
models (OM1a to OM5i; see Table~\ref{tab:tab4}) are superimposed on the CMDs,
arranged sequentially from left to right. For CAGB stars, we adopt the results
from \citet{suh2026}.

Figure~\ref{f8} depicts a CMD that plots absolute magnitudes in W3[12] against
K[2.2]$-$W3[12] colors for both OAGB and CAGB stars in the LMC and the Milky way.
OAGB stars consistently exhibit brighter absolute magnitudes in W3[12] across all
K[2.2]$-$W3[12] color ranges. The results indicate that the theoretical models
align reasonably well with the observed properties of OAGB and CAGB stars.

\citet{suh2026} found that CAGB stars in the LMC and the Milky Way exhibit
broadly similar properties in IR CMDs, although the LMC sample lacks the most
heavily dust-enshrouded CAGB stars. The contrast is even more pronounced for OAGB
stars, as the LMC contains only a small number of objects with high dust optical
depths, as discussed in Section~\ref{sec:lum-goagb}.

\subsection{IR 2CDs}\label{sec:ir2cd}

The upper panels of Figure~\ref{f9} show IR 2CDs for the LMC sample stars, while
the lower panels show the corresponding diagrams for the Galactic sample stars.
The upper-left panel of Figure~\ref{f9} presents a Spitzer-WISE-2MASS 2CD, with
W3[12]$-$S5[24] plotted against K[2.2]$-$W3[12] for the LMC sample stars.

The lower-left panel of Figure~\ref{f9} presents an IRAS 2CD, with
IA[25]$-$IA[60] plotted against IA[12]$-$IA[25] for Galactic sample stars with
IRAS counterparts. Following \citet{suh2024a}, this 2CD is divided into four
regions (R1--R4) to facilitate the classification of different types of objects.
Most OAGB stars, including all MW-OAGB\_IC-S objects, are located in regions R1
and R2. E-AGB stars with relatively thin dust envelopes are predominantly found
in R1, whereas TP-AGB stars with thicker dust envelopes occupy R2. The
theoretical models for OAGB stars provide a useful framework for interpreting the
distribution of the observed data points. Most PAGB stars and PNe are
concentrated in the far-right region (R3; \citealt{suh2015}). The upper-left
region (R4) contains various AGB stars with detached or complex dust envelopes,
such as visual carbon stars (\citealt{suh2024a}) and binary systems.

The upper-right panel of Figure~\ref{f9} shows a WISE-2MASS 2CD, with
W3[12]$-$W4[22] plotted against K[2.2]$-$W3[12] for the LMC sample stars, while
the lower-right panel presents the corresponding diagram for the Galactic sample
stars. These 2CDs provide a clearer separation between OAGB and CAGB stars. The
theoretical dust-shell models for both OAGB and CAGB stars reproduce the overall
distributions of the observed data reasonably well. The distributions of OAGB
stars on these 2CDs are broadly similar in the two galaxies, except that the LMC
sample lacks objects with redder K[2.2]$-$W3[12] colors (i.e., larger dust
optical depths, $\tau_{10} > 7$).

\section{Discussion: OAGB and CAGB stars in the LMC and the Milky
Way}\label{sec:comoc}

\citet{suh2026} showed that CAGB stars in the LMC and the Milky Way have broadly
similar distributions in IR SEDs, CMDs, and 2CDs, despite the absence of the most
heavily dust-enshrouded CAGB stars in the LMC sample. In contrast, the present
study reveals a much greater difference for OAGB stars: heavily obscured objects
with very large dust optical depths ($\tau_{10} > 7$) are absent in the LMC-OAGB
sample (see Sections~\ref{sec:lmc-agb} and \ref{sec:seds}), whereas 408 of the
19,126 MW-OAGB\_IC objects with available model parameters (2.1\%) reach these
optical depths, corresponding to the fifth group in Table~\ref{tab:tab4}. This
suggests that the LMC-OAGB sample does not extend to the most heavily obscured
stage seen in the Galactic population.

These heavily obscured OAGB stars are generally believed to be TP-AGB objects
representing the most massive and/or most evolved members of the OAGB population,
having evolved from relatively high-mass progenitors. CAGB stars, in contrast,
are thought to originate from intermediate-mass progenitors that undergo
sufficient third dredge-up to become carbon rich (see Section~\ref{sec:intro}).
The near-absence of extreme OAGB stars, together with the much smaller deficiency
of CAGB stars, suggests that the LMC is relatively deficient in the high-mass
progenitors required to produce the most extreme OAGB population.

Two metallicity-dependent effects specific to O-rich chemistry may also
contribute. First, the lower metallicity of the LMC (approximately half solar)
limits the refractory elements available for silicate dust formation, making it
more difficult for O-rich stars to develop the very thick silicate dust envelopes
characteristic of OH/IR stars. Second, fewer third dredge-up episodes are
required for the C/O ratio to exceed unity (e.g., \citealt{karakas2014}), causing
stars to become C-rich earlier and shortening the O-rich phase. By contrast,
carbon dust formation is largely insensitive to metallicity because it depends on
carbon freshly dredged up from the stellar interior rather than on the initial
abundance of refractory elements. This naturally explains why the deficiency of
CAGB stars in the LMC is much smaller than that of OAGB stars.

Together, a deficiency of high-mass progenitors, suppressed silicate dust
formation, and an earlier transition to the C-rich phase likely contribute to the
scarcity of heavily obscured OAGB stars in the LMC.

Extending a similar study to nearby galaxies such as M31 or M33, whose masses and
metallicities are more comparable to those of the Milky Way, would provide a
valuable test of these interpretations and help establish whether the observed
differences primarily reflect galaxy mass, metallicity, or both.

\begin{figure*}
\centering
\smallplotfour{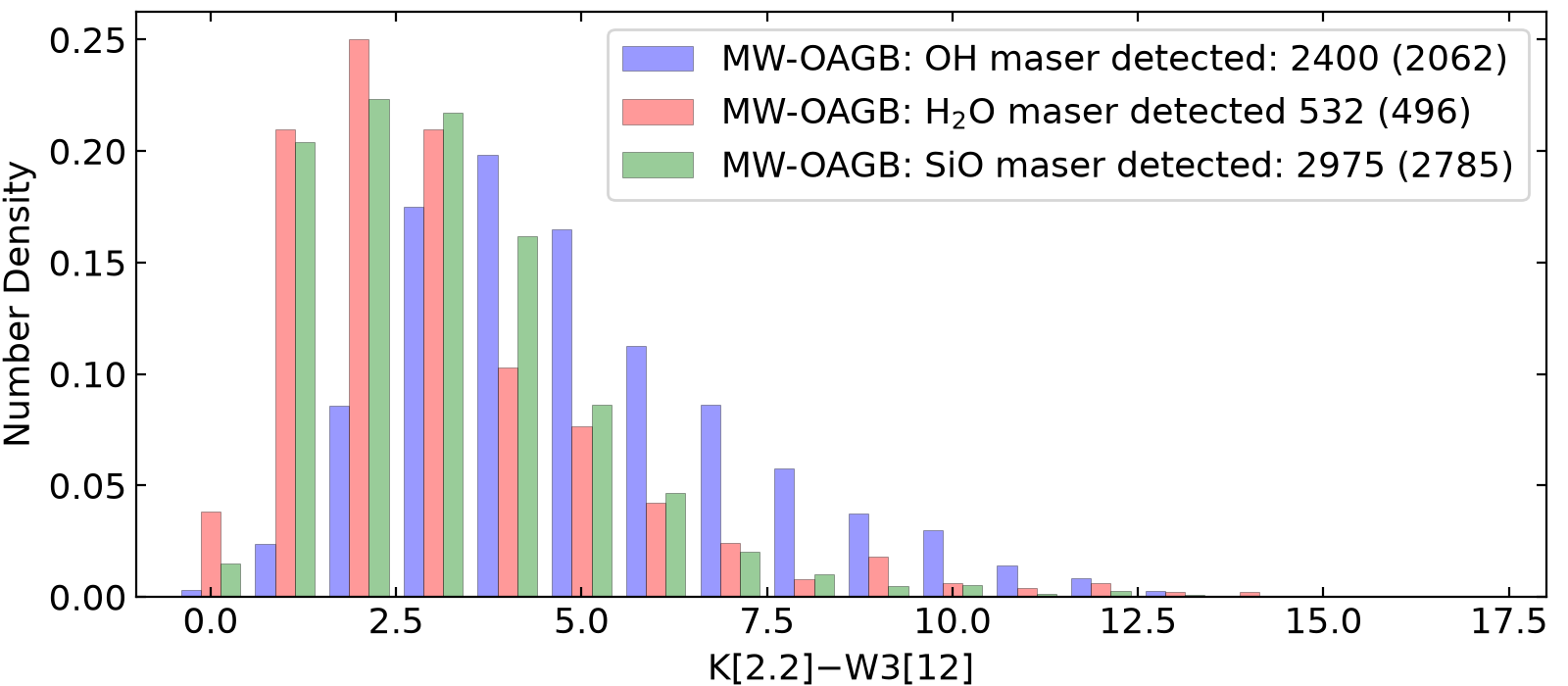}{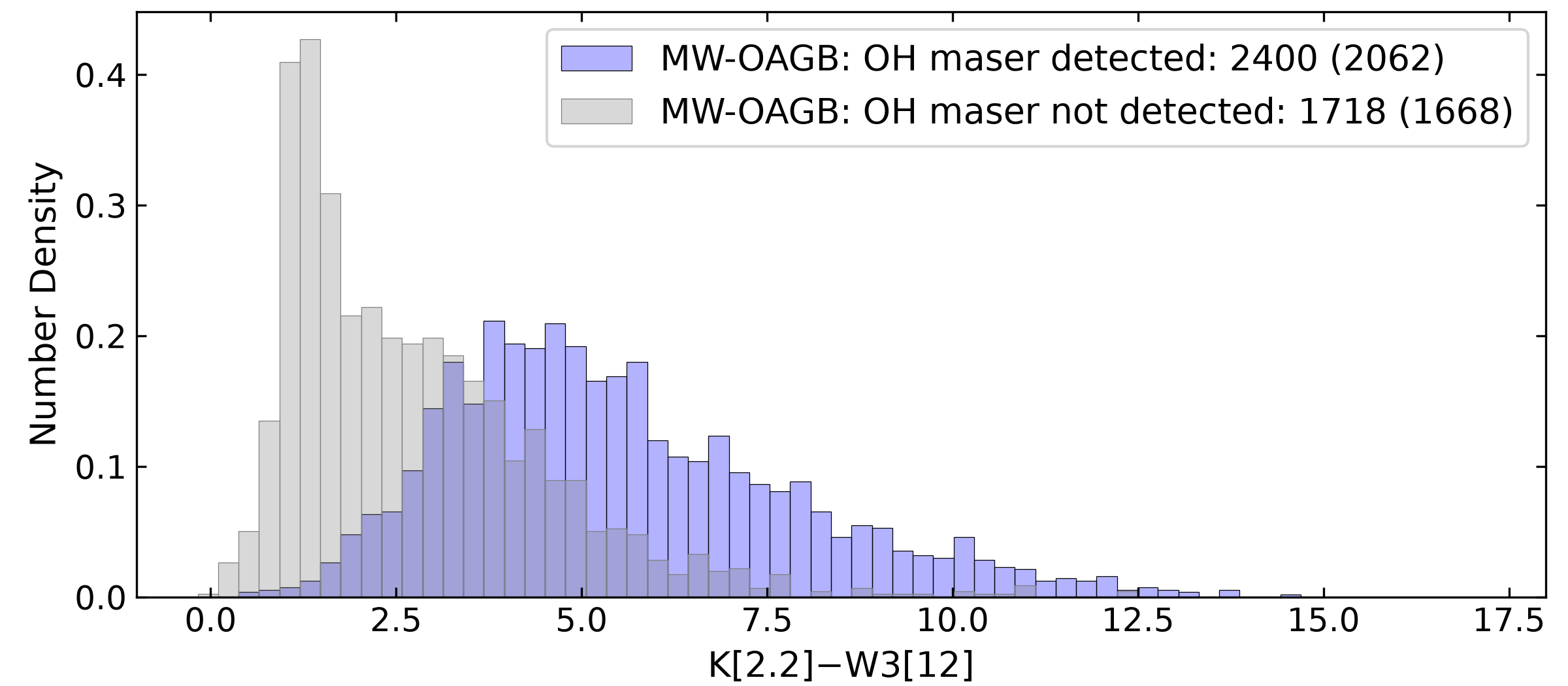}{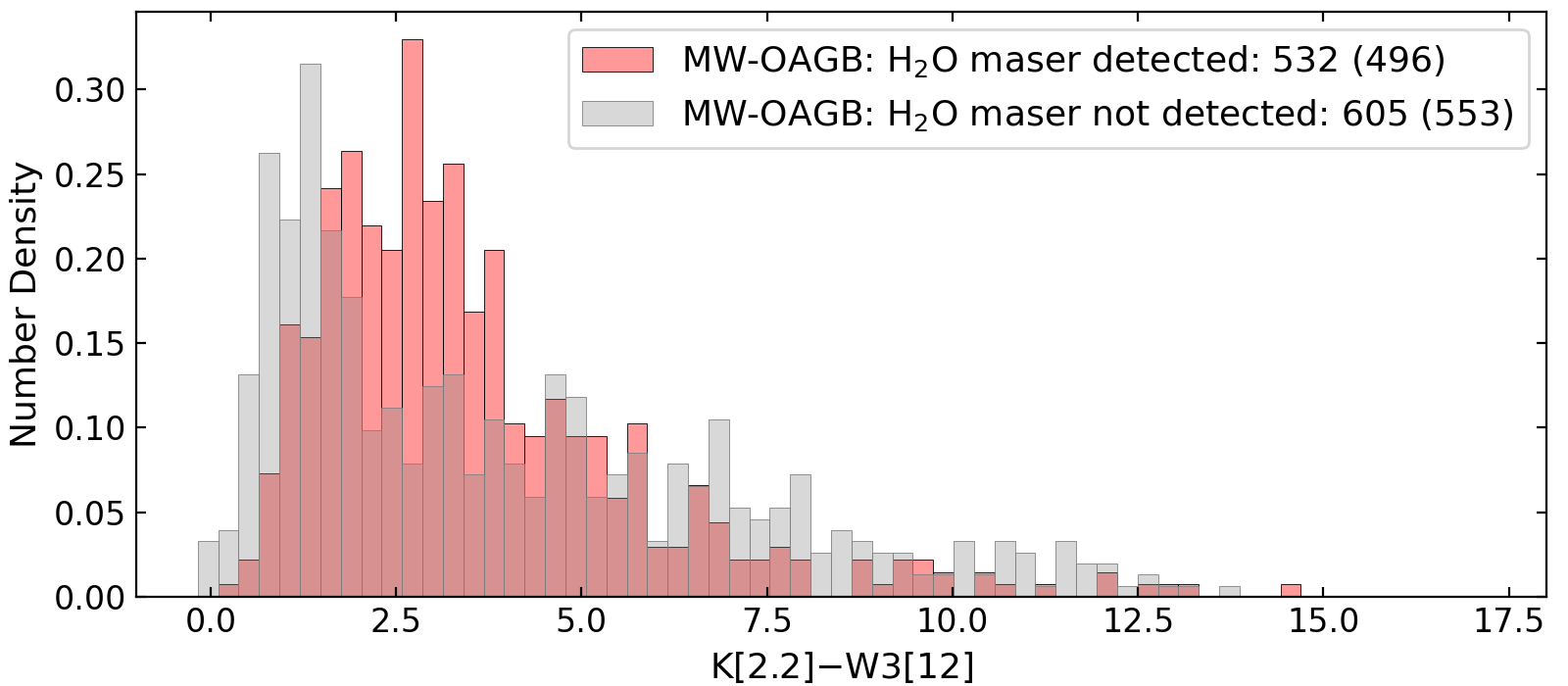}{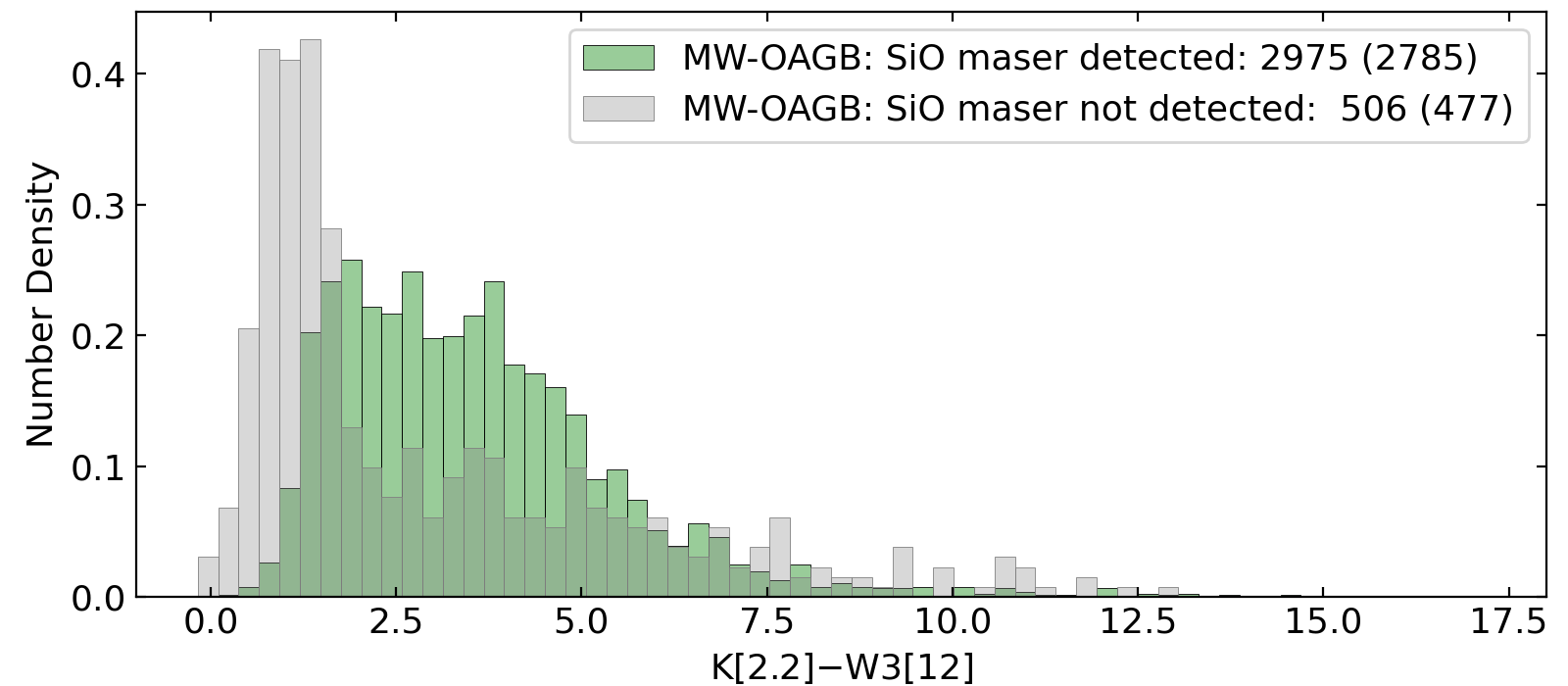}
\caption{Histograms of K[2.2]$-$W3[12] colors for maser sources identified as Galactic OAGB stars
(see Tables~\ref{tab:tab1} and \ref{tab:tab2}).
See Section~\ref{sec:maser} for details.} \label{f10}
\end{figure*}

\begin{figure*}
\centering
\smallplotfour{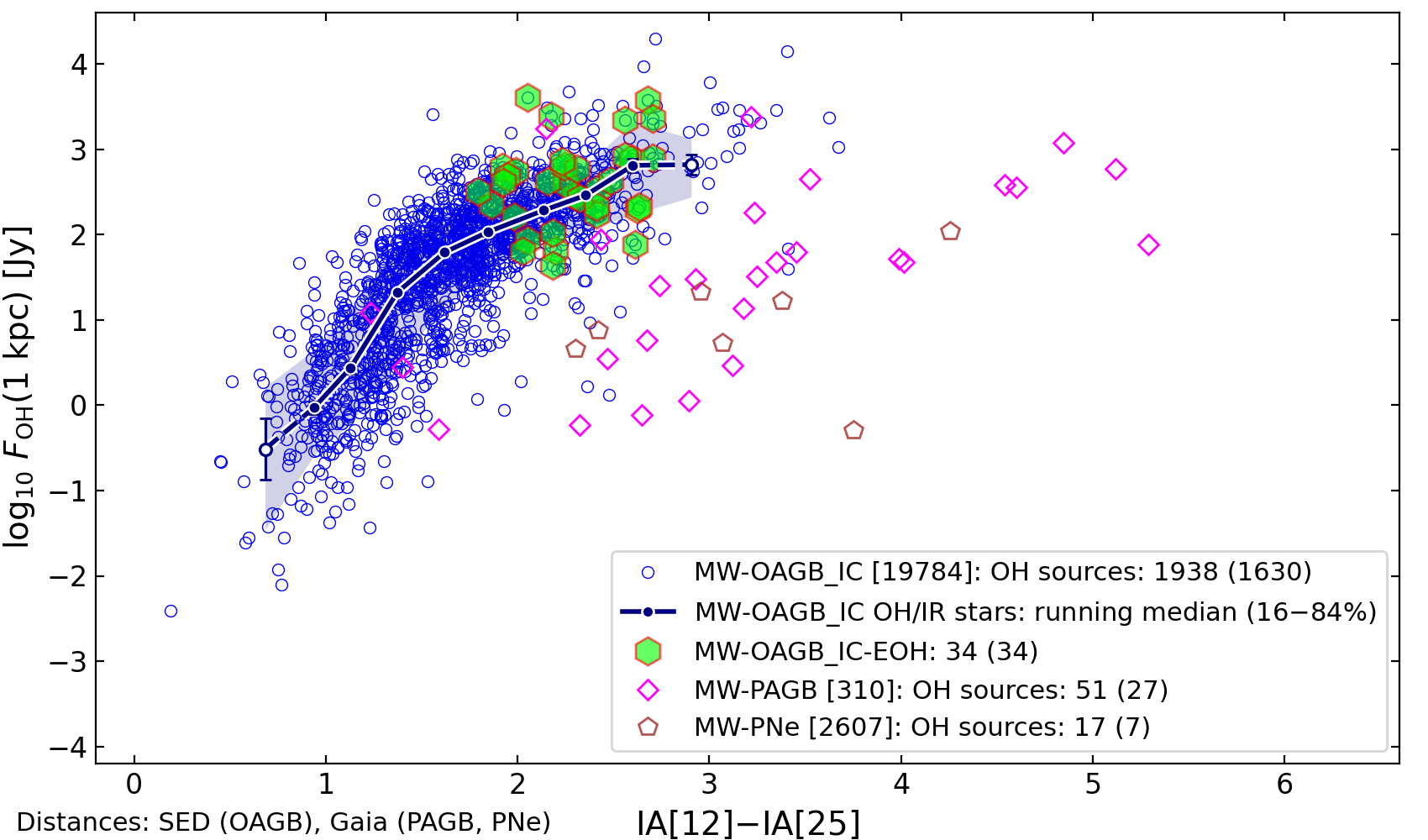}{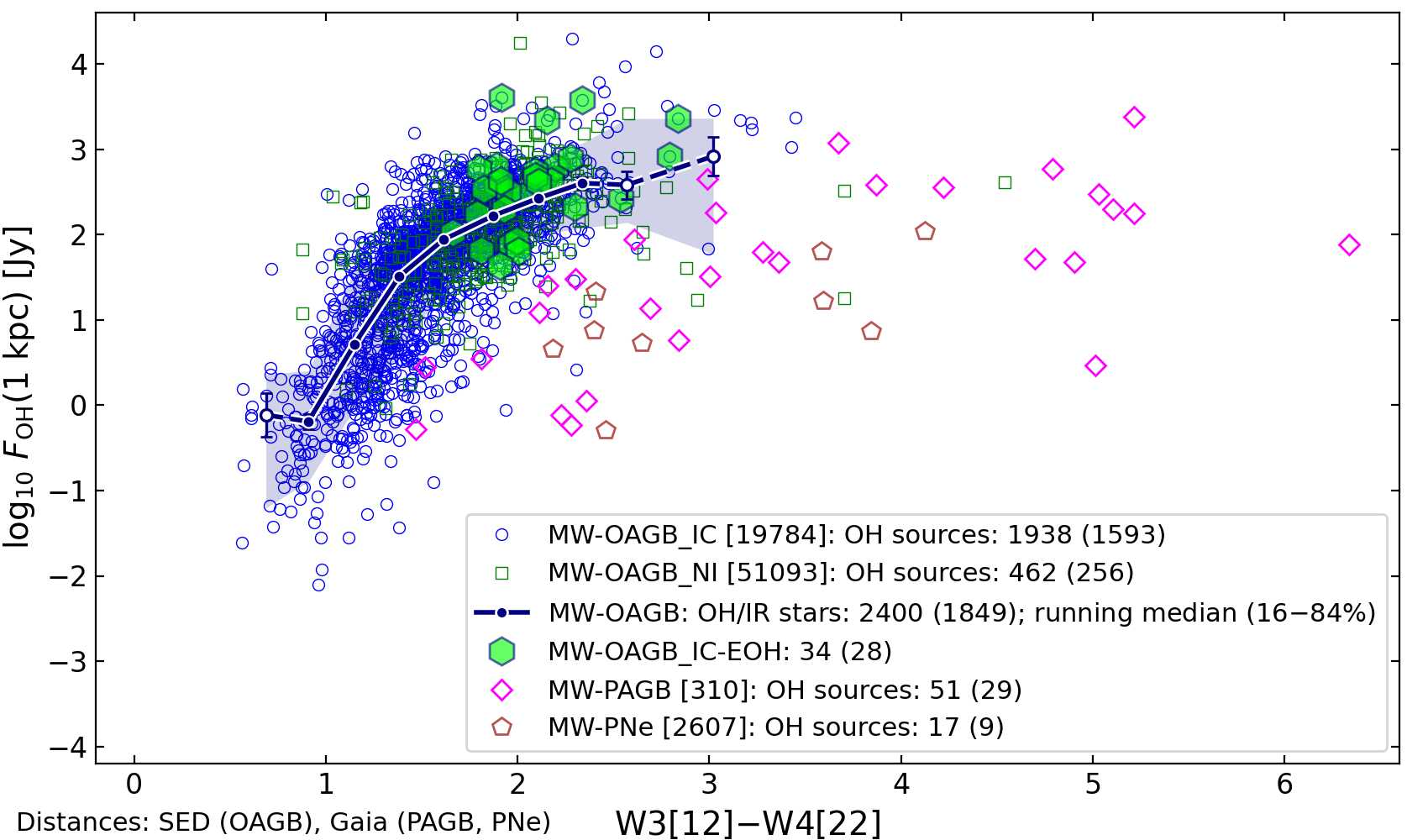}{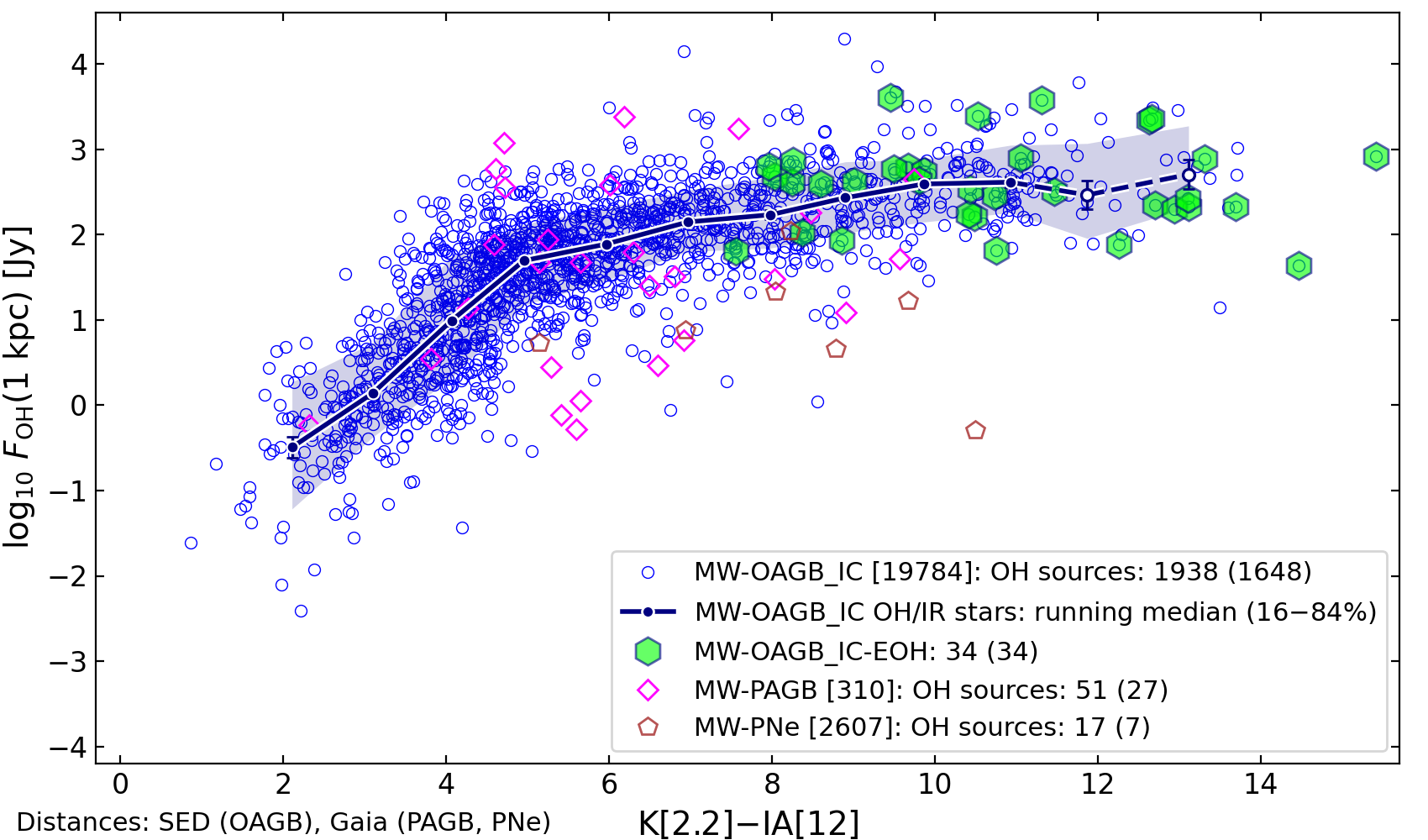}{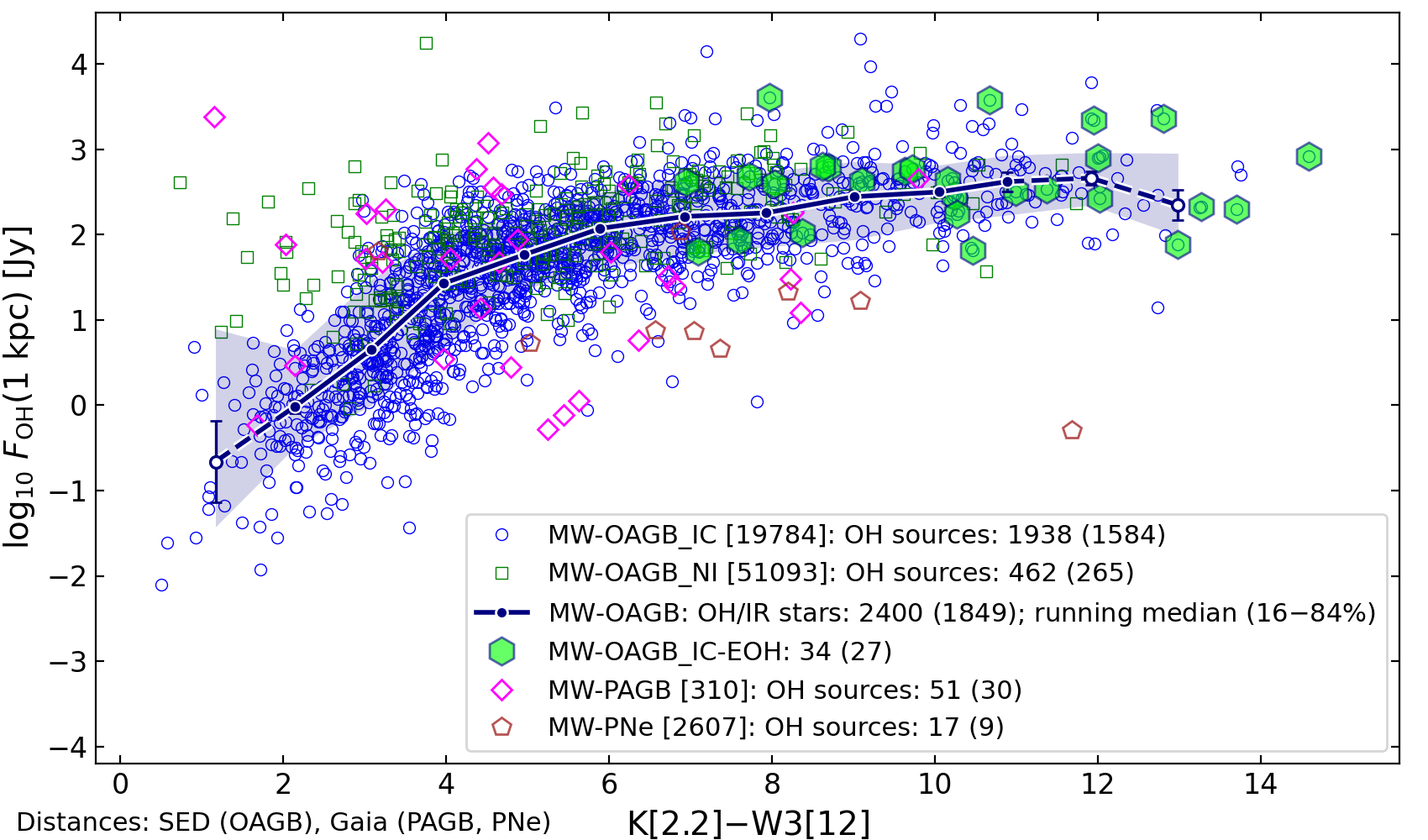}
\caption{Absolute OH maser fluxes (peak flux densities scaled to 1 kpc;
Equation~\ref{eq:1}) for OAGB stars, PAGB stars, and PNe in the Milky Way as a
function of IR colors. Only OH maser sources detected at 1612 MHz in the
database of \citet{engels2015} are included (Section~\ref{sec:omaser}); for OAGB
stars, only those with reliable SED fits (Section~\ref{sec:seds}) are plotted.
The solid lines show the running medians of the OH/IR stars (OAGB stars with OH
maser detections) with their 16th--84th percentile ranges; in the IA[12]$-$IA[25]
diagram the median is computed only for IA[12]$-$IA[25]~$\leq$~3, and
bins at either end of the color range with fewer than 30 stars are drawn
with open symbols and dashed segments. The total number of objects in each
subgroup is listed, with the numbers in parentheses representing the subset for
which the required data are available and therefore plotted. See
Section~\ref{sec:maser} for details.}\label{f11}
\end{figure*}

\section{Maser Emission in Galactic OAGB Stars}\label{sec:maser}

We examined the properties of OH, H$_2$O, and SiO maser sources associated with
OAGB stars, PAGB stars, and PNe in the Milky Way (see Section~\ref{sec:omaser};
Tables~\ref{tab:tab1} and \ref{tab:tab2}). Observations indicate that the
fraction of stars with detectable SiO and H$_2$O maser emission decreases as MIR
colors become redder, whereas OH maser emission does not follow this trend (e.g.,
\citealt{nyman1998}).

Figure~\ref{f10} displays histograms of K[2.2]$-$W3[12] colors for maser-emitting
OAGB stars; for the OH masers, the detections refer to the 1612 MHz line
(Section~\ref{sec:omaser}). At the onset of the AGB phase, SiO, H$_2$O, and OH
masers emerge and grow increasingly prominent as stars evolve from the E-AGB to
the TP-AGB phase. By the end of the AGB phase, SiO and H$_2$O masers decline
rapidly, while OH masers reach their highest incidence and remain detectable into
the PAGB phase (see the PAGB stars and PNe in Figure~\ref{f11}).

For the H$_2$O and SiO masers, only the information on whether a maser was
detected or not detected is available in the compiled catalogs, so their
properties are presented only in the histograms of Figure~\ref{f10}. Because the
underlying surveys are heterogeneous (Section~\ref{sec:omaser}), the relative
numbers of detections and non-detections in Figure~\ref{f10} should be read as
indicative rather than as unbiased detection rates for the OAGB population. For
the OH masers, however, the peak flux densities compiled by \citet{engels2015}
(Version 2.5, updated on March 19, 2023) allow a more quantitative analysis.

To compare the OH maser strengths of stars at different distances, we use the
absolute OH maser flux, $F_{\rm OH}(1\,{\rm kpc})$, defined as the observed peak
flux density of the OH 1612~MHz maser, $F_{\rm OH}$ \citep{engels2015}, scaled to
a distance of 1~kpc:
\begin{equation}
F_{\rm OH}(1\,{\rm kpc}) = F_{\rm OH} \left( \frac{D}{1\,{\rm kpc}} \right)^{2},
\label{eq:1}
\end{equation}
where $D$ is the distance derived from the OAGB model SEDs for OAGB stars (see
Section~\ref{sec:seds}) and the Gaia DR3 geometric distance
(\citealt{bailer-jones2021}) for PAGB stars and PNe.

Figure~\ref{f11} presents the absolute OH maser fluxes at 1612 MHz of OAGB stars,
PAGB stars, and PNe as a function of four IR colors. Of the 2400 MW-OAGB stars
with OH maser detections (1938 MW-OAGB\_IC and 462 MW-OAGB\_NI;
Table~\ref{tab:tab1}), 2397 have a measured peak flux density (the only
observation of the other three lists a negative flux density, which is not used),
and only those with reliable SED fits (Section~\ref{sec:seds}) are plotted; the
number of stars in each panel (1630--1849) is further reduced because the
photometry required for each color is not available for all of them.

To describe the typical absolute OH maser flux of OH/IR stars (OAGB stars with OH
maser detections) at a given color without imposing a functional form, we compute
running medians and 16th--84th percentile ranges of the OH/IR stars in color bins
of 1.0~mag for the colors involving the K band and 0.25~mag for the mid-IR
colors, merging adjacent bins where necessary so that each bin contains at least
10 stars. The uncertainties of the medians are estimated by bootstrap resampling.
For the IRAS colors, the medians are computed only for MW-OAGB\_IC objects, which
have IRAS counterparts; PAGB stars and PNe are not included in the medians. In
all diagrams, the bins at either end of the color range that contain fewer than
30 stars are drawn with open symbols and dashed segments to indicate the low
statistics.

The running medians of the absolute OH maser flux increase steeply with IR color
in all four diagrams of Figure~\ref{f11}, from $\log F_{\rm OH}(1\,{\rm kpc})
\approx -0.5$ at the bluest colors to above 2 within 2--3 mag, and none of them
shows a decrease at the red end. In the colors involving the K band, the medians
become nearly level for K[2.2]$-$W3[12] $>$ 8.5 and K[2.2]$-$IA[12] $>$ 8.5, at
weighted mean values of $\log F_{\rm OH}(1\,{\rm kpc}) = 2.51 \pm 0.03$ and $2.53
\pm 0.03$, respectively ($\approx$ 320--340 Jy at 1 kpc), with slopes of $0.048
\pm 0.030$ and $0.068 \pm 0.028$ dex mag$^{-1}$. In the mid-IR colors, the
medians continue to increase and reach $\log F_{\rm OH}(1\,{\rm kpc}) \approx
2.6$--2.8 at the colors of the MW-OAGB\_IC-EOH stars ($2.81 \pm 0.08$ at
IA[12]$-$IA[25] $\approx 2.6$ and $2.60 \pm 0.07$ at W3[12]$-$W4[22] $\approx
2.3$). The reddest bins, which contain only 10 stars with $2.75 < $
IA[12]$-$IA[25] $\leq 3$ and 16 and 18 stars with $2.5 < $ W3[12]$-$W4[22] $\leq
2.75$ and W3[12]$-$W4[22] $> 2.75$, have medians of $2.82 \pm 0.12$, $2.58 \pm
0.16$, and $2.92 \pm 0.22$, consistent with a plateau or a continued rise but
with low statistics.

A small number of objects with the reddest mid-IR colors show even larger
absolute OH fluxes. In particular, \textbf{17} OAGB objects with
IA[12]$-$IA[25]~$>$~3 lie well inside region R3 of Figure~\ref{f9}, which is
occupied by PAGB stars and PNe; such objects may have detached dust envelopes and
be in transition to the post-AGB phase (e.g., \citealt{suh2015};
\citealt{suh2024b}), and their distances derived from the OAGB model SEDs are the
least reliable. The running median in the IA[12]$-$IA[25] diagram is therefore
computed only for IA[12]$-$IA[25]~$\leq$~3; these \textbf{17} objects are still
plotted but are not included. No such restriction is needed for the other three
colors, where the OH/IR stars do not extend beyond the OAGB model sequences, and
all OH/IR stars are used.

The distribution in Figure~\ref{f11} is also shaped by selection effects.
Although the detection limits of the individual OH surveys differ
(Section~\ref{sec:omaser}), they are all limits in observed flux density,
typically $\sim$0.1~Jy, so the corresponding limit in absolute maser flux scales
as the square of the distance. Redder OAGB stars in our sample are on average
more distant, so only their brighter masers are detected, and the conversion to
absolute flux places them preferentially at the upper end of the range. Because
only detections are included, the running medians describe the detected OH/IR
stars, and the rise of the medians toward red colors is at least partly a
selection effect. The underlying trend is nevertheless expected on physical
grounds, as thicker envelopes provide both a larger OH column and more infrared
pump photons.

Furthermore, the absolute OH maser fluxes of the reddest OAGB stars should be
regarded as upper estimates. The SED distances of stars with $\tau_{10} \geq 1$
may be overestimated by up to 0.6 dex (Section~\ref{sec:sed-dist}), and since
$F_{\rm OH}$(1 kpc) scales as $D^{2}$, this corresponds to an overestimate of up
to 1.2 dex in absolute flux. The leveling-off of the medians at the reddest
K-band colors and their maximum at the colors of the MW-OAGB\_IC-EOH stars are
therefore robust as upper bounds, but the true absolute fluxes of these stars may
be lower. A few unidentified RSGs, which can have bright OH masers, may remain
among the brightest sources, although they are statistically negligible (see
Section~\ref{sec:gagb}).

The MW-OAGB\_IC-EOH stars lie at the colors where the median absolute OH maser
flux levels off (K-band colors) or reaches its highest values (mid-IR colors).
They are classified as extreme OH/IR stars with thick dust envelopes ($\tau_{10}
\geq 10$) and are likely nearing the end of the AGB phase. These stars are
thought to be intermediate-mass stars with masses exceeding approximately
4~$M_{\odot}$ and exhibit high mass-loss rates of up to $\dot{M} \sim
10^{-4}\,M_{\odot}\,{\rm yr}^{-1}$ (e.g., \citealt{justtanont2015}).

For OH maser sources, PAGB stars and PNe display redder IA[12]$-$IA[25] and
W3[12]$-$W4[22] indices compared to OAGB stars, likely due to the presence of
more detached dust envelopes surrounding PAGB stars and PNe
\citep{suh2015,suh2024b}. The absolute OH maser fluxes of the PAGB stars are
comparable to those of the OH/IR stars of similar colors, whereas the few PNe
with OH maser detections lie 1--2 dex below the running medians, although the
small numbers (51 PAGB stars and 17 PNe with OH maser detections, of which 27--30
and 7--9 have the required Gaia distances and colors) do not allow a firm
conclusion.

\begin{figure*}
\centering
\largeplotone{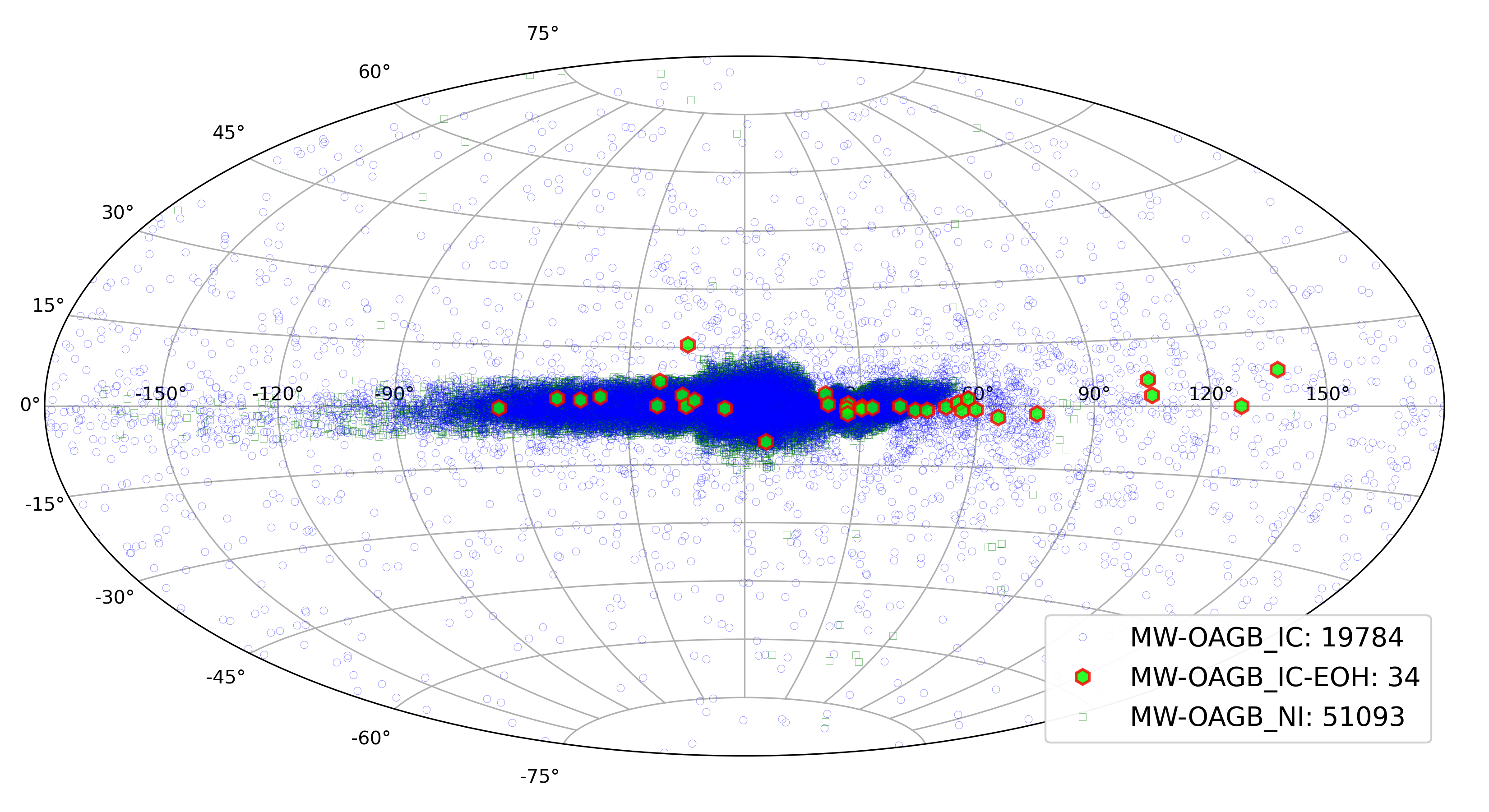}\caption{Spatial distribution of Galactic OAGB stars shown in Galactic coordinates.}
\label{f12}
\end{figure*}

\begin{figure*}
\centering
\smallplottwo{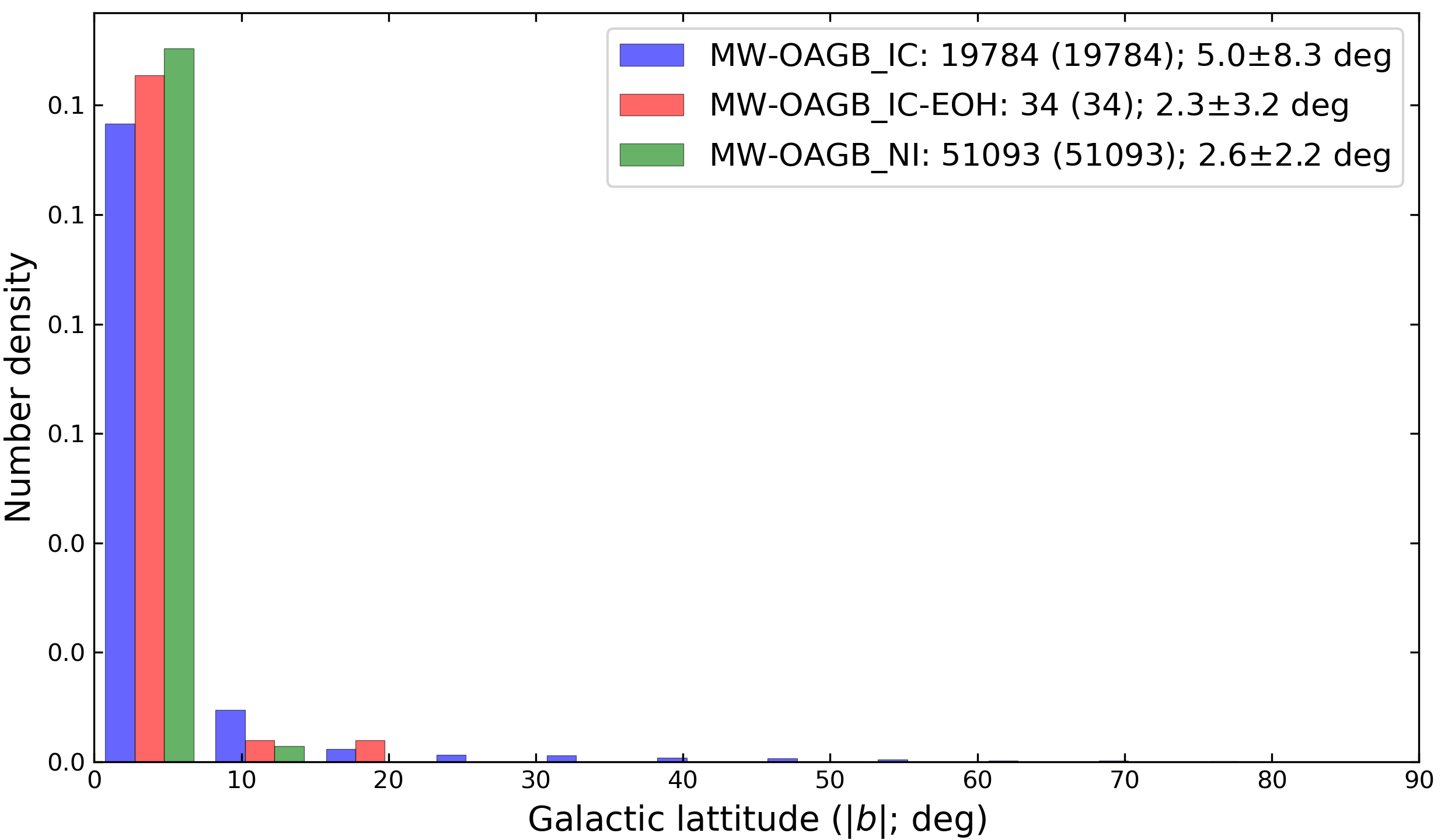}{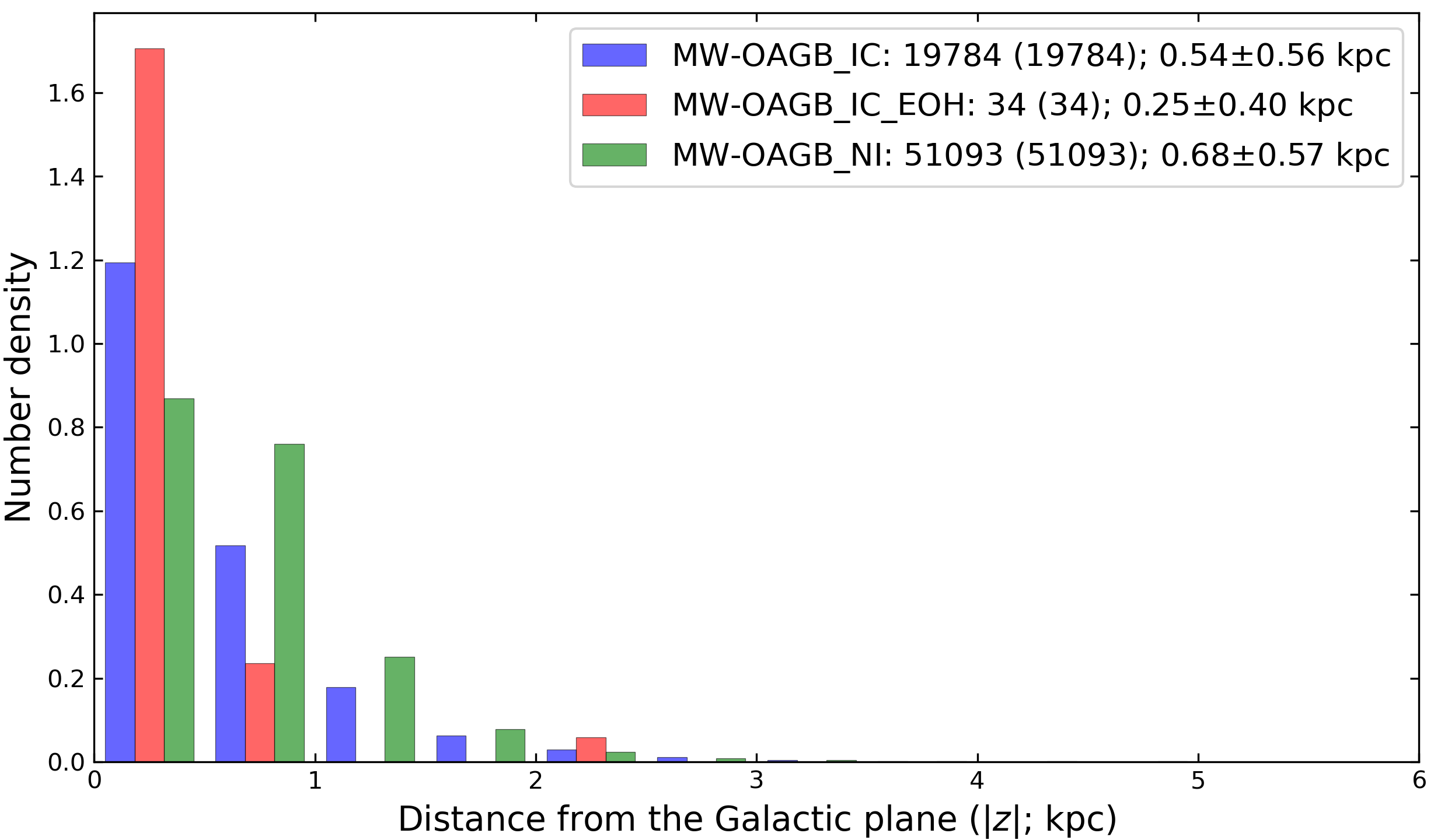}
\caption{Number distributions of Galactic latitudes and distances from the Galactic plane for various subgroups of Galactic OAGB stars.
(See Section~\ref{sec:spatial}).}
\label{f13}
\end{figure*}

\section{Spatial distribution of Galactic OAGB stars}\label{sec:spatial}

Figure~\ref{f12} shows the spatial distributions of Galactic OAGB stars in
Galactic coordinates. Figure~\ref{f13} presents the number distributions of
Galactic latitudes and distances from the Galactic plane for various OAGB
subgroups, with mean values and standard deviations provided for each subgroup.

Overall, MW-OAGB\_IC objects exhibit a stronger concentration toward the Galactic
plane compared to MW-OAGB\_NI objects. Our analysis further reveals that
MW-OAGB\_IC-EOH stars display a particularly pronounced concentration toward the
Galactic plane relative to most OAGB stars.

\section{IR properties of pulsating OAGB stars}\label{sec:pul}

Almost all AGB stars are observed as LPVs, in the collective sense adopted in
Section~\ref{sec:intro} for OSARGs, SRVs, and Mira variables (e.g.,
\citealt{swu13}), and are surrounded by circumstellar dust envelopes (e.g.,
\citealt{suh2021}). While many SRVs and a substantial fraction of OSARGs are
believed to be AGB stars, Mira variables are generally regarded as exclusively
associated with the AGB phase (e.g., \citealt{hofner2018}). As AGB stars become
more evolved (or are initially more massive), they tend to display larger
pulsation amplitudes, longer pulsation periods, and increased mass-loss rates
(\citealt{debeck2010}; \citealt{justtanont2015}; \citealt{jimenez2021};
\citealt{engels2024}).

\begin{figure*}
\centering
\smallplotfour{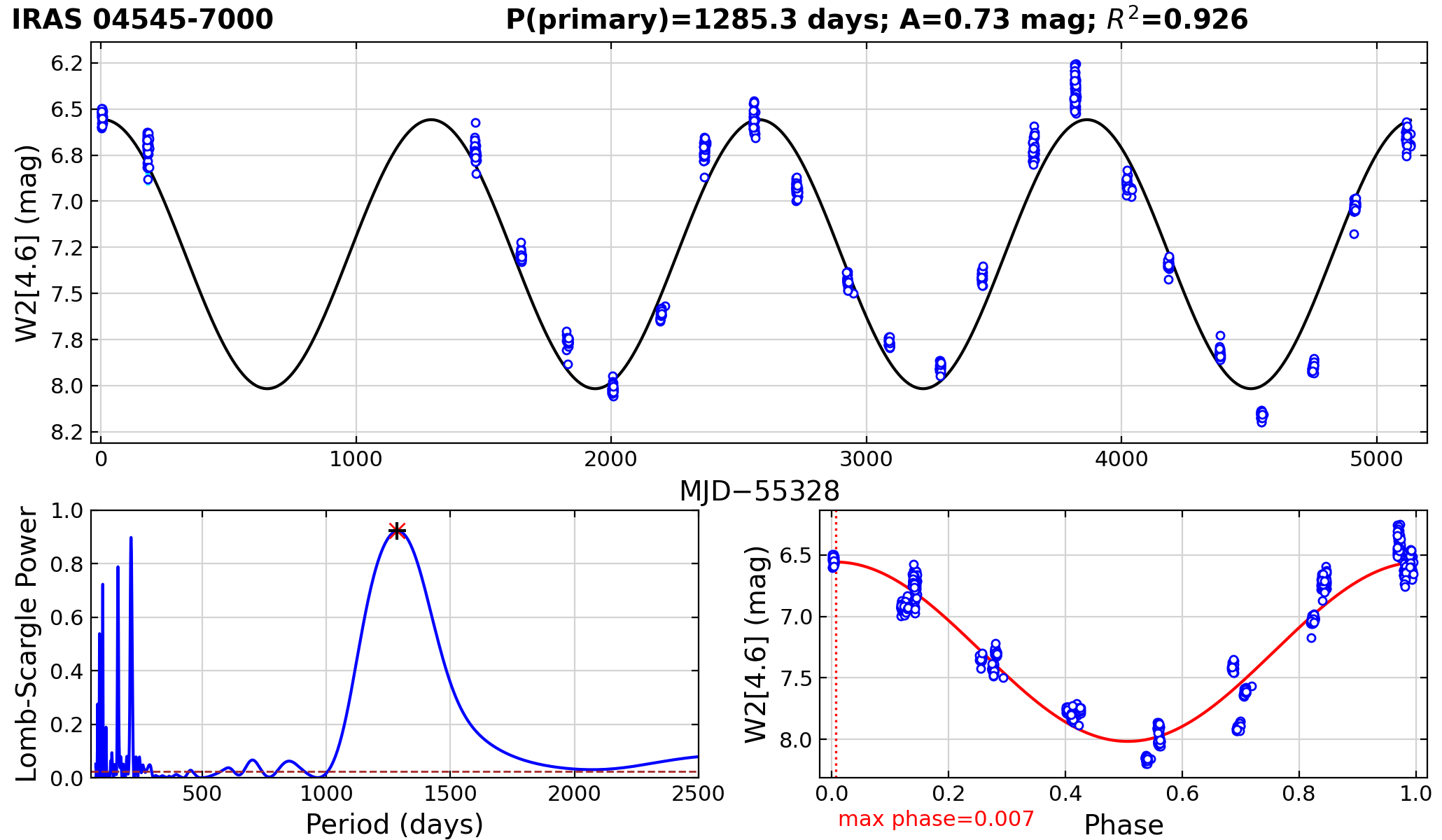}{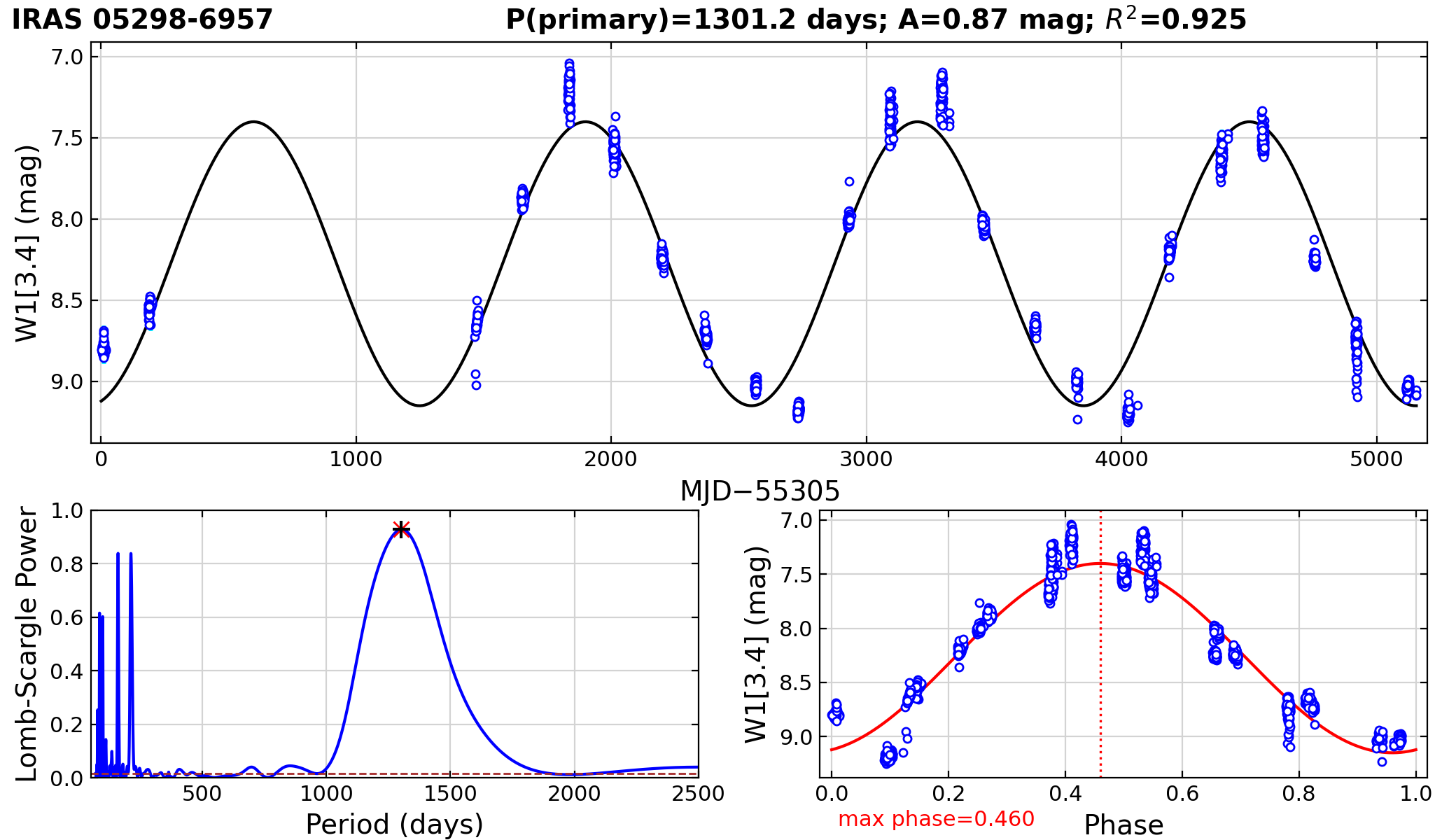}{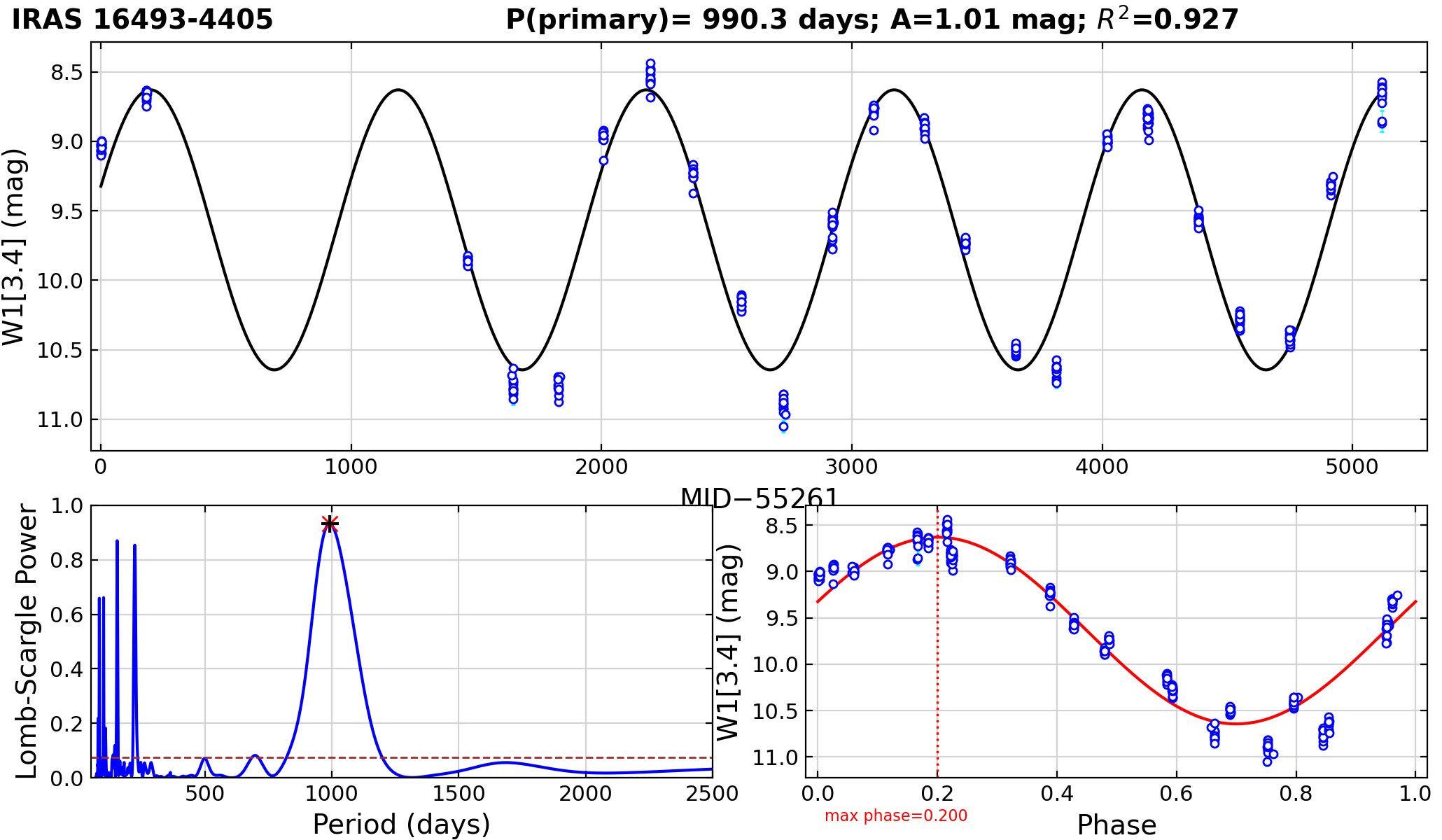}{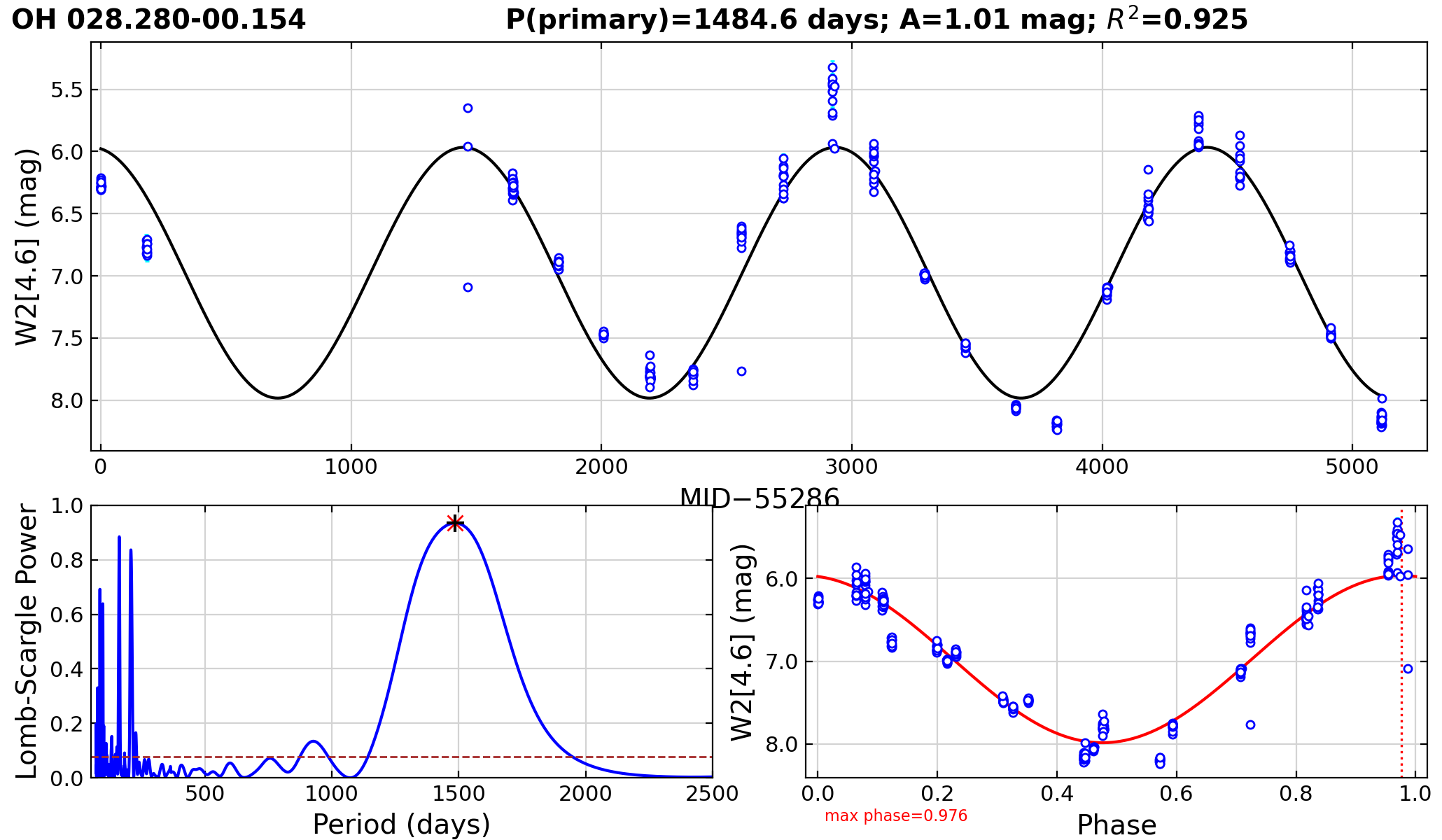}
\caption{Lomb–Scargle periodograms derived from WISE light curves for four representative long-period OH/IR stars
(top two panels: LMC; bottom two panels: Milky Way) whose periods were previously unknown according to OGLE-III or AAVSO.
These objects are candidates for newly identified Mira variables.
See Section~\ref{sec:neo}.} \label{f14}
\end{figure*}

\subsection{Finding IR variations from WISE data}\label{sec:neo}

\citet{suh2021} examined the variability of Galactic AGB stars in the W1[3.4] and
W2[4.6] bands using NEOWISE-R data from WISE. \citet{suh2025} further explored
the variability of LMC-CAGB stars with a new sample stars and updated NEOWISE-R
data.

Our analysis focuses on a newly identified sample of OAGB stars in the LMC and
the Milky Way (see Table~\ref{tab:tab1}), excluding those classified as Mira
variables with known periods according to AAVSO or OGLE-III. The sample consists
of 12,894 LMC-OAGB stars and 4145 MW-OAGB stars. We examine their variability in
W1[3.4] and W2[4.6] over the past 16 years, utilizing AllWISE multi-epoch
photometry from 2009–2010 and NEOWISE-R data from the 2024 final data release.
The NEOWISE-R data includes 21 epochs, with two observations per year from 2014
to 2023 and one additional observation in 2024.

To identify Mira-like variations in the WISE light curves, we use a simple
sinusoidal light curve model and employ the Lomb-Scargle periodogram, a widely
used statistical tool for detecting periodic signals (e.g.,
\citealt{zechmeister2009}; \citealt{vanderPlas2018}). The periodograms are
computed using the Astropy implementation, with further details on the algorithm
available in \citet{vanderPlas2018} and \citet{suh2021}. For the sample stars, we
generated light curves using WISE data and computed Lomb-Scargle periodograms.
Unlike light curves from other observations (e.g., OGLE), the WISE data that were
observed at regular six-month intervals often produce periodograms with multiple
peaks of similar power. This makes it challenging to determine accurate pulsation
periods based solely on WISE data (see \citealt{suh2021}).

To ensure robust results, we applied stringent selection criteria, including a
minimum of 200 observations, a coefficient of determination ($R^2$) greater than
0.8, and a Lomb–Scargle power exceeding 0.8. Among the 12,894 LMC-OAGB stars,
reliable pulsation parameters could be derived from the WISE light curves for
only 9 sources. Likewise, among the 4145 MW-OAGB stars, reliable pulsation
parameters were obtained for only 116 sources.

Figure~\ref{f14} presents Lomb–Scargle periodograms derived from the WISE light
curves for four representative long-period OH/IR stars (top two panels: LMC;
bottom two panels: Milky Way) whose periods were previously unknown in OGLE-III
or AAVSO catalogs. These OAGB stars are potential candidates for newly identified
Mira variables based on WISE observations.

\begin{figure*}
\centering
\smallplotfour{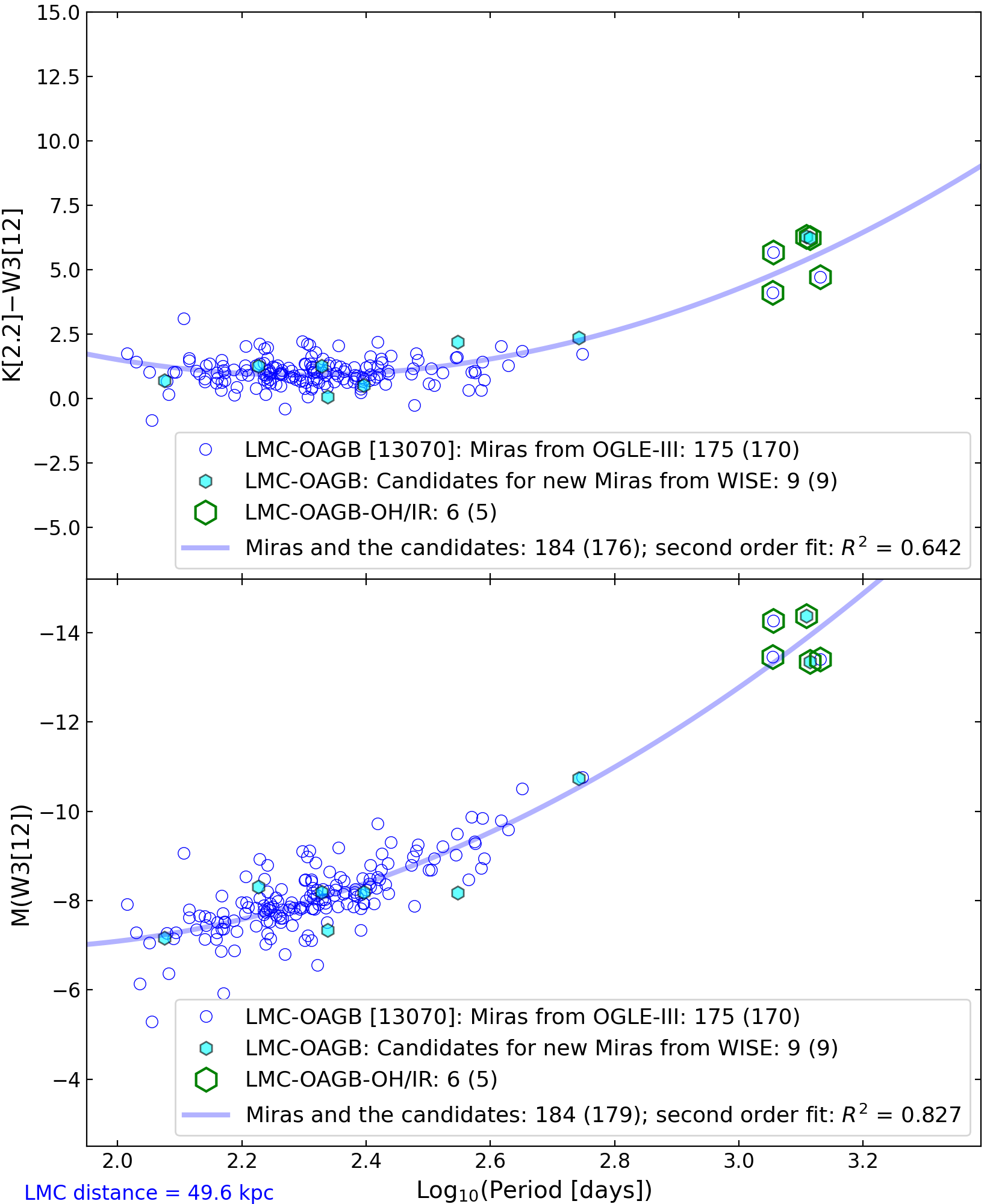}{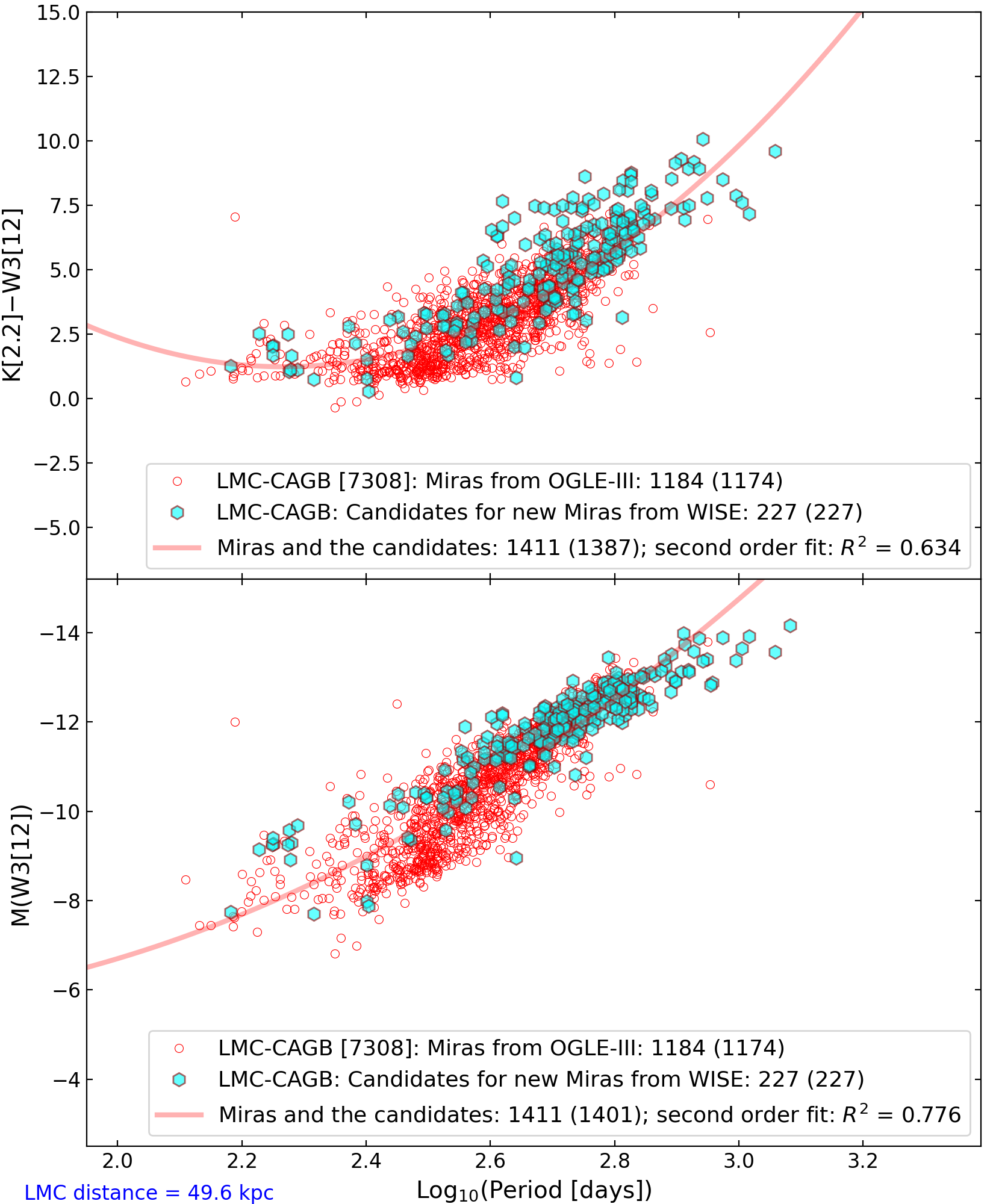}{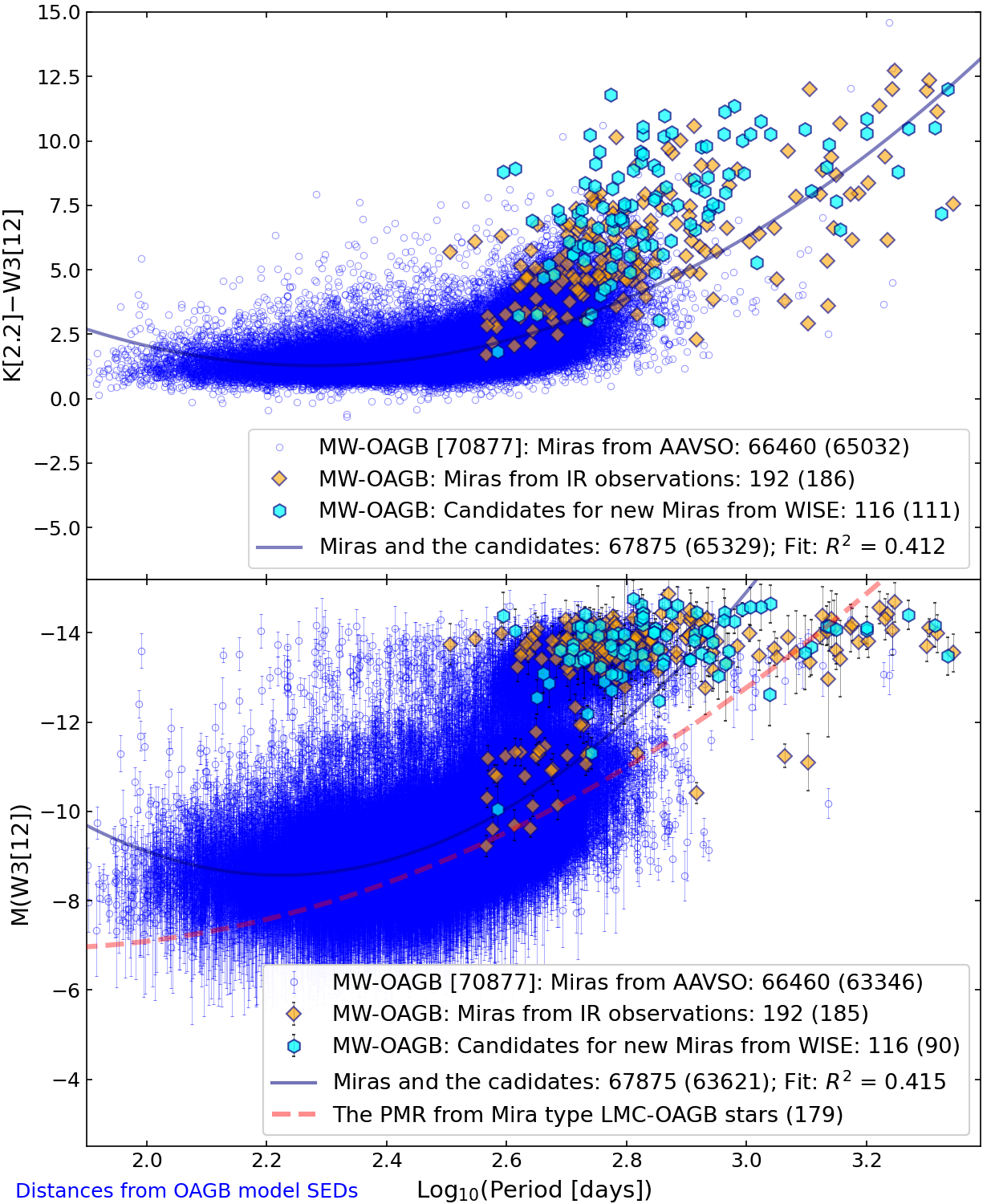}{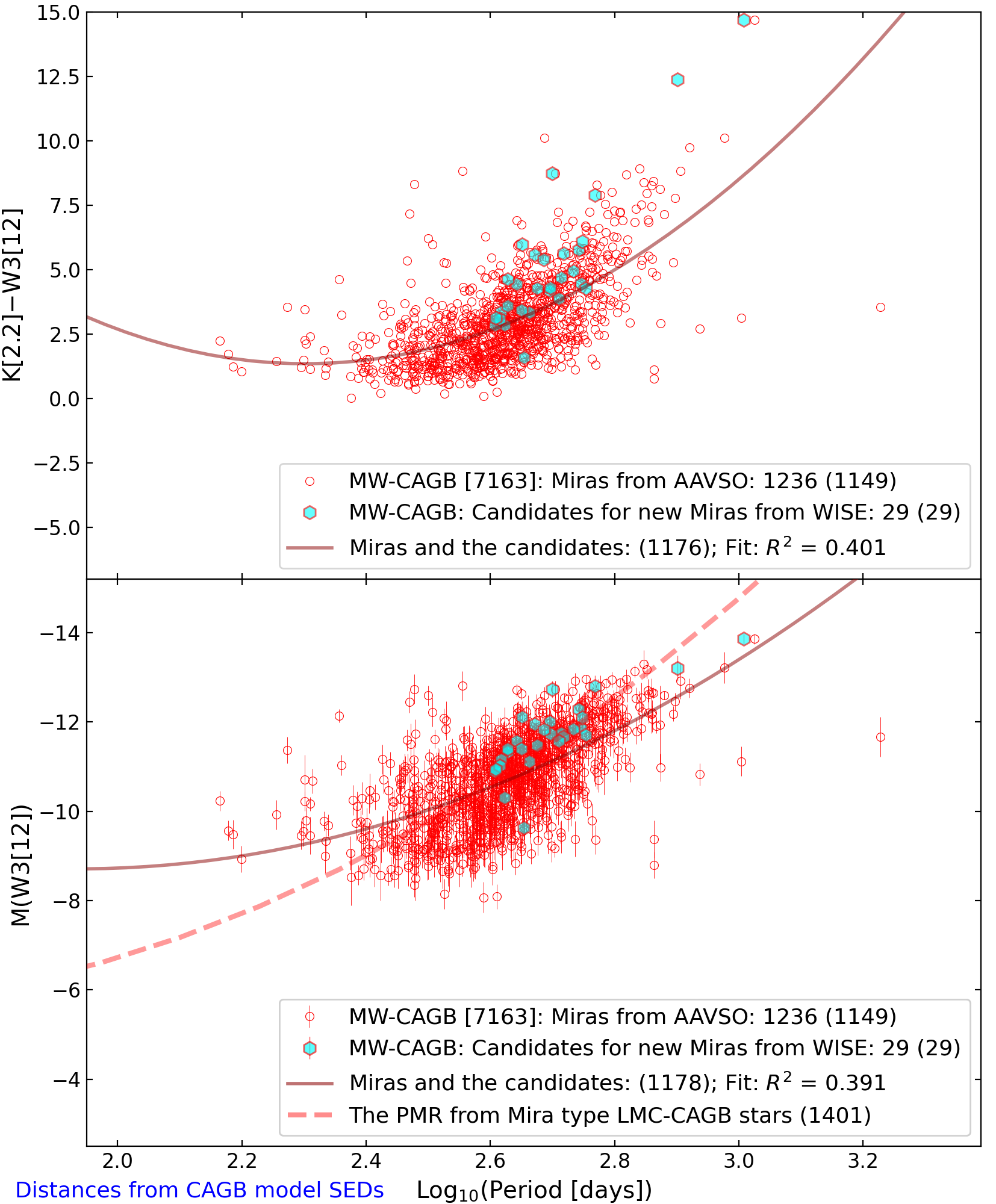}\caption{Period-color and period-magnitude relations
for Mira-type AGB stars in the LMC (upper panels) and the Milky
Way (lower panels), including known Miras from AAVSO and OGLE-III, along with
candidate Miras identified from WISE data (see Section~\ref{sec:pmr} for details).
The total number of objects in each subgroup is listed, with the numbers in parentheses
representing the subset for which the required data are available and therefore plotted.
In period-magnitude relation diagrams, the vertical error bars represent the uncertainties in the distances derived from AGB model SEDs.
See Section~\ref{sec:pmr}.}
\label{f15}
\end{figure*}

\subsection{Period-color and Period-magnitude relations}\label{sec:pmr}

There are 175 OGLE-III Mira variables in the sample of 13,070 LMC-OAGB stars.
Extensive photometric data for Galactic OAGB stars are available in the optical
bands. Within the sample of 70,877 Galactic OAGB stars, 66,460 Mira variables
have been identified through AAVSO with periods. Since most pulsation periods
listed in the AAVSO database are derived from optical-band observations, the
majority of these objects are early-phase AGB stars with relatively thin dust
envelopes. However, data in the IR bands remain available for only a limited
number of OAGB stars.

For Galactic OAGB stars with thick dust envelopes, variability is best
characterized through observations in the IR or radio bands. \citet{jimenez2021},
for instance, analyzed IR variability in the Arecibo sample of OH/IR stars and
determined pulsation periods for 345 sources. In this study, pulsation periods
for 192 Mira variables derived from IR data were compiled from
\citet{engels1983}, \citet{ks2010}, \citet{jimenez2021}, and \citet{etoka2022},
excluding duplicates already listed in AAVSO. The 192 Miras from the IR
observations belong to the MW-OAGB\_IC subgroup.

The left panels of Figure~\ref{f15} show period–color relations (PCRs) and
period–magnitude relations (PMRs) for Mira-type OAGB stars in the LMC (upper
panels) and the Milky Way (lower panels), including known Miras from AAVSO and
OGLE-III, along with candidate Miras identified from WISE data. The plots
highlight the relationships between K[2.2]$-$W3[12] colors, absolute magnitudes
in W3[12], and pulsation periods. While there is noticeable scatter, clear trends
in the PCRs and PMRs are evident among the Mira variables. The objects in the LMC
show better coefficients of determination ($R^2$) for PCRs and PMRs than Galactic
objects.

Lower parts of the four panels on Figure~\ref{f15} show PMRs for Mira-type AGB
stars in the LMC (upper panels) and the Milky Way (lower panels), including known
Miras from AAVSO and OGLE-III, along with candidate Miras identified from WISE
data. The literature on PMRs for Miras in our Galaxy is relatively limited,
primarily due to the uncertainties in distance measurements. In contrast, there
is abundant literature on PMRs for a large sample of Miras in the LMC, thanks to
its proximity and well-determined distance (e.g., \citealt{suh2020};
\citealt{iwanek2021}). Therefore, the alternative distances of OAGB stars in our
Galaxy derived from the PMRs for Miras in the LMC can be useful to be compared
with Gaia DR3 distances.

Out of the 13,070 LMC-OAGB stars (see Table~\ref{tab:tab1}), 175 Mira variables
were identified in the OGLE-III catalog (see Section~\ref{sec:lmc-agb}) and 9
objects were identified as candidates for new Miras from the WISE data (see
section~\ref{sec:neo}). The upper-left panel of Figure~\ref{f15} shows the PMR in
the W3[12] band for these Mira-type OAGB stars in the LMC. The PMR derived from
the second-order (quadratic) fit is:
\begin{equation}
\mathrm{M(W3[12])} = -4.03 (\log_{10} P)^{2} + 14.5 \log_{10} P - 19.9 \pm 0.51.\label{eq:2}
\end{equation}

Compared to the PMR for OAGB stars in the Milky Way (see the lower-left panel of
Figure~\ref{f15}), the PMR for LMC-OAGB stars shows a much stronger correlation,
primarily due to the well-determined distance to the LMC.

It is meaningful to compare the PCRs and PMRs of OAGB stars obtained in this work
with those of CAGB stars presented by \citet{suh2026}. The right panels of
Figure~\ref{f15} show the PCRs and PMRs for Mira-type CAGB stars in the LMC
(upper panels) and the Milky Way (lower panels), including known Miras from AAVSO
and OGLE-III, as well as candidate Miras identified from WISE data.

As discussed in Section~\ref{sec:irpro}, OAGB and CAGB stars in the LMC and the
Milky Way exhibit broadly similar distributions in IR SEDs, CMDs, and 2CDs.
However, the LMC samples lack the most heavily obscured objects, particularly
among OAGB stars, where no stars with very large dust optical depths ($\tau_{10}
> 7$) are found. The PCRs and PMRs derived here show the same trend.

Although the PCRs and PMRs are broadly similar for the LMC and the Milky Way, the
LMC contains far fewer long-period Mira variables among OAGB stars. Because
long-period Mira variables generally have large dust optical depths, this result
is consistent with the absence of heavily obscured OAGB stars in the LMC. In
contrast, the difference is much smaller for CAGB stars, likely reflecting both
the greater similarity of their IR properties in the two galaxies (see
Section~\ref{sec:comoc}) and the more complete identification of Mira-type CAGB
stars in the LMC through large-scale surveys such as OGLE-III and WISE.

\begin{figure}
\smallplotthree{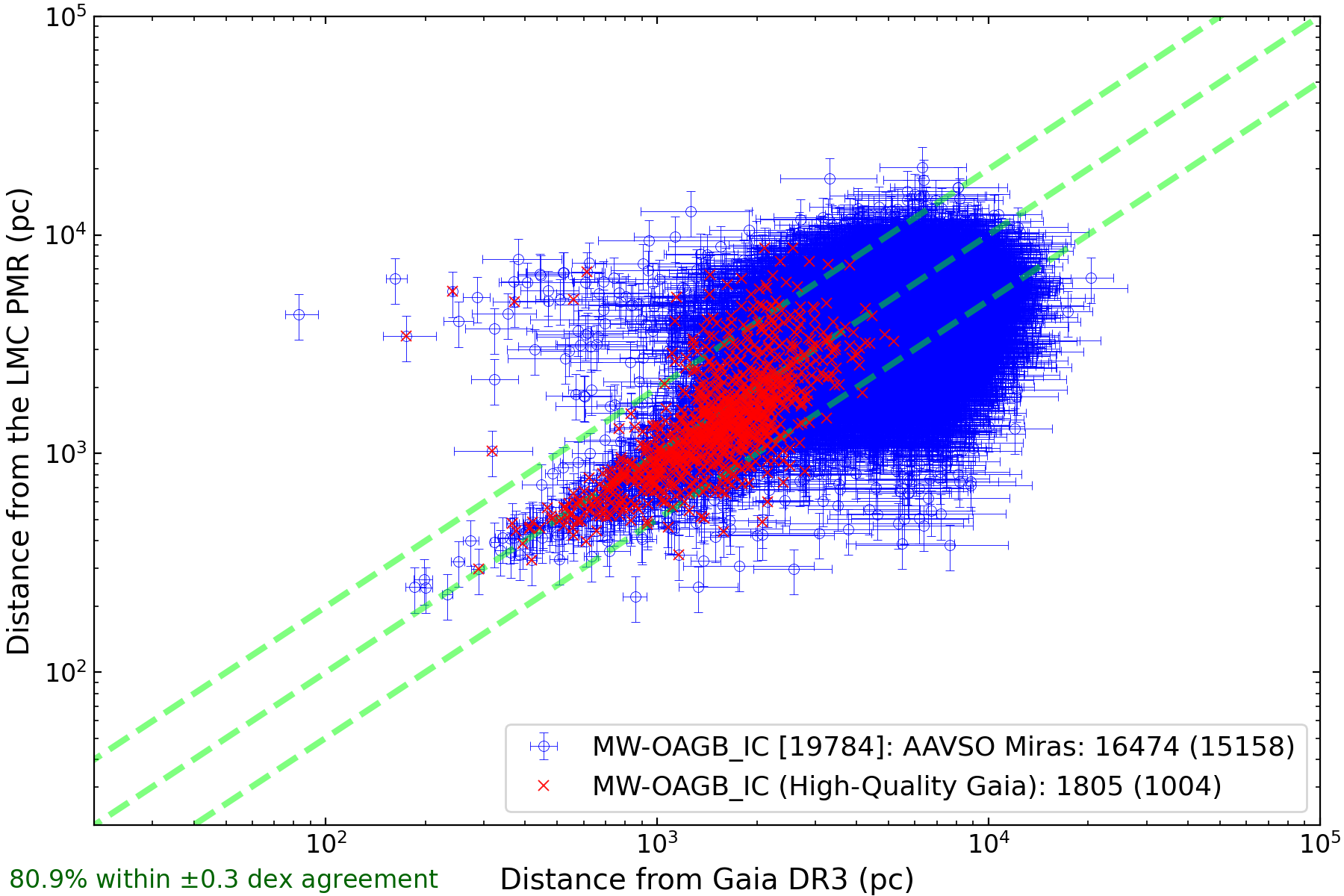}{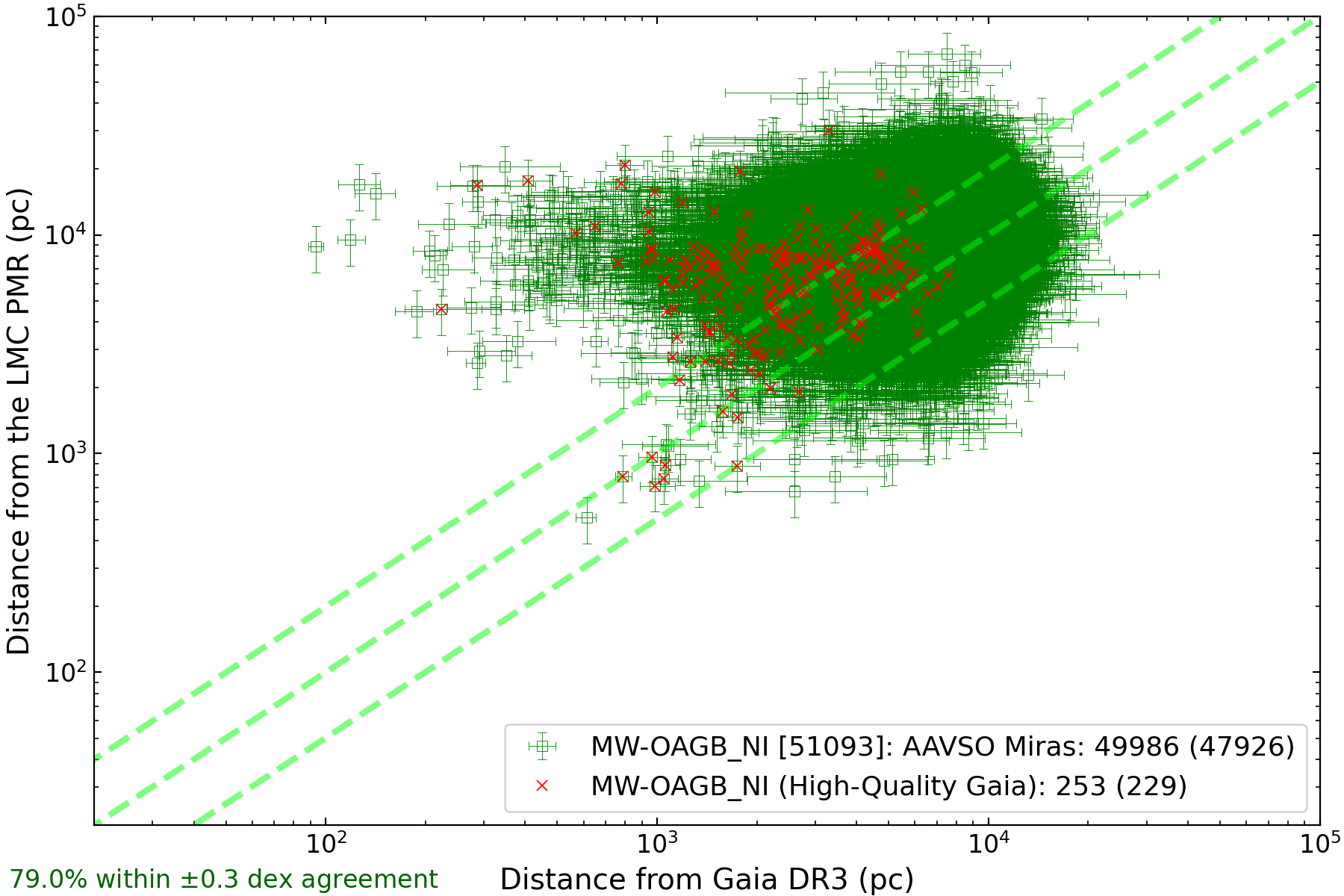}{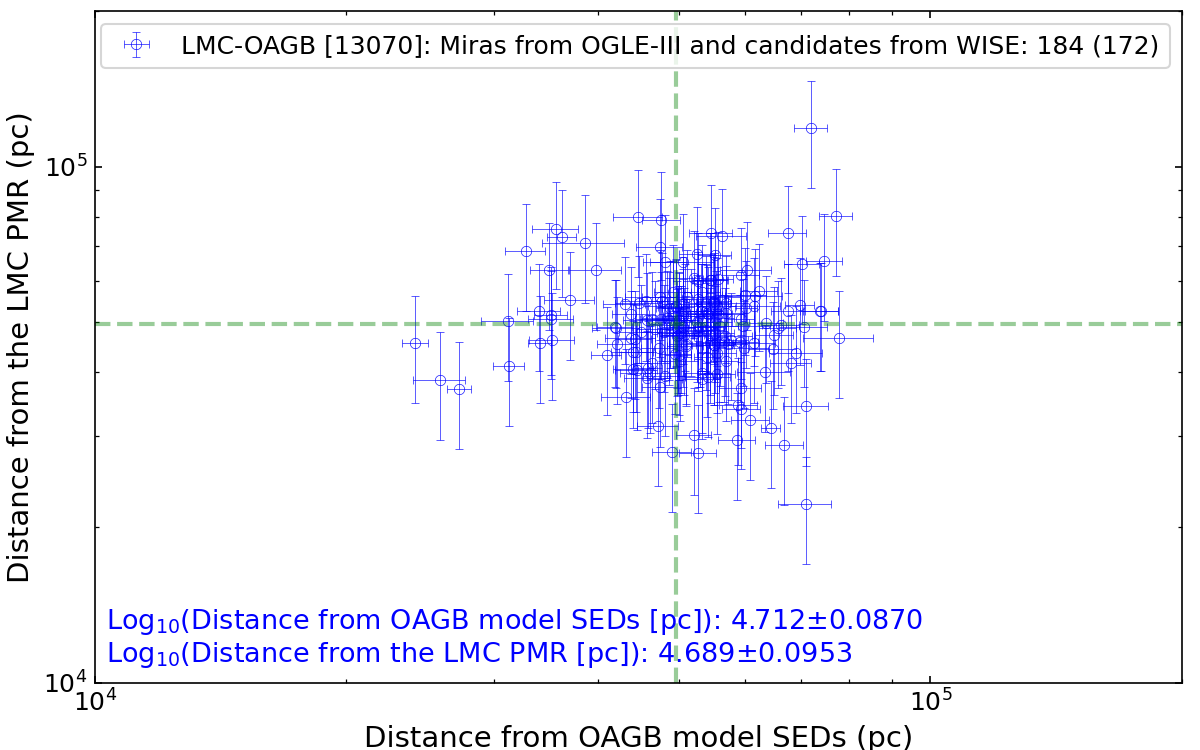}\caption{The top two panels compare distances
derived from the LMC based PMR with those from Gaia DR3 for Mira-type Galactic OAGB stars.
The bottom panel compares distances derived from the LMC-based PMR with
those obtained from OAGB model SEDs for Mira-type LMC-OAGB stars.
Red crosses in the upper two panels mark sources with high-quality Gaia astrometry (see Section~\ref{sec:dist-ext}).
The total number of objects in each subgroup is listed, with the numbers in parentheses
representing the subset for which the required data are available and therefore plotted.
See Section~\ref{sec:dpmr}.}
\label{f16}
\end{figure}

\subsection{Distances from the LMC-based PMR}\label{sec:dpmr}

For the 66,460 Galactic OAGB stars identified as Miras with known periods from
AAVSO, the top two panels of Figure~\ref{f16} compare the distances derived from
the LMC-based PMR (Equation~\ref{eq:2}) with the Gaia DR3 distances. The vertical
error bars for the PMR distances represent the uncertainties associated with the
LMC-based PMR.

Among the 15,158 MW-OAGB\_IC objects (upper panel), 80.9\% fall within $\pm$0.3
dex agreement between the two distance estimates. Restricting the comparison to
the stars with high-quality Gaia astrometry (Section~\ref{sec:dist-ext}) raises
this to 89.8\%. When the samples are compared over the same distance range, here
defined by the PMR distance, the agreement improves from 60.1\% to 93.9\% within
2 kpc but decreases from 84.1\% to 78.1\% beyond 2 kpc. Within 2 kpc, most of the
discrepant stars in the full sample have Gaia and SED distances larger than their
PMR distances, indicating that the PMR underestimates the luminosity of these
objects; they may be misclassified Miras or have inaccurate periods in the AAVSO
database, and almost all of them are removed by the quality cut. Beyond 2 kpc, in
contrast, the astrometric selection does not improve the agreement.

Of the 47,926 MW-OAGB\_NI objects (middle panel), 79.0\% fall within this range.
Here the astrometric selection does not improve the overall agreement: for the
stars with high-quality Gaia astrometry it falls to 43.0\%. Within 2 kpc the
agreement improves from 40.8\% to 100\%, although only 11 high-quality stars lie
in this range, whereas beyond 2 kpc it decreases from 79.0\% to 40.2\%. For both
subgroups, the poorer agreement beyond 2~kpc after the quality cut reflects the
limitations of the Gaia data rather than of the PMR distances: at these distances
the parallaxes are comparable to or smaller than their uncertainties, so that the
selection by \texttt{parallax\_over\_error} preferentially retains stars with
underestimated Gaia distances, and many of the high-quality stars, particularly
those of MW-OAGB\_NI, may lie in crowded fields at low Galactic latitude, where
blending can yield formally precise but unreliable parallaxes (see
Section~\ref{sec:sed-dist}).

The PMR distances therefore show the same pattern as the SED distances when
compared with Gaia distances (see the top panels of Figure~\ref{f7}), although
the two are not fully independent, since both rest on the luminosity scale
calibrated with the LMC sample.

For Mira-type LMC-OAGB stars, the bottom panel of Figure~\ref{f16} compares
distances derived from the LMC-based PMR (Equation~\ref{eq:2}) with those
obtained from the OAGB model SEDs. Both methods reproduce the overall trend with
the true distances. These results further support the reliability of the distance
estimates derived from the OAGB model SEDs.

\begin{table}
\centering
\caption{Columns in the catalog of Galactic OAGB stars$^a$ (See Table~\ref{tab:tab1})\label{tab:tab5}}
\begin{tabular}{lll}
\hline \hline
Name  &	Unit	& Description	\\
\hline
Number	        &		& Object number$^a$\\
SIMBAD\_id 	    &		& Main identifier from SIMBAD\\
IRAS$^b$	    &		& IRAS PSC or FSC number\\
RAdeg	        & deg   & Right ascension \\
DEdeg	        & deg   & Declination \\
Gaia\_id	    &       & Gaia DR3 identifier \\
AllWISE\_id	    &       & AllWISE identifier \\
dist\_gaia$^c$  & pc   & Distance from Gaia DR3\\
dist\_sed	    & pc   & Distance from OAGB model SEDs\\
tau10           &      & $\tau_{10}$ from OAGB model SEDs\\
dist\_pmr	    & pc   & Distance from the LMC PMR\\
A\_Type         &      & Variable type from AAVSO\\
A\_Period       & day  & Pulsation period from AAVSO\\
WISE\_Period    & day  & Pulsation period from WISE\\
OH$^d$              &      & OH 1612 MHz maser source\\
OH\_flux$^e$    & Jy   & OH maser peak flux density\\
H$_2$O$^d$          &      & H$_2$O maser source\\
SiO$^d$             &      & SiO maser source\\
\hline
\end{tabular}
\begin{flushleft}
$^a$The catalog consists of two files: one for MW-OAGB\_IC objects
(19,784 rows and 18 columns) and another for MW-OAGB\_NI objects (51,093 rows and 17
columns). The position data (RAdeg, DEdeg; J2000) are from SIMBAD.
$^b$Only for MW-OAGB\_IC objects.
$^c$Distance from Gaia DR3 (\citealt{bailer-jones2021}).
$^d$The OH maser sources are categorized as follows: D: detected at 1612 MHz, N: observed at 1612 MHz but not detected,
blank: not observed at 1612 MHz in the database of \citet{engels2015};
for H$_2$O and SiO: D: detected, N: not detected, blank: not observed.
$^e$Peak flux density of the OH 1612 MHz maser from \citet{engels2015},
the maximum over all 1612 MHz detections of the source; blank when the source is not detected at 1612 MHz in that database
or when no usable (positive) flux density is listed.
\end{flushleft}
\end{table}

\section{A catalog of OAGB stars in the Milky Way}\label{sec:catalog-oagb}

In this study, we assembled a new sample of Galactic OAGB stars and conducted a
detailed analysis of their properties. We present a new catalog of 70,877
Galactic OAGB stars, comprising 19,784 MW-OAGB\_IC objects and 51,093 MW-OAGB\_NI
objects (see Tables~\ref{tab:tab1} and \ref{tab:tab2}). The catalog includes
valuable information such as distance estimates derived from Gaia DR3 parallaxes,
OAGB model SED fitting, and the LMC-based PMR, as well as maser emission data.

Table~\ref{tab:tab5} lists the catalog contents. The catalog consists of two
files: one for MW-OAGB\_IC objects (19,784 rows and 18 columns) and another for
MW-OAGB\_NI objects (51,093 rows and 17 columns). The full catalog data will be
available on doi:
\dataset[10.5281/zenodo.22266771]{https://zenodo.org/uploads/22266771}.

\section{summary}\label{sec:sum}

We investigated the properties of OAGB stars in the LMC and the Milky Way using a
comprehensive analysis of visual, IR, and radio data. Datasets from Gaia DR3,
2MASS, IRAS, ISO, Akari, MSX, WISE, AAVSO, SIMBAD, and the maser emission
database enabled a detailed comparison between observational data and theoretical
models. The comparison involves the use of various SEDs, CMDs, and 2CDs.

Gaia DR3 provides valuable parallax-based distance estimates for many stars
(\citealt{bailer-jones2021}). However, for pulsating OAGB stars, particularly
those in the TP-AGB phase, extended circumstellar dust envelopes often limit the
availability and accuracy of Gaia DR3 distance estimates. In such cases,
distances derived from comparisons between observed and theoretical model SEDs
may offer an alternative and potentially more reliable means of estimating
distances.

For the LMC, we derived a set of best-fitting OAGB models by fitting the observed
SEDs with radiative transfer models, taking advantage of the galaxy's
well-established distance. For Galactic OAGB stars, whose Gaia DR3 parallaxes are
often uncertain, distances were estimated by fitting the observed SEDs with the
OAGB models validated using the LMC sample. For the entire sample of OAGB stars,
the best-fitting OAGB model SEDs (characterized by $\tau_{10}$) and the
corresponding scaling distances (along with their uncertainties) were determined
using a least-squares optimization method, based on the overall shape and
normalization of the observed SEDs.

These distances rest on an empirical relation between SED shape and luminosity
calibrated on the LMC sample, which lacks the most heavily obscured objects,
plausibly through a combination of fewer high-mass progenitors and metallicity
effects (Section~\ref{sec:comoc}). Since AGB stars of all masses pass through
optically thin phases and low-mass stars are far more numerous, the calibrated
luminosities are weighted toward them, and the distances of higher-mass Galactic
OAGB stars in such phases may be underestimated (see Section~\ref{sec:limit}).
Nevertheless, in the LMC, OAGB stars, as well as CAGB stars (\citealt{suh2026}),
with larger dust optical depths clearly show larger luminosities. Therefore,
despite its limitations, the SED fit method remains a useful and reasonable way
to estimate distances for large samples of Galactic AGB stars, given that
alternative methods are either unavailable for most of the sample or have more
serious limitations.

In addition to distances derived from fitting observed SEDs with theoretical OAGB
models, distance estimates were also obtained for Galactic Mira-type OAGB stars
using the PMR calibrated with Mira-type OAGB stars in the LMC. Both the SED-based
and the PMR distances agree well with the Gaia DR3 distances where the latter are
reliable, but the Gaia parallaxes become uncertain beyond 2--3 kpc and therefore
provide only a limited test at large distances. The two estimates are consistent
with each other, although they are not fully independent, since both rest on the
luminosity scale calibrated with the LMC sample.

Utilizing the theoretical distances derived from the OAGB model SEDs, combined
with extinction data (Section~\ref{sec:dist-ext}), enables the determination of
absolute magnitudes and unreddened colors over a wide range of wavelengths for
Galactic OAGB stars. Comparisons of absolute luminosity distributions and CMDs
based on the two distance estimates show that the SED-based distances yield a far
more coherent set of absolute magnitudes for a large sample of Galactic OAGB
stars. This is not a fully independent test, since the expected luminosity range
is itself defined by the OAGB models; it does show, however, that the Gaia-based
magnitudes scatter well beyond the range spanned by the LMC-OAGB stars, whose
distance is independently known.

We presented a variety of CMDs and 2CDs for OAGB stars, together with theoretical
radiative transfer models that reproduce the observed IR distributions well.
\citet{suh2026} found that CAGB stars in the LMC and the Milky Way exhibit
broadly similar IR SEDs, CMDs, and 2CDs, although the LMC sample lacks the most
heavily dust-enshrouded CAGB stars. In contrast, we find a much more pronounced
difference for OAGB stars: heavily obscured objects (i.e., those with large dust
optical depths, $\tau_{10} > 7$) are absent from the LMC-OAGB sample, whereas
they constitute 2.1\% of the MW-OAGB\_IC objects.

These heavily obscured OAGB stars are generally believed to be TP-AGB objects
representing the most massive and/or most evolved members of the OAGB population.
Their absence in the LMC-OAGB sample may reflect a relative deficiency of
high-mass progenitor stars. In addition, the LMC's lower metallicity likely
suppresses silicate dust formation and promotes an earlier transition to the
C-rich phase, further reducing the number of heavily obscured OAGB stars.

We examined the characteristics of OH, H$_2$O, and SiO maser sources associated
with Galactic OAGB stars. All three masers emerge at the start of the AGB phase
and become more common as the stars evolve. However, SiO and H$_2$O masers fade
rapidly by the end of the OAGB phase, while OH masers reach their highest
incidence and remain detectable into the PAGB phase. MW-OAGB\_IC-EOH objects,
which are extreme OH/IR stars with thick dust envelopes and are near the end of
the AGB phase, emit the strongest OH maser fluxes.

We investigated the IR variability of OAGB stars using WISE photometric data
spanning the past 16 years. By conducting a detailed examination of the WISE
light curves, we identified reliable pulsation parameters for 9 LMC-OAGB stars
and 116 Galactic OAGB stars, which may serve as candidates for new Mira
variables. We analyzed the IR properties of known Mira variables among OAGB
stars, finding that their IR colors become increasingly red, and their absolute
magnitudes brighten with longer pulsation periods.

For Galactic OAGB stars identified as Miras from the AAVSO and WISE data,
alternative distances derived from the LMC-based PMR were compared with Gaia DR3
distances. The results show that the overall distributions and correlations are
consistent with those obtained from the OAGB model SEDs.

Finally, we introduced a new catalog of Galactic OAGB stars, providing enhanced
reliability and comprehensive information. This includes distances derived from
Gaia DR3 and model SEDs for OAGB stars, variability parameters from AAVSO and
WISE light curves, and maser emission data. The catalog comprises 70,877 MW-OAGB
stars, categorized into 19,784 MW-OAGB\_IC objects and 51,093 MW-OAGB\_NI objects
within our Galaxy.

\begin{acknowledgments}
I thank the anonymous reviewer for constructive comments and suggestions. This
research was supported by Basic Science Research Program through the National
Research Foundation of Korea (NRF) funded by the Ministry of Education
(RS-2022-NR075638). This research has made use of the SIMBAD database and VizieR
catalogue access tool, operated at CDS, Strasbourg, France. This research has
made use of the NASA/IPAC Infrared Science Archive, which is operated by the Jet
Propulsion Laboratory, California Institute of Technology, under contract with
the National Aeronautics and Space Administration.
\end{acknowledgments}

\section*{software}

A significant portion of the computations and figures presented in this paper
were performed using Python. The analysis extensively utilized the \texttt{NumPy}
\citep{van der Walt2011}, \texttt{Pandas} \citep{McKinney2010}, \texttt{Astropy}
\citep{astropy2013, astropy2018, astropy2022}, \texttt{Matplotlib}
\citep{hunter2007}, and \texttt{SciPy} \citep{virtanen2020} libraries.
\texttt{TOPCAT} \citep{taylor2005} was used for most table-related tasks.


\label{lastpage}

\begin{thebibliography}{}
\bibitem[Abrahamyan et al.(2015)]{abrahamyan2015} Abrahamyan, H. V., Mickaelian,
    A. M., \& Knyazyan, A. V. 2015, A\&C,10, 99
\bibitem[Andriantsaralaza et al.(2022)]{andriantsaralaza2022} Andriantsaralaza,
    M., Ramstedt, S., Vlemmings, W. H. T., \& De Beck, E. 2022, A\&A, 667, A74
\bibitem[Astropy Collaboration et al.(2013)]{astropy2013} Astropy Collaboration,
    Robitaille, T. P., Tollerud, E. J., et al. 2013, A\&A, 558, A33
\bibitem[Astropy Collaboration et al.(2018)]{astropy2018} Astropy Collaboration,
    Price-Whelan, A. M., Sip\H{o}cz, B. M., et al. 2018, AJ, 156, 123
\bibitem[Astropy Collaboration et al.(2022)]{astropy2022} Astropy Collaboration,
    Price-Whelan, A. M., Lim, P. L., et al. 2022, ApJ, 935, 167
\bibitem[Bailer-Jones et al.(2021)]{bailer-jones2021} Bailer-Jones, C. A. L.,
    Rybizki, J., Fouesneau, M., Demleitner, M., \& Andrae, R. 2021, AJ, 161, 147
\bibitem[Beichman et al.(1988)]{beichman1988} Beichman, C. A., Neugebauer, G.,
    Habing H., Clegg, P. E., \& Chester, T. C. 1988, IRAS Catalogs and Atlases:
    Explanatory Supplement, NASA RP-1190 (Washington: NASA)
\bibitem[B\'eguin et al.(2024)]{beguin2024} B\'eguin, E., Chiavassa, A., Ahmad,
    A., Freytag, B., \& Uttenthaler, S. 2024, A\&A, 690, A125
\bibitem[Bhattacharya et al.(2024)]{bhattacharya2024} Bhattacharya, R., Medina,
    B. M., Pihlström, Y. M., et al. 2024, ApJ, 969, 109
\bibitem[Bovy et al.(2016)]{bovy2016} Bovy, J., Rix, H.-W., Green, G. M.,
    Schlafly, E. F., \& Finkbeiner, D. P. 2016, ApJ, 818, 130
\bibitem[Chiavassa et al.(2018)]{chiavassa2018} Chiavassa, A., Freytag, B., \&
    Schultheis, M. 2018, A\&A, 617, L1
\bibitem[Cho \& Kim(2012)]{cho2012} Cho, C.-Y., \& Kim, J. 2012, AJ, 144, 129.
\bibitem[Cho et al.(2017)]{cho2017} Cho, C.-Y., Cho, S.-H., Kim, S., Kim, J.,
    Yoon, D.-H., \& Hwang, J. 2017, ApJS, 232, 13.
\bibitem[Cohen et al.(2003)]{cohen2003} Cohen, M., Wheaton, W. A., \& Megeath, S.
    T. 2003, AJ, 126, 1090
\bibitem[De Beck et al.(2010)]{debeck2010} De Beck, E., Decin, L., de Koter, A.,
    et al. 2010, A\&A, 523, A18
\bibitem[Deguchi et al.(2010)]{deguchi2010} Deguchi, S., Shimoikura, T., \&
    Koike, K. 2010, PASJ, 62, 525.
\bibitem[Deguchi et al.(2012)]{deguchi2012} Deguchi, S., Sakamoto, T., \&
    Hasegawa, T. 2012, PASJ, 64, 4.
\bibitem[Dehaes et al.(2007)]{dehaes2007} Dehaes, S., Groenewegen, M. A. T.,
    Decin, L., et al. 2007, MNRAS, 377, 931
\bibitem[Di Criscienzo et al.(2016)]{dicriscienzo2016} Di Criscienzo, M.,
    Ventura, P., Garc\'ia-Hern\'andez, D. A. et al. 2016, MNRAS, 462, 395
\bibitem[Drimmel et al.(2003)]{drimmel2003} Drimmel, R., Cabrera-Lavers, A., \&
    L\'opez-Corredoira, M. 2003, A\&A, 409, 205
\bibitem[Egan et al.(2003)]{egan2003} Egan, M. P., Price, S. D., \& Kraemer, K.
    E. 2003, The Midcourse Space Experiment Point Source Catalog v2.3 (2003
    October) Air Force Research Lab. Tech. Rep. AFRL-VS-TR-2003-1589
    (Springfield, VA: NTIS)
\bibitem[Elia et al.(2017)]{elia2017} Elia, D., Molinari, S., Schisano, E., et
    al. 2017, MNRAS, 471, 100
\bibitem[Engels et al.(1983)]{engels1983} Engels, D., Kreysa, E., Schultz, G. V.,
    \& Sherwood, W. A. 1983, A\&A, 124, 123
\bibitem[Engels \& Bunzel(2015)]{engels2015} Engels, D., \& Bunzel, F. 2015,
    A\&A, 582, A68
\bibitem[Engels et al.(2024)]{engels2024} Engels, D., Etoka, S.,
    Jim\'{e}nez-Esteban, F. et al. 2024 IAU Symp., 376, 328
\bibitem[Etoka \& Engels(2022)]{etoka2022} Etoka, S. \& Engels, D. 2022, in
    European VLBI Network Mini-Symposium and Users’ Meeting 2021 (EVN2021), 12
\bibitem[Frew et al.(2013)]{frew2013} Frew, D. J., Boji{\v{c}}i{\'c}, I. S., \&
    Parker, Q. A. 2013, MNRAS, 431, 2
\bibitem[González-Santamaría et al.(2021)]{gonzales2021} González-Santamaría, I.,
    Manteiga, M., Manchado, A., et al. 2021, A\&A, 656, A51
\bibitem[Green et al.(2019)]{green2019} Green, G. M., Schlafly, E., Zucker, C.,
    Speagle, J. S., \& Finkbeiner, D. 2019, ApJ, 887, 93
\bibitem[H{\"o}fner \& Olofsson(2018)]{hofner2018} H{\"o}fner, S., \& Olofsson,
    H. 2018, A\&A Rev., 26, 1
\bibitem[Hunter (2007)]{hunter2007} Hunter, J. D. 2007, CSE, 9, 90
\bibitem[Iben \& Renzini(1983)]{iben1983} Iben, I., \& Renzini, A. 1983, ARA\&A,
    21, 271
\bibitem[Iwanek et al.(2021)]{iwanek2021} Iwanek, P., Soszyński, I., Kozłowski,
    S., 2021, ApJ, 919, 99
\bibitem[Jarrett et al.(2011)]{jarrett2011} Jarrett, T. H., Cohen, M., Masci, F.
    et al. 2011, ApJ, 735, 112
\bibitem[Jiménez-Arranz et al.(2023)]{Jiménez-Arranz2023} Jiménez-Arranz, Ó.,
    Romero-Gómez, M., Luri, X., et al. 2023, A\&A, 669, A91
\bibitem[Jim\'{e}nez-Esteban et al.(2021)]{jimenez2021} Jim\'{e}nez-Esteban, F.
    M., Engels, D., Aguado D. S., Gonz\'{a}lez, J. B., Garc\'{i}a-Lario, P.,
    2021, MNRAS, 506, 6051
\bibitem[Jones et al.(2017)]{jones2017} Jones, O. C., Woods, P. M., Kemper, F. et
    al. 2017, MNRAS, 470, 3250
\bibitem[Justtanont et al.(2015)]{justtanont2015} Justtanont, K., Barlow, M. J.,
    Blommaert, J., et al. 2015, A\&A, 578, A115
\bibitem[Karakas(2014)]{karakas2014} Karakas, A. I. 2014, MNRAS, 445,
    347
\bibitem[Kerber et al.(2003)]{kerber2003} Kerber, F., Mignani, R., Guglielmetti,
    F., \& Wicenec, A. 2003, A\&A, 408, 1029
\bibitem[Kim et al.(2010)]{kim2010} Kim, J., Cho, S.-H., \& Oh, C. S., Byun,
    D.-Y. 2010, ApJS, 188, 209.
\bibitem[Kim et al.(2013)]{kim2013} Kim, J., Cho, S.-H., \& Kim, S. J. 2013, AJ,
    145, 22.
\bibitem[Kohoutek(2001)]{kohoutek2001} Kohoutek, L. 2001, A\&A, 378, 843
\bibitem[Kwok et al.(1997)]{kwok1997} Kwok, S., Volk, K., \& Bidelman, W. P.
    1997, ApJS, 112, 557
\bibitem[Kwon \& Suh(2010)]{ks2010} Kwon, Y.-J., \& Suh, K.-W. 2010, JASS, 27,
    279
\bibitem[Kwon \& Suh(2012)]{ks2012} Kwon, Y.-J., \& Suh, K.-W. 2012, JKAS, 45,
    139
\bibitem[Lallement et al.(2022)]{lallement2022} Lallement, R., Vergely, J. L.,
    Babusiaux, C., \& Cox, N. L. J. 2022, A\&A, 661, A147
\bibitem[Lan{\c c}on \& Wood(2000)]{lancon2000} Lan{\c c}on, A., \& Wood, P. R.
    2000, \aaps, 146, 217
\bibitem[Le Bertre(2005)]{LeBertre2005} Le Bertre, T., Tanaka, M., Yamamura, I.,
    Murakami, H., \& MacConnell, D. J. 2005, PASP, 117, 199
\bibitem[Lebzelter et al.(2023)]{lebzelter2023} Lebzelter, T., Mowlavi, N.,
    Lecoeur-Taibi, I., et al. 2023, A\&A, 674, A15
\bibitem[López Martí et al.(2025)]{lopezmarti2025} López Martí, B.,
    Jiménez-Esteban, F. M., Engels, D., \& García-Lario, P. 2025, A\&A, 698, A109
\bibitem[McDonald et al.(2025)]{mcdonald2025} McDonald, I., Srinivasan, S.,
    Scicluna, P., et al. 2025, MNRAS, 541, 516
\bibitem[Marshall et al.(2006)]{marshall2006} Marshall, D. J., Robin, A. C.,
    Reyl\'e, C., Schultheis, M., \& Picaud, S. 2006, A\&A, 453, 635
\bibitem[McKinney (2010)]{McKinney2010} McKinney, W. 2010, in Proc. 9th Python in
    Science Conf., Data Structures for Statistical Computing in Python, ed. S.
    van der Walt \& J. Millman, 51
\bibitem[Mainzer et al.(2014)]{mainzer2014} Mainzer, A., Bauer, J., Cutri, R. M.,
    et al. 2014, ApJ, 792, 30
\bibitem[Messineo et al.(2018)]{messineo2018} Messineo, M., Habing, H. J.,
    Sjouwerman, L. O., Omont, A., \& Menten, K. M. 2018, A\&A, 619, 35
\bibitem[Miszalski et al.(2008)]{miszalski2008} Miszalski, B., Parker, Q. A.,
    Acker, A., et al. 2008, MNRAS, 384, 525
\bibitem[Murakami et al.(2007)]{murakami2007} Murakami, H., Baba, H., Barthel,
    P., et al. 2007, PASJ, 59, 369
\bibitem[Nakagawa et al.(2026)]{nakagawa2026} Nakagawa, A., Kurayama, T., Sudou,
    H., \& Orosz, G. 2026, Galaxies, 14, 26
\bibitem[Nyman et al.(1998)]{nyman1998} Nyman, L.-A., Hall, P. J., \& Olofsson,
    H. 1998, A\&AS, 127, 185
\bibitem[Parker et al.(2006)]{parker2006} Parker, Q. A., Acker, A., Frew, D. J.,
    et al. 2006, MNRAS, 373, 79
\bibitem[Parker et al.(2016)]{parker2016} Parker, Q. A., Boji{\v{c}}i{\'c}, I.
    S., \& Frew, D. J. 2016, J. Phys. Conf. Ser., 728, 032008
\bibitem[Pietrzyński et al.(2019)]{pietrzynski2019} Pietrzyński, G., Graczyk,
    D., Gallenne, A., et al. 2019, Nature, 567, 200
\bibitem[Ramos-Medina et al.(2018)]{ramos-medina2018} Ramos-Medina, J., Sánchez
    Contreras, C., García-Lario, P., et al. 2018, A\&A,611, A41
\bibitem[Riebel et al.(2012)]{riebel2012} Riebel, D., Srinivasan, S., Sargent,
    B.,\& Meixner, M. 2012, ApJ, 753, 71
\bibitem[Rimoldini et al. (2023)]{rimoldini2023} Rimoldini, L., Holl, B., Gavras,
    P., et al. 2023, A\&A, 674, A14
\bibitem[Rizzo et al.(2021)]{rizzo2021} Rizzo, J. R., Cernicharo, J., \&
    García-Miró, C. 2021, ApJS, 253, 44.
\bibitem[Skowron et al.(2021)]{skowron2021} Skowron, D. M., Skowron, J.,
    Udalski, A., et al. 2021, ApJS, 252, 23
\bibitem[Soszy{\'n}ski et al.(2009)]{sus09} Soszy{\'n}ski, I., Udalski, A.,
    Szyma{\'n}ski, M.~K., et al. 2009,  AcA, 59, 239
\bibitem[Soszy{\'n}ski et al.(2013)]{swu13} Soszy{\'n}ski, I., Wood, P. R.,
    Udalski, A. 2013, ApJ, 779, 167
\bibitem[Srinivasan et al.(2009)]{srinivasan2009} Srinivasan, S., Meixner,
    M., Leitherer, C., et al. 2009, AJ, 137, 4810
\bibitem[Stroh et al.(2019)]{stroh2019} Stroh, M. C., Pihlstr\"{o}m, Y. M.,
    Sjouwerman, L. O., et al. 2019, ApJS, 244, 25
\bibitem[Suh(1999)]{suh1999} Suh, K.-W. 1999, MNRAS, 304, 389
\bibitem[Suh(2000)]{suh2000} Suh, K.-W. 2000, MNRAS, 315, 740
\bibitem[Suh(2004)]{suh2004} Suh, K.-W. 2004, ApJ, 615, 485
\bibitem[Suh(2015)]{suh2015} Suh, K.-W. 2015, ApJ, 808, 165
\bibitem[Suh(2016)]{suh2016} Suh, K.-W. 2016, ApJ, 819, 61
\bibitem[Suh(2020)]{suh2020} Suh, K.-W. 2020, ApJ, 891, 43
\bibitem[Suh(2021)]{suh2021} Suh, K.-W. 2021, ApJS, 256, 43
\bibitem[Suh(2024a)]{suh2024a} Suh, K.-W. 2024a, ApJS, 272, 16
\bibitem[Suh(2024b)]{suh2024b} Suh, K.-W. 2024b, JKAS, 57, 123
\bibitem[Suh(2025)]{suh2025} Suh, K.-W. 2025, JKAS, 58, 131
\bibitem[Suh(2026)]{suh2026} Suh, K.-W. 2026, ApJ, 996, 26
\bibitem[Suh \& Kwon(2013)]{sk2013} Suh, K.-W., \& Kwon, Y.-J. 2013, ApJ, 762,
    113
\bibitem[Sylvester et al.(1999)]{sylvester1999} Sylvester, R. J., Kemper, F.,
    Barlow, M. J., et al. 1999, A\&A, 352, 587
\bibitem[Taylor(2005)]{taylor2005} Taylor, M. B. 2005, in ADASS XIV, edited
    by P. Shopbell, M. Britton, \& R. Ebert, vol. 347 of ASP Conf. Ser., 29
\bibitem[VanderPlas(2018)]{vanderPlas2018} VanderPlas, J. T. 2018, ApJS, 236, 16
\bibitem[van der Walt et al.(2011)]{van der Walt2011} van der Walt, S., Colbert,
    S. C., \& Varoquaux, G. 2011, CSE, 13, 22
\bibitem[Vergely et al.(2022)]{vergely2022} Vergely, J. L., Lallement, R., \&
    Cox, N. L. J. 2022, A\&A, 664, A174
\bibitem[Ventura et al.(2018)]{ventura2018} Ventura, P., Karakas, A., Dell'Agli,
    F., Garc\'ia-Hern\'andez, D. A., \& Guzman-Ramirez, L. 2018, MNRAS, 475, 2282
\bibitem[Virtanen et al.(2020)]{virtanen2020} Virtanen, P. et al., 2020,
    Nat. Methods, 17, 261
\bibitem[Wang \& Chen(2019)]{wang2019} Wang, S., \& Chen, X. 2019, ApJ, 877, 116
\bibitem[Watson et al.(2006)]{watson2006} Watson, C., Henden, A. A., \& Price, A.
    2006, SASS, 25, 47 (VizieR Online Data Catalog: B/vsx/vsx)
\bibitem[Wu et al.(2018)]{wu2018} Wu, Y. W., Natsunaga, N., Burns, R. A., \&
    Zhang, B. 2018, MNRAS, 473, 3325.
\bibitem[Wu et al.(2022)]{wu2022} Wu, Y. W., Zhang, B., Li, J., \& Zheng, X.-W.
    2022, MNRAS, 516, 1881.
\bibitem[Yung et al.(2014)]{yung2014} Yung, B. H. K., Nakashima, J.-I., \&
    Henkel, C. 2014, ApJ, 794, 81.
\bibitem[Yoon et al.(2014)]{yoon2014} Yoon, D.-H., Cho, S.-H., Kim, J., Yun, Y.,
    \& Park, Y.-S. 2014, ApJS, 221, 15.
\bibitem[Zechmeister \& K{\"u}rster (2009)]{zechmeister2009} Zechmeister, M., \&
    K{\"u}rster, M. 2009, A\&A, 496, 577
\end{thebibliography}
\end{document}